%% file: main_current.tex
\documentclass[11pt]{article}
\usepackage[letterpaper,margin=1in]{geometry}
\usepackage[T1]{fontenc}
\usepackage{lmodern,microtype}
\usepackage{amsmath,amssymb,amsthm,mathtools,bm}
\usepackage{aliascnt}
\usepackage{booktabs,array,enumitem}
\usepackage{needspace}
\usepackage{xcolor}
\usepackage{tikz}
\usetikzlibrary{arrows.meta,calc,positioning}
\usepackage{caption}
\usepackage{todonotes}
\definecolor{linkblue}{RGB}{22,52,89}
\usepackage[colorlinks=true,linkcolor=linkblue,citecolor=linkblue,urlcolor=linkblue,bookmarksnumbered=true]{hyperref}
\hypersetup{pdftitle={Addressable transversal CCZ gates on almost good quantum LDPC codes and nearly optimal magic state protocols},pdfsubject={Transversal CCZ gates and magic state protocols from balanced products; conditional on uniform subfield product expansion},pdfkeywords={quantum LDPC codes, transversal CCZ, product expansion, balanced products, cup products, cellular sheaves}}
\usepackage[nameinlink,noabbrev]{cleveref}
\usepackage{fancyhdr}
\setlist{topsep=4pt,itemsep=2pt,parsep=0pt,leftmargin=*}
\allowdisplaybreaks[2]

\definecolor{accentc}{HTML}{0B6B6B}
\definecolor{goldc}{HTML}{996B12}
\definecolor{dira}{HTML}{2563EB}
\definecolor{dirb}{HTML}{C2410C}
\definecolor{dirc}{HTML}{7C3AED}
\definecolor{goodc}{HTML}{15803D}
\definecolor{badc}{HTML}{B32020}
\definecolor{panelc}{HTML}{F4F4EF}
\definecolor{linec}{HTML}{C9C6BD}
\definecolor{mutedc}{HTML}{5D6871}
\tikzset{
  fbox/.style={draw=linec,fill=panelc,rounded corners=1.5mm,align=center,
               font=\small\sffamily,inner sep=5pt},
  flow/.style={-{Stealth[length=2.6mm]},draw=mutedc,line width=0.9pt},
  lab/.style={font=\small\sffamily,text=black},
  sml/.style={font=\footnotesize\sffamily,text=mutedc},
  smk/.style={font=\footnotesize\sffamily,text=black}
}

\newif\ifmarkedchanges
\markedchangesfalse

\newtheorem{theorem}{Theorem}[section]
\newaliascnt{lemma}{theorem}
\newtheorem{lemma}[lemma]{Lemma}
\aliascntresetthe{lemma}
\newaliascnt{proposition}{theorem}
\newtheorem{proposition}[proposition]{Proposition}
\aliascntresetthe{proposition}
\newaliascnt{corollary}{theorem}
\newtheorem{corollary}[corollary]{Corollary}
\aliascntresetthe{corollary}
\newaliascnt{assumption}{theorem}
\newtheorem{assumption}[assumption]{Assumption}
\aliascntresetthe{assumption}
\theoremstyle{definition}
\newaliascnt{definition}{theorem}
\newtheorem{definition}[definition]{Definition}
\aliascntresetthe{definition}
\newaliascnt{condition}{theorem}
\newtheorem{condition}[condition]{Condition}
\aliascntresetthe{condition}
\newaliascnt{example}{theorem}
\newtheorem{example}[example]{Example}
\aliascntresetthe{example}
\theoremstyle{remark}
\newaliascnt{remark}{theorem}
\newtheorem{remark}[remark]{Remark}
\aliascntresetthe{remark}
\crefname{assumption}{Assumption}{Assumptions}
\Crefname{assumption}{Assumption}{Assumptions}
\crefname{condition}{Condition}{Conditions}
\Crefname{condition}{Condition}{Conditions}
\newcommand{\F}{\mathbb F}
\newcommand{\C}{\mathbb C}
\newcommand{\Z}{\mathbb Z}
\newcommand{\eps}{\varepsilon}
\newcommand{\one}{\mathbf 1}
\newcommand{\calA}{\mathcal A}
\newcommand{\calV}{\mathcal V}
\newcommand{\calK}{\mathcal K}
\newcommand{\calS}{\mathcal S}
\newcommand{\shf}{\mathcal F}
\newcommand{\Tan}{\operatorname{Tan}}
\newcommand{\polylog}{\operatorname{polylog}}
\newcommand{\CCZ}{\operatorname{CCZ}}
\DeclareMathOperator{\RS}{RS}

\DeclareMathOperator{\Span}{span}

\DeclareMathOperator{\im}{im}
\DeclareMathOperator{\Tr}{Tr}
\DeclareMathOperator{\ev}{ev}
\DeclareMathOperator{\subrank}{subrank}
\newcommand{\srk}{\mathbf{s}}
\newcommand{\gb}[1]{\boldsymbol{#1}}
\DeclareMathOperator{\diag}{diag}
\newcommand{\ket}[1]{\lvert #1\rangle}
\newcommand{\bra}[1]{\langle #1\rvert}
\newcommand{\braket}[2]{\langle #1\,|\,#2\rangle}

\newcommand{\bwt}[1]{\lvert #1\rvert_{\mathrm{blk}}}

\newcommand{\inner}[2]{\langle #1,#2\rangle}
\title{Addressable transversal CCZ gates on almost good quantum LDPC codes and nearly optimal magic state protocols
\\[0.3em]}
\author{%
  Shi Jie Samuel Tan\textsuperscript{\textdagger,\textdaggerdbl}\qquad
  Guanyu Zhu\textsuperscript{\textdaggerdbl}\\[0.4em]
  \small\textsuperscript{\textdagger}University of Maryland\\
  \small\textsuperscript{\textdaggerdbl}IBM Quantum
}
\hypersetup{pdfauthor={Shi Jie Samuel Tan and Guanyu Zhu}}
\date{\today}
\makeatletter
\renewcommand{\@maketitle}{\begin{center}
{\LARGE\@title\par}\vspace{0.7em}
{\normalsize\begin{tabular}[t]{c}\@author\end{tabular}\par}\vspace{0.5em}
{\normalsize\@date\par}
\end{center}\vspace{0.2em}}
\makeatother

\begin{document}
\maketitle
\begingroup
\renewcommand{\thefootnote}{}
\footnotetext{The authors are listed in alphabetical order.}
\endgroup
\begin{abstract}
Assuming uniform subfield product expansion, we construct quantum
low-density parity-check (qLDPC) codes with
distances $\Omega_\eps(N/\log^2N)$ and at least $N^{1-\eps}$
individually addressable logical CCZ gates on pairwise disjoint
triples of logical qubits, for every fixed $0<\eps<1$.
With all logical qubits outside these triples initialized to $\ket0$,
any subset of the corresponding logical CCZ gates can be implemented with a transversal physical CCZ circuit. As
applications, we obtain two nearly optimal magic state protocols:
CCZ distillation with arbitrarily small overhead exponent $\gamma \rightarrow 0$
and error suppression of nearly linear order, and a magic state fountain that
directly prepares $N^{1-\eps}$ encoded CCZ magic states with nearly linear distance $\Omega_\eps(N/\log^2N)$. The
construction uses balanced products of expanding graph covers. The
assumption concerns Reed--Solomon codes on three
subfields of characteristic two and pairwise coprime extension degrees,
a local hypothesis studied in a companion manuscript.
We design the expanding graph covers to admit many common classical
codewords, which yield logical operators whose pairings select the
logical CCZ gates independently.
Without the product expansion assumption, the graph covers also
yield classical LDPC codes over a fixed field with almost linear
dimension and linear distance that satisfy a multiplication property:
the coordinatewise product of any three codewords lies in a second
LDPC code of linear distance.
\end{abstract}

{\small\tableofcontents}

\input{sections/intro}
\input{sections/prelims}
\input{sections/expander}
\input{sections/trilinear}
\input{sections/distance}
\input{sections/code_params}

\input{sections/appendix_reed_solomon}
\input{sections/appendix_fourier_transform}
\input{sections/appendix_bloch}
\input{sections/appendix_balanced_product_cup_product}

\bibliographystyle{alpha}
\bibliography{mybib_merge,ref}

\end{document}

%% file: sections/intro.tex
\section{Introduction}
Recent developments in quantum low-density parity-check (qLDPC) codes enable low-overhead
quantum memory
\cite{fiberbundlecode21,hastingswr21,BE,Bravyi:2024wc,BhardwajEtAl2026Mitten,Hong2026,Kasai2026Orthogonality},
including constant-rate families with linear distance
\cite{pkldpc22,Quantum_tanner,guefficient22,dhlv23}.
Lin and Hsieh~\cite{lh22} obtain another such family from two-sided
lossless expanders with a free group action, constructed
in~\cite{HsiehEtAl2025Lossless}.
A key remaining challenge is to achieve low spacetime overhead for
universal fault-tolerant logical gates.

Logical Clifford operations on high-rate qLDPC codes can be
implemented through code surgery
\cite{cohen22,CowtanBurton2024Surgery,cross2024improved,williamson2024low,swaroop2024universal,IdeEtAl2025Homological,CowtanEtAl2026Parallel,HeEtAl2025Extractors,BaspinEtAl2025Surgery, chang2026constanttime}.
Code symmetries also give automorphism, transversal, and fold-transversal logical
gates
\cite{Quintavalle:2022fold,breuckmann2022fold,BerthusenEtAl2025Automorphism,MalcolmEtAl2026Computing,SayginelEtAl2025Automorphism, chakraborty2026no,
KohEtAl2026Automorphism},
while chain maps describe logical gates and homomorphic measurements
\cite{huang2023homomorphic,BenhemouBerthusen2026ChainMaps,SahayLongKhemani2026ChainMaps}.
For product codes, explicit bases of logical operators can help describe
how gates and measurements act on the encoded logical qubits
\cite{xu2024fast,Lee2026LogicalSpectroscopy,ZhengZhengJiangXu2026}.
Code switching between product codes gives access to complementary
gate sets \cite{li2025transversal} and supports single-shot
implementations of gates on selected logical qubits
\cite{tan2025single,golowich2025constant}.

While product codes are natural candidates for high-rate qLDPC
codes, their structure can restrict non-Clifford gates
\cite{burtonbrowne21}.
For example, 2D hypergraph product (HGP) codes of distance at least three
admit no non-Clifford logical gate implemented solely by single-qubit
physical gates \cite[Theorem~39]{fu2025no}. On the other hand, transversal non-Clifford gates, including CCZ and T gates, have been constructed for three-dimensional color and
surface codes \cite{Bombin:2007eh,Kubica:2015br,Vasmer2019}. 
The physical origin of such gates was later understood as sweeping gauged symmetry-protected topological (SPT) defects or equivalently higher symmetries in topological quantum field theory (TQFT) \cite{Yoshida_gate_SPT_2015,  Yoshida2017387,   Webster_gates_2018, barkeshli2023codimension, barkeshli2022higher}. The TQFT connection then inspired a formalism of transversal logical CCZ gates in 3D surface codes and color codes using cup products in algebraic topology \cite{barkeshli2023codimension}, which was then applied to high-rate qLDPC codes built on 3-manifolds in \cite{zhu2023non}.
The cup-product formalism can be extended to codes built on more general chain complexes beyond  manifold triangulations, including hypergraph, homological and balanced product codes
\cite{GL,lin2024transversal,breuckmann2024cups, zhu2025topological,zhu2025transversal, tan2025single, gulshen2025quantum, li2025poincar, LiEtAl2026Cohomological, LLL, GTZ}.
Finite length constructions have also been shown to support these gates for
magic state preparation
\cite{jacob2025single,menon2025magic,TiewBreuckmann2026,MenonEtAl2026Factories}.
More general cohomology operations, including higher cup products and
higher Pontryagin powers, extend this approach to logical gates
beyond multi-controlled $Z$ gates \cite{Hsin2024:classifying}.
Further extensions include a twisted version of the higher
Pontryagin power \cite{BreuckmannEtAl2026CupsII}.

For the purpose of our work, we focus on the number of logical CCZ gates that qLDPC codes
can implement simultaneously and select independently.
The three-qubit CCZ gate acts by
$\CCZ\ket{x,y,z}=(-1)^{xyz}\ket{x,y,z}$ for $x,y,z\in\F_2$.
For three code blocks, the \emph{logical CCZ yield} $\srk$ counts
disjoint matched triples of logical qubits, one from each block,
on which CCZ gates can be implemented simultaneously.
These gates are \emph{addressable} if exactly the CCZ gates on
any chosen subset of the fixed triples can be implemented.
Our construction requires all logical qubits outside the fixed
set of $\srk$ matched triples to be initialized to $\ket0$.
Li, Li, and Liu~\cite{LLL} obtain nearly linear distance and a
nontrivial transversal logical CCZ gate.
Golowich, Tamo, and Zhu~\cite{GTZ} obtain almost linear addressable
yield and polynomial subsystem distance, with stabilizers of constant weight.

In this paper, we use balanced product codes to combine large
distance with many addressable logical CCZ gates. The balanced
product identifies cells of the product complex that are related
by the group action, reducing the code length. We must retain
sufficiently many logical qubits and construct cup product pairings
that distinguish their matched triples, while keeping the field
and local check sizes fixed. We achieve this by designing expanding
graph covers whose Tanner codes share many multiplicative character
codewords, and by constructing linear maps that return the coefficients
of the chosen cup products. An information set then
allows each matched logical triple to be selected independently.
The resulting codes have bounded stabilizer weight and almost linear
addressable yield. Their nearly linear distance bound assumes a local
product expansion condition. The graph covers also give classical
LDPC codes over a fixed field with almost linear dimension, linear
distance, and a multiplication property, independently of this assumption.

In addition to graph expansion, the distance proof must account
for cancellations between local codewords placed along different
coordinate directions. Product expansion ensures that their sum
can be expressed using few nonzero lines relative to its number
of nonzero entries.

Let $F$ be a finite field of characteristic two. For a nonempty
$J\subseteq\{1,2,3\}$, let $S_i$, $i\in J$, be finite nonempty
coordinate sets. A direction-$i$ line in $\prod_{j\in J}S_j$
is obtained by fixing all coordinates except the $i$th.
In two directions, these lines are rows and columns. Given linear codes
$C_i\subseteq F^{S_i}$, let
$\mathcal Q_i\subseteq F^{\prod_{j\in J}S_j}$ consist of
the arrays whose restriction to every direction-$i$ line
is a codeword of $C_i$.
Write $|A|$ for the number of nonzero entries of an array $A$
and $|A|_i$ for the number of direction-$i$ lines on which
$A$ is not identically zero.
For $\rho>0$, the tuple $(C_i)_{i\in J}$ is \emph{$\rho$-product expanding}
\cite{KalachevPanteleev2023} if
every $A\in\sum_{i\in J}\mathcal Q_i$ has a decomposition
$A=\sum_{i\in J}A_i$, with $A_i\in\mathcal Q_i$, such that
\begin{equation}\label{eq:pe}
 \sum_{i\in J}|S_i|\,|A_i|_i\le\rho^{-1}|A|.
\end{equation}
Each nonzero line in $A_i$ contributes its full length $|S_i|$
to the left-hand side. The bound is required for at least one decomposition
of $A$; other decompositions may use more lines whose contributions
cancel.
For a single code, this condition reduces to relative distance
at least $\rho$.

Our construction uses Reed--Solomon local codes. For a nonempty
$S\subseteq F$ and an integer $1\le k\le|S|$, the
Reed--Solomon code $\RS(S,k)$ consists
of the evaluation vectors $(f(c))_{c\in S}$ of polynomials
$f\in F[U]$ of degree below $k$.
\begin{assumption}[Uniform subfield product expansion]\label{ass:pe}
For every power $P\ge2$ of two and every $0<\nu<1/2$, there is
$\rho=\rho(P,\nu)>0$ with the following property.
Let $m_1,m_2,m_3$ be pairwise coprime positive integers, let
$S_i\subseteq\F_{P^{m_i}}$ have the same positive size $n$,
and let $F$ be any finite field containing these subfields.
For every choice of integers $\nu n\le k_i\le(1-\nu)n$,
every nonempty subtuple of $(\RS(S_1,k_1),\RS(S_2,k_2),\RS(S_3,k_3))$
over $F$ is $\rho$-product expanding. The constant is independent
of $n$, the extension degrees, the evaluation sets, and $F$.
\end{assumption}

The distance bounds for our code construction require both the local codes and their
duals. The dual code consists of the vectors orthogonal to every
codeword under the coordinate dot product over $F$.
For the Reed--Solomon codes in \cref{ass:pe}, the same
product expansion parameter $\rho$ applies after independently
taking duals and rescaling coordinates by nonzero elements of $F$.
Indeed, the dual of $\RS(S_i,k_i)$ is a coordinate rescaling of
$\RS(S_i,n-k_i)$ (\cref{app:classical}), and the complementary
dimensions satisfy the same rate bounds. Tensor products of
coordinate rescalings preserve both entry supports and line
supports in \eqref{eq:pe}.
Because $\rho$ is independent of the local field and code length,
we can fix the required graph expansion before choosing these
parameters (\cref{sec:final-parameters}).

We defer the proof of \cref{ass:pe} to a companion manuscript that is currently in preparation~\cite{PE}.\footnote{
The proof was generated entirely using AI and is
available in the
\href{https://github.com/guanyuzhu2014/Uniform-product-expansion---theory-and-Lean-certification/}
{companion repository}, together with Lean formalizations of
selected algebraic lemmas. We are revising the exposition and
extending the Lean formalization, which does not yet cover the full
argument.
}
We note that Bafna and Vyas formulate a related conjecture for RS codes on multiplicative subgroups of
pairwise coprime orders over a prime field~\cite[Conjecture~4.2]{BV}.
Our assumption concerns subfield evaluation sets and requires a
product expansion constant independent of the containing field $F$,
so their conjecture does not directly imply \cref{ass:pe}.

\Needspace{16\baselineskip}
\begin{theorem}[Main theorem]\label{thm:main}
Assume \cref{ass:pe}.
For every $0<\eps<1$ there are constants $c_\eps,w_\eps>0$ and an infinite family of triples of qubit CSS codes, each block of length $N\to\infty$, such that
\[
 \calK_h\ge N^{1-\eps},\qquad
 D_{X,h},D_{Z,h}\ge c_\eps\frac{N}{(\log N)^2},\qquad h=1,2,3.
\]
Here $D_{X,h}$ and $D_{Z,h}$ are the $X$ and $Z$ distances, and
$\calK_h$ is the number of encoded qubits in block $h$.
Each block has stabilizer generators of weight at most $w_\eps$,
with each physical qubit in at most $w_\eps$ generators.
There exist fixed logical bases, a set of
$\srk\ge N^{1-\eps}$ disjoint matched triples of logical qubits,
one qubit from each block, and a fixed transversal layer of
physical CCZ gates. Each physical gate acts on one qubit from each
block, and each qubit participates in at most one gate.
For every subset $I\subseteq\{1,\ldots,\srk\}$, performing a
corresponding subset of these physical gates implements precisely
the logical CCZs indexed by $I$ on the selected triples of logical
qubits, when all logical qubits outside the $\srk$ selected triples
are initialized to $\ket0$. That is, the $\srk$ matched logical CCZ gates
are individually \emph{addressable} within this fixed transversal layer.
\end{theorem}

Once the field and local construction data for the expanders are fixed as in \cref{prop:algorithm}, a randomized algorithm constructs the check matrices, cochains representing the selected logical qubits, and the transversal gate description in expected
time $N^{O_\eps(1)}$. We leave the fault-tolerant state preparation of the required logical qubits for the transversal logical CCZ gates to future work. 

The codes also give protocols for distilling the CCZ magic state
$\ket{\CCZ}:=\CCZ\ket{+++}$.
We measure distillation overhead by the expected number of noisy
input states consumed per output state. A protocol has overhead
exponent $\gamma$ if this cost is $O(\log^\gamma(1/\delta))$
when each output state has infidelity at most $\delta$, starting from a fixed
sufficiently small input error~\cite{BH}. Outside the qLDPC setting,
recent constructions achieve constant overhead magic state distillation under ideal
Clifford operations~\cite{Wills2024,nguyen2024good, GolowichGuruswami2025}.

Our codes combine arbitrarily small positive overhead exponents
with stabilizer checks of bounded weight. The distillation guarantee
assumes ideal Clifford operations and fault-tolerant implementations
are discussed in \cref{sec:open-questions}.
\Needspace{14\baselineskip}
\begin{theorem}[Nearly optimal magic state distillation]\label{thm:distill}
Assume \cref{ass:pe} and fix $0<\eps<1$.
The codes of \cref{thm:main} give CCZ magic state distillation
protocols with error suppression of order at least
$D_{\mathrm{dist}}:=\min_h D_{Z,h}=\Omega_\eps(N/\log^2N)$
per round. The protocols assume ideal Clifford operations and
independent input states with infidelity at most $p$, below a
threshold that may depend on the chosen code. For each fixed code,
the joint output infidelity relative to $\ket{\CCZ}^{\otimes\srk}$,
conditional on acceptance, is $O_N(p^{D_{\mathrm{dist}}})$ as $p\to0$.
Each round initializes the logical qubits in the $\srk$ selected
triples to $\ket+$ and all other logical qubits to $\ket0$.
It uses at most $N$ noisy CCZ states to implement the transversal
gate layer and accepts only when the measured stabilizer syndrome
is trivial. After ideal Clifford decoding, an accepted round
produces $\srk\ge N^{1-\eps}$ CCZ magic states.
Iterating the routine with fixed code blocks gives overhead exponent
$\gamma\le\eps\,(1+o(1))$, where $o(1)\to0$ as $N\to\infty$
along the family. Choosing $\eps$ sufficiently small therefore
gives arbitrarily small positive overhead exponents.
\end{theorem}

Transversal logical CCZ gates also prepare many encoded magic states \textit{in parallel} 
directly from logical stabilizer states. Such protocols are
called \emph{magic state fountains}
\cite{zhu2023non,zhu2025topological,zhu2025transversal, ZhuKobayashiHsin2026,ZhuKobayashiHsin2026}.
The \emph{magic rate} $r_M:=\srk/N$ is the number of states
prepared per application divided by the length of each code block.
For our construction, one transversal layer of physical CCZ gates gives the following guarantee.
\Needspace{10\baselineskip}
\begin{theorem}[Nearly optimal magic state fountain]\label{thm:fountain}
Assume \cref{ass:pe} and fix $0<\eps<1$.
The codes of \cref{thm:main} give a magic state fountain that
prepares $\srk\ge N^{1-\eps}$ encoded CCZ magic states per
application of the transversal physical CCZ gates, assuming ideal
logical stabilizer state preparation and ideal physical gates.
Initialize the logical qubits in the $\srk$ selected triples to
$\ket+$ and all other logical qubits to $\ket0$, then implement
the physical gates corresponding to all selected logical triples.
The output states remain encoded in the three CSS code blocks,
whose full code Pauli distances are $\Omega_\eps(N/\log^2N)$.
No additional concatenation with a distillation code or decoding
is required.
\end{theorem}

A related construction uses gauging measurements of logical
symmetries to prepare encoded CZ  magic states with almost linear yield ($\Theta(N^{1-\epsilon})$) in codes based on non-Abelian qLDPC codes and the corresponding 
twisted sheaf gauge theory~\cite{SGT}.

\paragraph{Related work.}
The following table compares the check weights, distances,
and logical CCZ yields of several constructions over qubits.
Here $N$ is the length of each code block, and logical yield
counts disjoint matched triples on which CCZ gates can be
implemented simultaneously, with all logical qubits outside
the selected triples initialized to $\ket0$.
A dagger marks a subsystem distance bound protecting a designated
set of logical qubits; unmarked distances protect all logical qubits
of the code.
\begin{center}
\small
\begin{tabular}{@{}llll@{}}
\toprule
Construction & Check weight & Distance & Logical CCZ yield\\\midrule
Golowich--Lin \cite{GL} & $O(\polylog N)$ & $\Omega(N^{1/3}/\polylog N)$ & $\Omega_\eps(N^{1-\eps})$\\
Zhu \cite{zhu2025topological} & $O(1)$ & $\Omega(N^{1/2})^{\dagger}$ & $\Omega(N^{1/2})$\\
Zhu \cite{zhu2025transversal} & $O(1)$ & $\Omega(N^{2/3})$ & $\Omega(N^{1/3})$\\
Golowich--Tamo--Zhu \cite{GTZ} & $O_\eps(1)$ & $\Omega_\eps(N^{(1-\eps)/3})^{\dagger}$ & $\Omega_\eps(N^{1-\eps})$\\
Li--Li--Liu \cite{LLL} & $O(1)$ & $\Omega(N/\log^2N)$ & $\Omega(1)$\\
Here (\cref{ass:pe}) & $O_\eps(1)$ & $\Omega_\eps(N/\log^2N)$ & $\Omega_\eps(N^{1-\eps})$\\
\bottomrule
\end{tabular}
\end{center}
Note that a large logical CCZ yield alone does not imply that arbitrary
subsets of the logical gates can be selected.

The subsystem distance bounds in~\cite{zhu2025topological,GTZ}
apply to selected logical qubits. Logical operators acting on the
remaining logical qubits may have low weight. The $\Omega(N^{2/3})$ distance bound for the construction in~\cite[Corollary~3.1 and Section~V.B]{zhu2025transversal} applies to the full code. The $\Omega(N^{1/2})$ and $\Omega(N^{1/3})$ yield in \cite{zhu2025topological} and \cite{zhu2025transversal}  count disjoint logical triples within a larger circuit of overlapping CCZ gates.
Li, Li, and Liu construct codes with a nontrivial logical CCZ gate,
but do not establish a yield that grows with code length. Their constant-rate guarantee applies to the three blocks combined \cite[Section~5.3]{LLL}.

Golowich, Tamo, and Zhu also obtain addressability and a decoder
\cite[Theorem~1.1 and Appendix~A]{GTZ}. Their subsystem distance
$d$ and yield $\srk$ satisfy $d\srk/N\to\infty$ as $N\to\infty$
for $0<\eps<1/4$. By comparison, taking independent copies of codes
that each support one logical CCZ gives $d\srk\le N$, since
each copy has length at least $d$.

Product expansion also underlies the construction of Bafna and
Vyas~\cite[Corollary~4.5]{BV}, which admits transversal CCZ gates
on the full logical space under their product expansion conjecture.
Their construction uses tensor products of punctured Reed--Solomon
codes over a growing prime field, giving stabilizer generators whose weights are sublinear in the code length for one Pauli type and linear for the other. The subsystem construction has gauge generators of sublinear weight for both $X$ and $Z$ types. Assuming a variant of the
Bafna--Vyas product expansion conjecture, Bafna, Li, and
Nguyen~\cite{BLN} use the distance framework of Dinur,
Lin, and Vidick~\cite{DLV} to obtain good quantum locally
testable codes.

Our construction uses common multiplicative
character codewords on expanding graph covers, together with
linear maps that return the coefficients of the chosen cup products
on the balanced product complex.
The quantum distance bound uses \cref{ass:pe};
the classical codes with a multiplication property are unconditional.

\subsection{Our techniques}

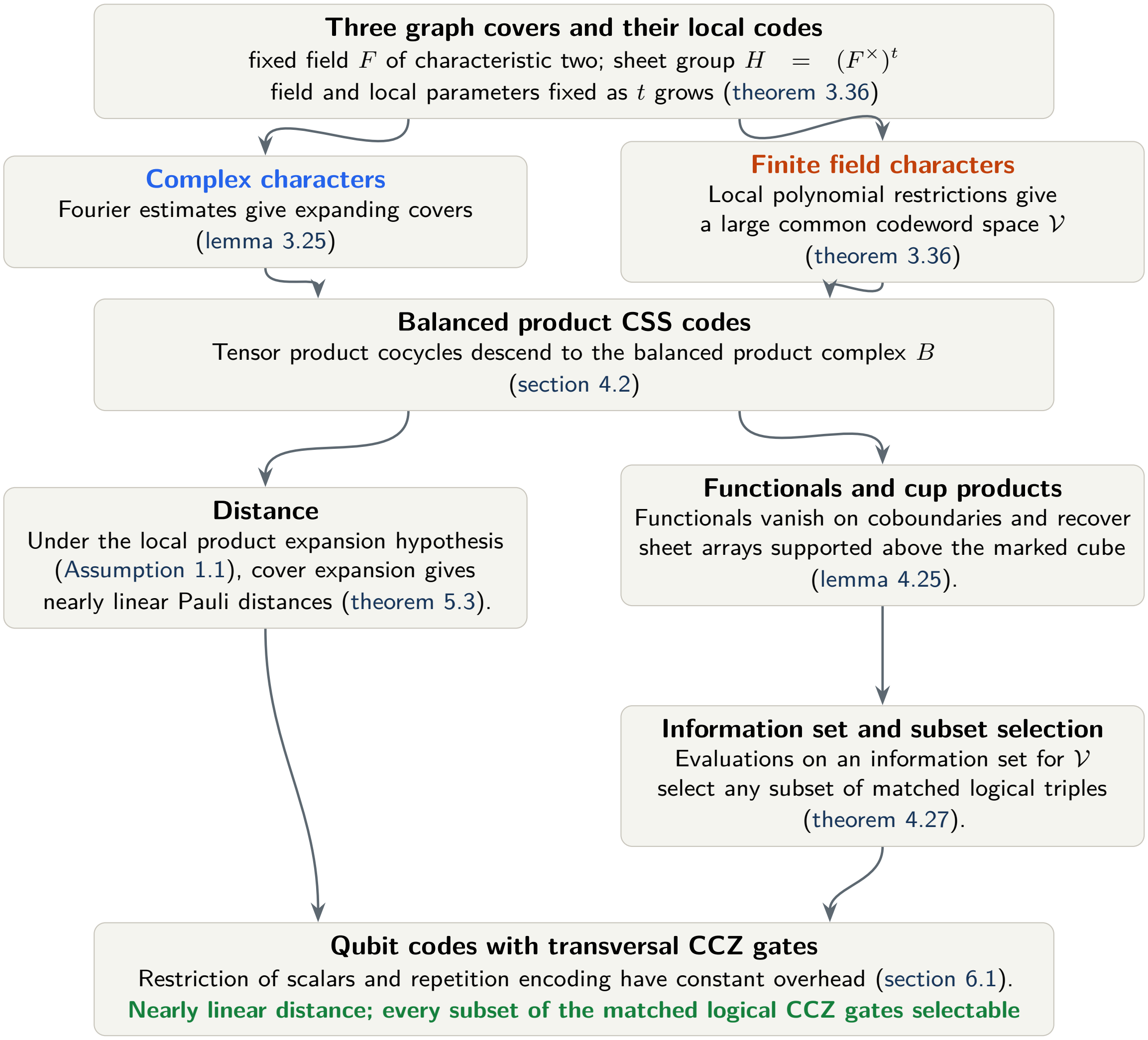
\begin{figure}[htbp]
\centering
\begin{tikzpicture}[x=1cm,y=1cm]
\node[fbox,text width=12.4cm] (top) at (0,0)
  {\textbf{Three graph covers and their local codes}\\
   {\footnotesize fixed field $F$ of characteristic two; sheet group $H=(F^\times)^t$}\\
   {\footnotesize field and local parameters fixed as $t$ grows (\cref{gp:theorem})}};
\node[fbox,text width=6.6cm] (spectral) at (-4.1,-2)
  {\textbf{\color{dira}Complex characters}\\
   {\footnotesize Fourier estimates give expanding covers\\
    (\cref{gp:spectrum})}};
\node[fbox,text width=6.6cm] (characters) at (4.1,-2)
  {\textbf{\color{dirb}Finite field characters}\\
   {\footnotesize Local polynomial restrictions give\\
    a large common codeword space $\calV$\\
    (\cref{gp:theorem})}};
\node[fbox,text width=12.4cm] (codes) at (0,-3.9)
  {\textbf{Balanced product CSS codes}\\
   {\footnotesize Tensor product cocycles descend to the balanced product complex $B$\\
    (\cref{cg:balanced})}};
\node[fbox,text width=6.6cm] (distance) at (-4.1,-6.6)
  {\textbf{Distance}\\
   {\footnotesize Under the local product expansion hypothesis\\
    (\cref{ass:pe}), cover expansion gives\\
    nearly linear Pauli distances (\cref{db:distance}).}};
\node[fbox,text width=6.6cm] (pairing) at (4.1,-6.3)
  {\textbf{Functionals and cup products}\\
   {\footnotesize Functionals vanish on coboundaries and recover\\
    sheet arrays supported above the marked cube\\
    (\cref{cg:keyidentity}).}};
\node[fbox,text width=6.6cm] (selection) at (4.1,-9.5)
  {\textbf{Information set and subset selection}\\
   {\footnotesize Evaluations on an information set for $\calV$\\
    select any subset of matched logical triples\\
    (\cref{cg:diagonal}).}};
\node[fbox,text width=12.4cm] (final) at (0,-12.2)
  {\textbf{Qubit codes with transversal CCZ gates}\\
   {\footnotesize Restriction of scalars and repetition encoding have constant overhead (\cref{db:binary}).}\\
   {\footnotesize\color{goodc}\textbf{Nearly linear distance; every subset of the matched logical CCZ gates selectable}}};
\draw[flow] (top.south) ++(-2.2,0) to[out=-90,in=90] (spectral.north);
\draw[flow] (top.south) ++(2.2,0) to[out=-90,in=90] (characters.north);
\draw[flow] (spectral.south) to[out=-90,in=90] ([xshift=-3.4cm]codes.north);
\draw[flow] (characters.south) to[out=-90,in=90] ([xshift=3.4cm]codes.north);
\draw[flow] (codes.south) ++(-2.2,0) to[out=-90,in=90] (distance.north);
\draw[flow] (codes.south) ++(2.2,0) to[out=-90,in=90] (pairing.north);
\draw[flow] (pairing) -- (selection);
\draw[flow] (distance.south) to[out=-90,in=90] ([xshift=-3.4cm]final.north);
\draw[flow] (selection.south) to[out=-90,in=90] ([xshift=3.4cm]final.north);
\end{tikzpicture}
\caption{Construction of the codes and their logical gates.
Complex characters establish expansion, while finite field
characters supply the common codewords used to construct cocycles
on the balanced product complex. Functionals that vanish on coboundaries
make the resulting phase invariant under $X$ stabilizers.
Recovering the sheet coefficients and evaluating
on an information set makes the pairing diagonal, allowing
independent selection of matched logical triples. The distance
bound uses \cref{ass:pe}. The gate selection uses fixed codes,
logical bases, and physical gate locations, with all logical qubits
outside the fixed set of matched triples initialized to $\ket0$.}
\label{fig:pipeline}
\end{figure}

Tensor product constructions provide explicit logical operators
and cup product pairings that determine logical CCZ gates~\cite{GL,GTZ}.
Forming the balanced product reduces the code length. We must
show that the resulting code still encodes sufficiently many
logical qubits and supports independently selectable gates
\cite[Section~1.2]{GTZ}.
\Cref{fig:pipeline} summarizes the construction.

\subsubsection{Expanding covers with many common codewords}

To obtain both large distance and many logical CCZ gates, we need
expanding graph covers whose Tanner codes contain a large common
space of explicit codewords. An edge assignment is a Tanner
codeword when its values around each vertex belong to a prescribed
local code. We use Reed--Solomon local codes of small dimension
relative to their length, so that evaluations of products of three
local polynomials lie in a Reed--Solomon code of rate below one half.
In this regime, counting the Tanner constraints gives no positive
dimension guarantee~\cite{GL,GTZ}.

\paragraph{Common codewords over a fixed field.}
The covers $\Gamma_1,\Gamma_2,\Gamma_3$ have sheet group
$H=(F^\times)^t$, where $F$ is a
fixed finite field of characteristic two. Their common bipartite
base graph has $t$ vertices on each side. Each base vertex and edge has
copies indexed by $H$; the index $x\in H$ is its sheet label.
The field and local parameters remain fixed as $t$ grows, ensuring constant stabilizer
weight after conversion to qubits.

Affine line restrictions of low total degree polynomials give
large Tanner code subspaces
\cite[Lemmas~4.1 and~5.2]{GL}.
Over a fixed field, a fixed total degree bound gives only
polynomially many independent polynomials in $t$, whereas the
number of sheets grows exponentially.
We therefore use the full basis of multiplicative characters,
namely group homomorphisms $\chi:H\to F^\times$, as candidates:
a character assigns the value $\chi(x)$ to every lifted edge
with sheet label $x$. We impose degree bounds on the characters'
local restrictions. At each base vertex--edge incidence, a group
element multiplies a lifted vertex's sheet label to give the incident
edge's sheet label. We choose these incidence labels so that the
character values around each vertex are, up to a common nonzero
factor, products of powers of affine factors.
Frobenius identities give low degree polynomial
representatives for many exponents on the evaluation subfields.
Column multipliers rescale the entries of the local evaluation vectors.
An averaging argument chooses these multipliers so that a large
set of characters satisfies every local constraint in all three
covers.

\paragraph{Expansion of the same covers.}
The same incidence labels must also give expanding covers.
Fourier decomposition using complex characters of $H$ reduces
adjacency to matrices indexed by the base vertices. Adapting the
character sum approach of \cite{LuWanWangZhang2013}, we use Wan's
bound \cite[Corollary~2.2]{Wan} to control the matrix entries where
character sum cancellation occurs. Random choices of the exponents
defining the incidence labels bound the number of vertices where
this cancellation fails, and
expansion of the base graph bounds the contribution of edges
between those vertices. Organizing the probability estimate by
character order allows the local parameters to remain fixed.
Thus complex characters prove expansion, while finite field
characters supply the common codewords. \Cref{gp:theorem}
establishes these two properties simultaneously.
For any fixed $\zeta>0$, it gives a choice of field and local
parameters for which the common characters, indexed by $\calA$,
span a space
\begin{equation}\label{intro:common-space}
 \calV=\Span_F\{\chi_\alpha:\alpha\in\calA\},
 \qquad \srk=|\calA|=\dim_F\calV\ge |H|^{1-\zeta}.
\end{equation}
We fix such a choice for the logical gate construction below.

These covers also yield classical LDPC codes of almost linear
dimension and linear distance over a fixed field. Coordinatewise
products of three codewords from any one of these codes lie in a
second LDPC code of linear distance
(\cref{gp:classical-multiplication}), independently of \cref{ass:pe}.
Related constructions give these dimension and distance bounds
over a growing alphabet with polylogarithmic parity-check weights
\cite[Section~5, Lemmas~3.31 and~5.2]{GL}, or use a fixed field
and bounded parity-check weights with almost linear dimension
and almost linear distance for both codes
\cite[Theorem~4.1 and proof of Corollary~5.1]{GTZ}.
Classical constructions with Reed--Solomon local constraints and
multiplication also include codes on high dimensional expanders
\cite{DinurLiuZhang2023} and algebraic expander codes
\cite{KoppartyTamo2026}.

\subsubsection{Balanced product codes and addressable logical gates}
\label{sec:techniques-cup}

The common codewords in $\calV$ must give $\srk$
logical qudits in each of three CSS blocks, together with
a pairing that selects any subset of matched triples.
For now, each qudit has computational basis
$\{\ket a:a\in F\}$; we convert to qubits at the end of the construction.
In the cochain description of a CSS code, cocycles label strings
satisfying the $Z$ checks, and $X$ stabilizers identify strings
differing by a coboundary. Thus a \emph{cohomology class}, meaning
a cocycle modulo coboundaries, labels an encoded computational basis
state. We construct explicit cocycles and a trilinear pairing
on their cohomology classes.

\paragraph{Cocycles on the balanced product complex.}
On $\Gamma_i$, a graph 0-cocycle assigns local polynomials whose
scaled evaluations agree at each edge. The accepted character $\chi_\alpha$
gives such a cocycle with edge values $\chi_\alpha(x)$ on sheet $x$.
A graph 1-cochain assigns a field value to each edge. Every such
cochain is a cocycle, since a graph has no 2-cells.
Mark one base edge $b_i$ in each graph. In block $h$, tensor the
1-cochain supported above $b_h$, with values $\chi_\alpha(x)$,
with these 0-cocycles in the other two directions.
The tensor product differential therefore makes the result a
1-cocycle on the Cartesian product (\cref{cg:graph-sec}).
Its sheet dependence is
$\chi_\alpha(x_1)\chi_\alpha(x_2)\chi_\alpha(x_3)
=\chi_\alpha(x_1x_2x_3)$.
We may therefore identify sheet triples with the same product,
obtaining a 1-cocycle $\Phi_h(\chi_\alpha)$ on the balanced
product complex $B$: the restriction maps commute with sheet translations,
so the differential preserves these invariant cochains.
Each cell retains one sheet label
$x=x_1x_2x_3$, reducing the sheet count from $Q^3$ to $Q$,
where $Q=|H|$ (\cref{cg:balanced}). Extending linearly defines
$\Phi_h(G)$ for $G\in\calV$.

\paragraph{Cup products and detecting functionals.}
To compute the logical action, we first multiply the chosen
cocycles using the cup product, denoted by $\smile$. This product sums products of
coefficients on complementary faces of a cube. The local pairings
come from code multiplication, as in~\cite{GL,GTZ,tan2025single}.
Here polynomial multiplication induces coordinatewise
multiplication of the evaluation vectors.
The degree bounds in the graph construction ensure that these
pairings respect the local constraints.
The restriction maps and pairings are the same on every lift,
so the cup product descends to $B$
(\cref{cg:cup-sec}).
For $G_1,G_2,G_3\in\calV$, the cup product of the $\Phi_h(G_h)$
is supported over the marked base cube $(b_1,b_2,b_3)$, with
value $G_1(x)G_2(x)G_3(x)$ on sheet $x$
(\cref{cg:marked-product}).

A nonzero cup product could still be a coboundary, representing
the zero cohomology class. We therefore need functionals that
vanish on coboundaries and recover every coefficient of a cochain
supported above the marked cube. Retaining the sheet coefficients
separately will let us select individual matched triples.
Related constructions use interpolation or code duality to obtain
pairings unchanged by stabilizers~\cite{GL,GTZ}.

Write $R=F[H]$ for the group algebra, whose basis vector
$\gb{x}$ records sheet $x\in H$.
For each graph, the triple cup product uses local codes spanned
by coordinatewise products of local codewords from the three blocks.
With these local codes, we construct a map from graph
1-cochains to $R$ that commutes with sheet translations,
vanishes on coboundaries, and returns the sheet coefficients of
cochains supported above the marked edge.
Fourier decomposition reduces this construction to showing that,
in every character component, the graph coboundary matrix remains
injective after deleting the marked-edge row. Random nonzero
multipliers at base vertex--edge incidences give this condition
simultaneously in all components (\cref{cg:dual-existence}).
In block $h$, these rescalings modify only direction $h$, where
the chosen factor is a graph 1-cochain and hence remains a
cocycle for every choice of restriction maps. The 0-cocycles in
the other two directions are unchanged.
Combining the three graph functionals on the balanced product
gives an $R$-valued
functional $\Lambda$ on 3-cochains that vanishes on coboundaries
and, on a cochain supported above the marked cube, returns its
sheet coefficient array.
Writing the pairing  $\langle\Lambda,u\rangle=\Lambda(u)$, we obtain
\begin{equation}\label{intro:keyid}
\big\langle\,\Lambda,\
\Phi_1(G_1)\smile\Phi_2(G_2)\smile\Phi_3(G_3)\big\rangle
=\sum_{x\in H}G_1(x)G_2(x)G_3(x)\,\gb{x}\in R
\end{equation}
by \cref{cg:keyidentity}. Composing $\Lambda$ with an $F$-linear
map $\ell:R\to F$ gives a scalar functional
$\Lambda_\ell=\ell\circ\Lambda$ that also vanishes on coboundaries.
By the Leibniz rule, a product with one coboundary and two
cocycles is a coboundary. For cocycles $f_h$ in block $h$, it follows that
$T_\ell(f_1,f_2,f_3):=\langle\Lambda_\ell,
f_1\smile f_2\smile f_3\rangle$
is unchanged when a coboundary is added to any input.
Thus $T_\ell$ is well defined on cohomology classes
(\cref{cg:invariance}).

\paragraph{Selecting logical triples.}
To select individual matched triples, choose an information set
for $\calV$~\cite[Claim~3.11]{GTZ}.
The evaluation matrix
$(\chi_\alpha(x))_{\alpha\in\calA,x\in H}$ has full row rank
$\srk$. Choose $\srk$ independent columns, indexed by
$\calS\subseteq H$, and change the row basis so that these columns form the
identity matrix. The resulting basis $(F_z)_{z\in \calS}$ satisfies
$F_z(z')=\one_{z=z'}$ for $z,z'\in \calS$, where
$\one_{\mathcal E}$ is $1$ when the condition $\mathcal E$ holds
and $0$ otherwise.
These functions may be nonzero outside $\calS$
(\cref{cg:infoset-sec}).
For $w\in H$, let $\ev_w:R\to F$ extract the coefficient of
$\gb{w}$.
For $I\subseteq \calS$, define
$\ell_I=\sum_{w\in I}\ev_w$ and $T_I:=T_{\ell_I}$.
Then \eqref{intro:keyid} gives, for $x,y,z\in \calS$,
\begin{equation}\label{intro:diag}
T_I\big(\Phi_1(F_x),\Phi_2(F_y),\Phi_3(F_z)\big)
=\sum_{w\in I}F_x(w)F_y(w)F_z(w)=\one_{x=y=z\in I}.
\end{equation}
Taking $I=\calS$ in \eqref{intro:diag} proves that no nonzero
$F$-linear combination of the cocycles $\Phi_h(F_z)$, $z\in \calS$,
is a coboundary.
Indeed, pairing a proposed dependence with the two inputs indexed
by $z$ in the other blocks extracts its $z$-coefficient, which
must vanish because the pairing annihilates coboundaries.
Their cohomology classes therefore extend to a basis defining
the logical qudits, with $\srk$ selected qudits in each block
(\cref{cg:diagonal}).
Only $\ell_I$ depends on the selected subset $I$; the codes and
logical bases remain fixed.

\paragraph{Distance and transversal implementation.}
For fixed field and local parameters, the qubit conversion below
has constant overhead. The final block length satisfies
\begin{equation}\label{intro:length}
 N=\Theta(t^3|H|),\qquad t=\Theta(\log N),
\end{equation}
by \eqref{eq:final-length} and $|H|=(|F|-1)^t$.
Choosing $0<\zeta<\eps$ in \eqref{intro:common-space} therefore
gives $\srk\ge N^{1-\eps}$ for all sufficiently large $t$.

Under \cref{ass:pe}, graph expansion and local product expansion
let us apply the cocycle and cycle
arguments of Dinur, Lin, and Vidick~\cite{DLV} to bound both
Pauli distances. Choose a cocycle whose support meets as few faces
as possible within its nonzero cohomology class. Product expansion
forces its support to contain many nearby faces, and graph expansion
then forces the total support to be large.
We verify these arguments for local coefficient matrices that
depend on the base vertex (\cref{db:dlv-compatibility}).
For each direction $i$, the graph of moves in that direction
splits into $4t^2$ copies of $\Gamma_i$, one for each pair of
fixed base vertices in the other two directions. The resulting
factor $t^{-2}$ in the relative distance bound is of order
$(\log N)^{-2}$ by \eqref{intro:length} (\cref{db:distance}).

Choose an $\F_2$-linear map $\varphi:F\to\F_2$ with $\varphi(1)=1$.
Restriction of scalars represents each physical qudit by
$\log_2|F|$ qubits and
expresses the phase $(-1)^{\varphi(T_I)}$ as a product of physical
CCZ gates. The binary expansion gives a common list of CCZ gates
for all subsets $I$, with only their activation depending on $I$.
Each qubit occurs in a bounded number of entries in this list,
so repetition encoding assigns a different qubit of its repetition
block to each entry, following~\cite[Lemmas~3.44--3.45]{GL}.
This fixes the codes and transversal gate locations, with constant
overhead because the field and local parameters are fixed.
Using a binary basis of $F$ containing $1$, we retain the logical
qubit associated with $1$ in each of the $\srk$ chosen logical
qudits and initialize all other logical qubits to $\ket0$.
Activating the appropriate locations then implements
the selected matched logical CCZ gates (\cref{addr:layer}).

\Cref{app:bloch} presents the Fourier decomposition and logical
pairing in Dirac's bra--ket notation, with a comparison to Bloch theory.

\clearpage
\subsection{Open questions and future directions}\label{sec:open-questions}
Several questions remain for this balanced product construction.

\paragraph{Logical inputs.}
Can the selected CCZ gates act on the full code space while
leaving all other logical qubits unchanged? We currently require
that all logical qubits outside the fixed set of $\srk$ matched
triples be initialized to $\ket0$;
removing this restriction requires controlling the trilinear form outside the selected
logical subspaces.

\paragraph{Choice of logical triples.}
Can a transversal CCZ act on any choice of one selected logical
qubit from each block, while retaining our distance bounds and
bounded stabilizer weight and degree? This stronger form of
addressability is known for other code families
\cite{he2025quantum,he2025asymptotically}.
This question allows logical qubits outside the fixed set of
$\srk$ matched triples to remain initialized to $\ket0$.

\paragraph{Distance and logical yield.}
Can this construction achieve linear distance and linear CCZ
yield with bounded stabilizer weight and degree?
The current distance estimate loses a factor of $t^2$ from the
connected components of the graphs of moves in a single direction. The gate
construction uses at most $Q=|H|$ characters on one marked base cube, while
$N=\Theta_\eps(t^3Q)$ by \eqref{eq:final-length}.
Even retaining all characters therefore leaves a factor of $t^3$
in the yield. Linear yield requires more independently selectable
logical triples or a change to the construction.

\paragraph{Decoding.}
Can local product expansion yield efficient decoding, including
with noisy syndrome measurements? Quantum Tanner codes have
efficient decoders and single-shot correction using one round of
noisy syndrome measurements~\cite{lzdecoding23,gusingleshot23}.
For finite instances, can the balanced product structure be used
to reduce decoding to simpler problems? Related reductions use
classical decoders for hypergraph product codes and certain
lifted product codes~\cite{QuintavalleCampbell2022ReShape,GolowichGuruswami2024Decoding},
and graph matching for bivariate bicycle
codes~\cite{TanEtAl2026Matching}.

\paragraph{Syndrome extraction.}
How can syndrome extraction reduce ancilla and measurement costs
while limiting fault propagation? For hypergraph product codes,
prior work analyzes single-ancilla extraction and partial or
adaptive syndrome measurements
\cite{TanStambler2025,BerthusenGottesman2024,BerthusenEtAl2025Adaptive}.
For neutral-atom implementations, the analysis must also account
for atom loss and its propagation through the extraction circuit
\cite{LiuEtAl2026PauliEnvelope}.

\paragraph{Fault-tolerant logical operations.}
Can the required logical states be prepared and logical Clifford
gates implemented fault tolerantly with low overhead?
Code switching between product codes provides access to
complementary gate sets
\cite{JochymOConnor2019,tan2025single,li2025transversal}.
Here it would need to transfer the states of the selected logical
qubits while controlling errors and resource costs. Distillation
must also account for fault-tolerant implementations of the ideal
Clifford operations assumed in \cref{thm:distill}.


\textit{Note added.---}
While completing this manuscript, we became aware of work by
Li, Li, Han, and Liu~\cite{LLHL}, who report asymptotically good
qLDPC codes and quantum locally testable codes with linearly many
addressable transversal logical CCZ gates. Their approach builds
on recent work on quantum locally testable
codes~\cite{GJ,BLN,CHLT}.\footnote{
In version~3 of~\cite{LLHL}, Proposition~3.1 attributes the
covering and spectral properties of its six-dimensional
arithmetic quotient to~\cite{BLN,GJ}.
However, \cite[Lemma~4.1]{BLN} treats a different quotient,
and \cite[Section~9]{GJ} uses a number-field construction.
The application of these results to the particular quotient
in~\cite{LLHL}, together with the character and transport
properties used in Lemmas~3.6--3.7, requires further
justification.
} In comparison, our construction has a much simpler structure and can be completely understood using standard 3D balanced product and Fourier transform language.

\subsection*{Acknowledgements}
We thank Louis Golowich, Virgile Guemard, Ting-Chun Lin, Quynh Nguyen, and Adam Wills for their helpful discussions. We are particularly grateful to Virgile Guemard for working with us on technical aspects of logical CCZ gates in balanced product codes, Quynh Nguyen for discussing a similar proof for this result, and Adam Wills for pointing us to the Bafna and Vyas work for the product expansion result that inspired us to reinvestigate this line of research.
SJST acknowledges support from the NUS Development
Grant. He thanks IBM for supporting his internship and Keller
Blackwell, Anqi Gong, Shubham Jain, and Adam Wills for their
companionship during that time.
GZ acknowledges support through the Co-design Center for
Quantum Advantage (C2QA), funded by the U.S. Department of Energy's
Office of Science as part of the National Quantum Information
Science Research Centers, under contract DE-SC0012704.

\subsection*{AI statement}
The authors began this project in March 2026 and developed a
framework for disjoint logical CCZ gates on balanced product codes using
Fourier transforms, graph lifts and information set. Our progress was initially limited
by the product expansion obstacle and the lack of an expanding
graph cover with the required properties. Using the construction
of Golowich and Lin~\cite{GL} as inspiration, the authors prompted
ChatGPT Astra and, through several iterations, constructed the
expanding graph cover used here. Astra then assisted the authors
in adapting the cup product and balanced product
framework that the authors have developed to this graph cover.
Claude Opus and Fable, together with ChatGPT Sol and Astra, also
assisted with proof writing and manuscript preparation. The
authors directed the research and take responsibility for all
results and the final manuscript.

%% file: sections/prelims.tex
\section{Preliminaries}\label{sec:prelim}
Throughout the construction, $F$ is a finite field of characteristic two.
Code and cochain spaces are finite-dimensional, and all
vector spaces have specified bases. Hamming weight $|x|$ counts
nonzero scalar entries. For an array $x=(x_\sigma)$ of coefficient
vectors, its block weight $\bwt{x}=|\{\sigma:x_\sigma\ne0\}|$
counts nonzero vectors.
Logarithms are natural unless a base is indicated.

\subsection{Reed--Solomon and Tanner codes}\label{pre:sheafsec}

The classical codes in our construction impose a Reed--Solomon
constraint on the edges incident to each vertex of a graph. Together with graph
expansion, their local distance will give a global distance bound,
while polynomial
multiplication will supply the pairings used for logical gates.

For the Reed--Solomon code $\RS(S,k)$ introduced before
\cref{ass:pe}, write $n=|S|$ and assume $1\le k\le n$.
A polynomial of degree below $k$ is determined by any $k$
evaluations and has at most $k-1$ roots unless it is zero.
Thus the code has dimension $k$ and minimum weight $n-k+1$;
this is the maximum distance separable (MDS) property.
A \emph{generalized RS code} is obtained by rescaling each evaluation
entry by a prescribed nonzero scalar. Its dual is a generalized RS code of
dimension $n-k$, with $\RS(S,0)=\{0\}$ when $k=n$.
\Cref{app:classical} collects these facts and the results about
information sets and systematic generator matrices used below.

For vectors $u,v\in F^S$, write $u\odot v$ for their coordinatewise
product, $(u\odot v)_c=u_cv_c$. For codes $C,C'\subseteq F^S$,
their \emph{Schur product} is
$C\odot C'=\Span_F\{u\odot v:u\in C,\ v\in C'\}$.
For $\sigma\in(F^\times)^S$, write
$\sigma\odot C=\{\sigma\odot u:u\in C\}$ and
$\sigma^{-1}=(\sigma(c)^{-1})_{c\in S}$.
Coordinate rescaling preserves supports, distances, and information
sets, and satisfies
$(\sigma\odot C)^\perp=\sigma^{-1}\odot C^\perp$.

Let $\Gamma$ be a finite loopless graph with an orientation on each edge; parallel
edges are allowed. The tail and
head of an edge are its initial and terminal endpoints; in the
bipartite graphs used later they lie on the minus and plus sides,
respectively. Let $E(v)$ be the set of edges incident to $v$.
We use an edge coloring by a finite set $S$ in which every color
appears exactly once at every vertex. It identifies $E(v)$ with $S$,
so each local coordinate corresponds to a specified incident edge.

\begin{definition}[System of local codes and Tanner code]\label{pre:sheaf}
A \emph{system of local codes} (\textit{sheaf}) $\shf$ on $\Gamma$ assigns to each
vertex $v$ a coefficient space $\shf_v=F^{d_v}$, where $d_v$ is a
nonnegative integer, and an injective $F$-linear generator map
$\shf_{v\to E(v)}:\shf_v\to F^{E(v)}$.
Its image is the local code at $v$.
Each edge $e$ has coefficient space $\shf_e=F$.
These spaces are called \emph{stalks}, and the coordinate maps
$\shf_{v\to e}:\shf_v\to\shf_e$ of the generator map are called
\emph{restriction maps}. Together they form a cellular sheaf on
$\Gamma$, in the formalism of systems of local codes used in~\cite{GL,GTZ}.
The associated \emph{Tanner code} is
\begin{equation}\label{pre:tanner}
 \Tan(\Gamma,\shf)
 =\{x\in F^{E(\Gamma)}:
 x|_{E(v)}\in\im\shf_{v\to E(v)}\text{ for every }v\}.
\end{equation}
The dual local code system $\shf^\perp$ uses
$(\im\shf_{v\to E(v)})^\perp$ at each vertex, with any fixed
injective $F$-linear generator map for that space. Orthogonality is with
respect to the coordinate dot product over $F$.
\end{definition}

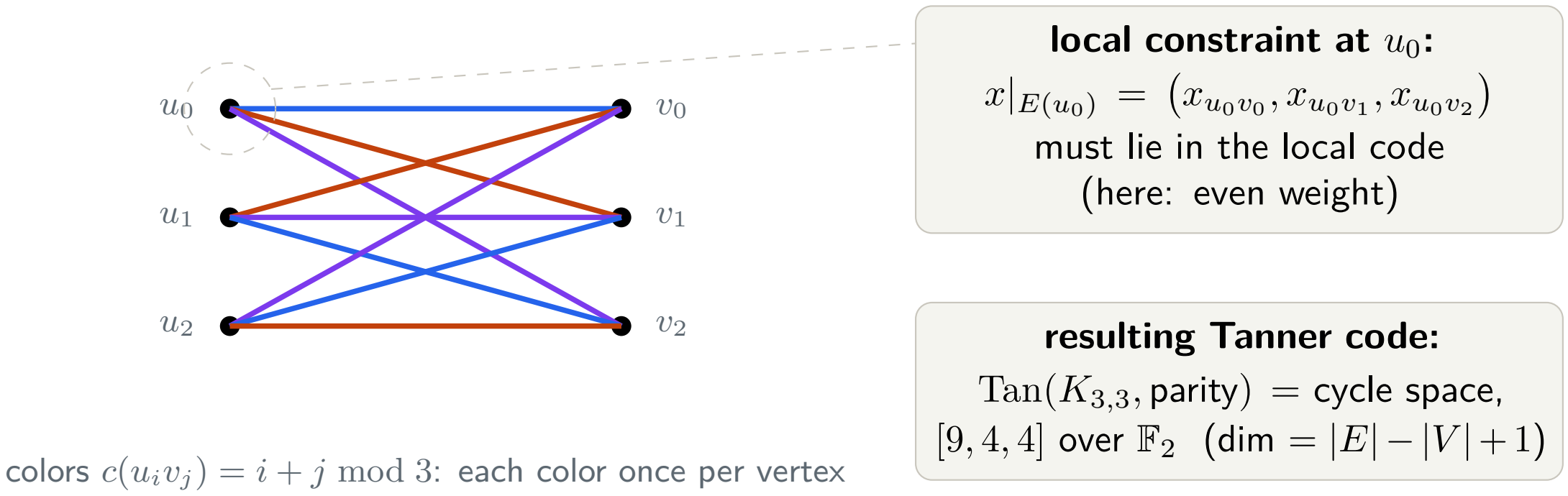
\begin{figure}[htbp]
\centering
\begin{tikzpicture}[x=1cm,y=1cm]
\foreach \i in {0,1,2}{
  \fill[black] (0,{2.0-1.0*\i}) circle(2.6pt);
  \node[sml,anchor=east] at (-0.18,{2.0-1.0*\i}) {$u_{\i}$};
  \fill[black] (3.6,{2.0-1.0*\i}) circle(2.6pt);
  \node[sml,anchor=west] at (3.78,{2.0-1.0*\i}) {$v_{\i}$};}
\foreach \i in {0,1,2}{\foreach \j in {0,1,2}{
  \pgfmathtruncatemacro{\cc}{mod(\i+\j,3)}
  \ifnum\cc=0 \def\col{dira}\fi
  \ifnum\cc=1 \def\col{dirb}\fi
  \ifnum\cc=2 \def\col{dirc}\fi
  \draw[\col,line width=1.4pt] (0,{2.0-1.0*\i}) -- (3.6,{2.0-1.0*\j});}}
\node[sml] at (1.8,-1.35) {colors $c(u_iv_j)=i+j\bmod 3$: each color once per vertex};
\draw[linec,dashed] (0,2.0) ellipse (0.42 and 0.42);
\draw[linec,dashed] (0.38,2.18) -- (6.3,2.6);
\node[fbox,text width=5.6cm,anchor=west] at (6.3,1.9)
 {\textbf{local constraint at $u_0$:}\\[2pt]
  $x|_{E(u_0)}=\big(x_{u_0v_0},x_{u_0v_1},x_{u_0v_2}\big)$\\[2pt]
  must lie in the local code\\ (here: even weight)};
\node[fbox,text width=5.6cm,anchor=west] at (6.3,-0.6)
 {\textbf{resulting Tanner code:}\\[2pt]
  $\Tan(K_{3,3},\text{parity})=$ cycle space,\\[1pt]
  $[9,4,4]$ over $\F_2$ \ (dim $=|E|-|V|+1$)};
\end{tikzpicture}
\caption{Edge colors and Tanner codes on $K_{3,3}$. Every
vertex applies one copy of the $[3,2]$ parity code to its three incident
edges, indexed by a proper edge coloring with three colors; the global code is the cycle
space of $K_{3,3}$ and has parameters $[9,4,4]$ over $\F_2$.
In \cref{gp:section}, the same construction uses lifted expander
graphs and rescaled Reed--Solomon local codes.}
\label{fig:k33}
\end{figure}

\Cref{fig:k33} illustrates Tanner constraints with parity local
codes. The same Sipser--Spielman construction~\cite{SS} will be
used with rescaled RS codes on the lifted graphs of \cref{gp:section}.

To use Tanner codewords in the quantum construction, we also need
the local vertex data that produce their edge values. A
\emph{cochain complex} consists of $F$-vector spaces $C^j$ and $F$-linear maps
$\delta^j:C^j\to C^{j+1}$ with $\delta^{j+1}\delta^j=0$.
Its elements are \emph{cochains}; the elements of $\ker\delta^j$
and $\im\delta^{j-1}$ are \emph{cocycles} and
\emph{coboundaries}, respectively. Their quotient is the
\emph{cohomology} $H^j=\ker\delta^j/\im\delta^{j-1}$.
For a graph local code system, the complex has two terms:
\begin{equation}\label{pre:graph-complex}
 \begin{gathered}
 C^0(\Gamma,\shf)=\bigoplus_v\shf_v
 \xrightarrow{\ \delta^0\ }
 C^1(\Gamma,\shf)=F^{E(\Gamma)},\\
 (\delta^0c)_e=\shf_{v\to e}(c_v)+\shf_{v'\to e}(c_{v'})
 \qquad(e=(v,v')).
 \end{gathered}
\end{equation}
All other terms are zero. A 0-cochain assigns a stalk vector to
each vertex, and its coboundary compares the values supplied by
the two endpoints of each edge. Thus a 0-cocycle has equal endpoint
values. Reading them gives its \emph{evaluated Tanner codeword}
\begin{equation}\label{pre:evaluated-tanner}
 c(z)\in\Tan(\Gamma,\shf),\qquad
 c(z)_e=\shf_{v\to e}(z_v)\quad(v\text{ an endpoint of }e),
 \qquad z\in Z^0(\Gamma,\shf):=\ker\delta^0.
\end{equation}
The map $z\mapsto c(z)$ is an isomorphism: injectivity of the
vertex generator maps gives a unique local stalk vector for each
Tanner codeword, and its endpoint values agree by the Tanner constraints.

We will also need edge functionals that vanish on all vertex
coboundaries. They are described by the dual local codes.
\begin{lemma}[Dual Tanner code]\label{pre:dualtanner}
Identify a linear functional on $C^1=F^{E(\Gamma)}$ with its
coefficient vector $\lambda$ under the coordinate dot product.
Then
\[
 (\im\delta^0)^\perp
 =\ker\big((\delta^0)^{\mathsf T}\big)
 =\Tan(\Gamma,\shf^\perp).
\]
\end{lemma}
\begin{proof}
For $\lambda\in F^{E(\Gamma)}$ and a 0-cochain $c$, regroup by vertex:
\begin{equation}\label{pre:dual-annihilator}
 \sum_e\lambda_e(\delta^0c)_e
 =\sum_v\sum_{e\in E(v)}\lambda_e\,\shf_{v\to e}(c_v).
\end{equation}
The vectors $c_v$ can be chosen independently. The expression
therefore vanishes for every $c$ exactly when, for every vertex $v$,
$\lambda|_{E(v)}$ is orthogonal to $\im\shf_{v\to E(v)}$, as required.
\end{proof}

Here $\Tan(\Gamma,\shf^\perp)$ applies duality to the local
codes before imposing the Tanner constraints. The orthogonal
complement $\Tan(\Gamma,\shf)^\perp$ of the global code is
generally a different space.

\begin{remark}[Generator matrices]\label{pre:vrow}
For the Reed--Solomon local code systems used later, the color set satisfies
$S\subseteq F$ and the vertex stalk is
$F[U]_{<k}$, the space of polynomials of degree below $k$.
The restriction to a color-$c$ edge sends $f$ to $\sigma_v(c)f(c)$,
where $\sigma_v(c)\in F^\times$ is a column multiplier.
In the monomial basis the generator map is $D_{\sigma_v}\mathsf V$, where
$\mathsf V_{c,j}=c^j$ for $0\le j<k$ and
$D_{\sigma_v}=\diag(\sigma_v(c))_{c\in S}$.
Thus $\mathsf V^{\mathsf T}D_{\sigma_v}$ is a row generator matrix,
and a single restriction map is the row
$\sigma_v(c)(1,c,\ldots,c^{k-1})$.
\end{remark}

\subsection{CSS codes and cohomology}

A Calderbank--Shor--Steane (CSS) code requires commuting $X$ and $Z$ checks. A cochain complex
$C^0\xrightarrow{\delta^0}C^1\xrightarrow{\delta^1}C^2$
provides them: the identity $\delta^1\delta^0=0$ is their
commutation condition. Each scalar entry of $C^1$ labels one
physical qudit with computational basis $\{\ket{x}:x\in F\}$.
Write $\mu=[F:\F_2]$ and
\begin{equation}\label{pre:field-trace}
 \Tr(a)=\Tr_{F/\F_2}(a)=\sum_{j=0}^{\mu-1}a^{2^j}
 \qquad(a\in F)
\end{equation}
for the field trace. For $a,b,x\in C^1$, the Pauli operators act by
$X(a)\ket{x}=\ket{x+a}$ and
$Z(b)\ket{x}=(-1)^{\Tr(b\cdot x)}\ket{x}$,
where $b\cdot x$ is the coordinate dot product over $F$.
The CSS check matrices $H_X=(\delta^0)^{\mathsf T}$ and $H_Z=\delta^1$
satisfy $H_XH_Z^{\mathsf T}=0$. Their stabilizer is generated by
$X(a)$ for $a\in\im\delta^0$ and $Z(b)$ for
$b\in\im\big((\delta^1)^{\mathsf T}\big)$. These are $F$-linear spaces, so all
field multiples of each check row are included.

The allowed computational strings are the \emph{cocycles}
$\ker\delta^1$. The $X$ checks identify strings differing by a
\emph{coboundary} in $\im\delta^0$. For $z\in\ker\delta^1$,
the corresponding computational logical basis state is
\begin{equation}\label{pre:coset-state}
 \ket{z+\im\delta^0}
 =\frac{1}{\sqrt{|\im\delta^0|}}
   \sum_{b\in\im\delta^0}\ket{z+b}.
\end{equation}
The encoded computational basis is therefore indexed by the \emph{first cohomology}
$H^1=\ker\delta^1/\im\delta^0$.
For a cocycle $a$, the operator $X(a)$ translates the label $[z]$
to $[z+a]$. A basis of $H^1$ thus specifies independent logical
translations, and the code encodes $\dim_F H^1$ qudits.
The minimum weight in $\ker\delta^1\setminus\im\delta^0$ is
$D_X$, called the cosystolic distance. For the other Pauli type,
transpose the maps: the minimum weight in
$\ker\big((\delta^0)^{\mathsf T}\big)\setminus\im\big((\delta^1)^{\mathsf T}\big)$ is $D_Z$, the
systolic distance. Elements of these latter kernel and image
are called cycles and boundaries, respectively.
We take the minimum over an empty set to be $+\infty$.
These distances count physical qudits. Restriction of scalars
represents each qudit by $\mu$ qubits; \cref{db:binary} gives
the resulting binary parameters.

\subsection{Logical CCZ gates and subrank}

A diagonal circuit preserves a CSS code if its phase is constant
on each coset in \eqref{pre:coset-state}. We express this condition
using trilinear forms, following~\cite[Section~3.8]{GL}
and~\cite{GTZ,lin2024transversal}.

\begin{definition}[CCZ gates over $F$]\label{def:cnz}
For $a\in F$, the diagonal unitary $\CCZ^a$ acts on three qudits by
\begin{equation}\label{eq:cnz}
 \CCZ^a\ket{x_1,x_2,x_3}
 =(-1)^{\Tr(a x_1x_2x_3)}\ket{x_1,x_2,x_3},
 \qquad x_1,x_2,x_3\in F.
\end{equation}
For $F=\F_2$ and $a=1$, this is the ordinary qubit CCZ gate.
\end{definition}

An implementation across three code blocks is \emph{transversal}
if its physical gates have pairwise disjoint supports and each
gate acts on one qudit from each block. After relabeling the
qudits within each block, the gates act on matching qudits,
in depth one~\cite[Definition~3.43]{GL}.

Let $C_h^1$ be the space of computational basis labels of block $h$, for
$h=1,2,3$. An $F$-trilinear form
$T:C_1^1\times C_2^1\times C_3^1\to F$ specifies a physical
circuit: at a triple $(c_1,c_2,c_3)$ of sites, apply
$\CCZ^{T(\one_{c_1},\one_{c_2},\one_{c_3})}$,
where $\one_c$ is the coordinate vector with value one at $c$
and zero elsewhere. On computational strings $x_h\in C_h^1$,
the resulting phase is $(-1)^{\Tr(T(x_1,x_2,x_3))}$.
The circuit has constant depth if each qudit participates in at
most a fixed number of nontrivial gates, independent of block length.
A greedy coloring of intersecting gate supports then partitions
the gates into a constant number of disjoint layers.
In \cref{db:binary}, restriction
of scalars and repetition encoding give a transversal
implementation over qubits.

\begin{definition}[Logical trilinear form and subrank]
\label{def:logical-subrank}
For three CSS complexes, let $H_h^1$ be the cohomology in degree
$1$ of block $h$, for $h=1,2,3$.
An $F$-trilinear form $T$ on their cocycle spaces
\emph{descends to cohomology} if its value depends only
on the three cohomology classes. Equivalently, it vanishes
whenever one argument is a coboundary and the other two are
cocycles. When $T$ descends, we also write $T$ for the induced form on
$H_1^1\times H_2^1\times H_3^1$.
Its \emph{subrank}, denoted $\subrank(T)$, is the largest integer
$s\ge0$ for which there are $F$-linear injections
$\phi_h:F^s\to H_h^1$ satisfying
\begin{equation}\label{eq:logical-subrank}
 T(\phi_1(x),\phi_2(y),\phi_3(z))=\sum_{j=1}^{s}x_jy_jz_j
 \qquad\text{for all }x,y,z\in F^s.
\end{equation}
\end{definition}

For a physical form $T:C_1^1\times C_2^1\times C_3^1\to F$
as above, the trace phase preserves the three code spaces
precisely when the restriction of $T$ to cocycles descends.
Descent makes the phase constant on each product of coboundary
cosets, as required by \eqref{pre:coset-state}.
Conversely, phase invariance and $F$-linearity give
$\Tr(aT(b,z_2,z_3))=0$ for every $a\in F$, every coboundary $b$,
and cocycles $z_2,z_3$. The trace pairing is nondegenerate:
if $u\in F$ satisfies $\Tr(au)=0$ for every $a\in F$, then $u=0$.
Hence $T(b,z_2,z_3)=0$; the same argument applies to the other inputs.
This relates logical trilinear forms to logical gates
as in~\cite[Lemma~3.42]{GL}.

Extend the images of the standard basis under each $\phi_h$
in \eqref{eq:logical-subrank} to a basis of $H_h^1$.
This identifies $s$ matched triples of logical qudits.
Initializing every other logical qudit
to $\ket0$ removes terms involving its basis label.
We reserve $\srk$ for the number of triples constructed in this
paper, so satisfying \eqref{eq:logical-subrank} with $s=\srk$ proves
$\subrank(T)\ge\srk$.
When $\mu$ is even, $\Tr(1)=0$. To obtain qubit CCZ gates
on binary logical inputs, \cref{db:trace-phase} chooses
$\gamma_{\mathrm{tr}}\in F$ with $\Tr(\gamma_{\mathrm{tr}})=1$
and defines $\varphi:F\to\F_2$ by
$\varphi(a)=\Tr(\gamma_{\mathrm{tr}}a)$ for $a\in F$.
Then $\varphi(1)=1$. Multiplying the physical gate coefficients by
$\gamma_{\mathrm{tr}}$ gives the phase $(-1)^{\varphi(T)}$.
For addressability, we construct descended forms $T_I$ on the
same logical subspaces, one for each subset $I$ of selected
triples (\cref{cg:diagonal}). The fixed transversal layer is
established in \cref{addr:layer}.

%% file: sections/expander.tex
\section{Lifted graphs and classical Tanner codes}\label{gp:section}

We construct three expanding graph covers whose Tanner codes contain
a large common space of explicit codewords over a fixed finite field
$F$ of characteristic two. The local Reed--Solomon codes have small
dimension, giving large local distance and controlling the polynomial
degrees in the later cup product construction. However, counting the
local constraints gives no positive dimension guarantee
(\cref{gp:constraint-count}). We therefore choose the covers and their
local codes together, so that expansion is compatible with a large
space of codewords.

The group $H$ indexes the sheets of each cover. Our candidate
codewords are multiplicative characters, namely group homomorphisms
$f:H\to F^\times$:
the lifted edge $(e,x)$ receives the value $f(x)$. To satisfy the
Tanner constraints, the values around each vertex, divided by the
local column multipliers, must be evaluations of a low degree
polynomial. Frobenius identities provide such polynomials for a
large set of exponents. An averaging argument
then gives one choice of multipliers for which many characters
satisfy all the local constraints in all three covers.

Expansion is proved using complex characters of the same group.
The covers are determined by group labels assigned to vertex--edge
incidences. Fourier decomposition reduces adjacency on each cover
to matrices indexed by the base vertices. Character sum cancellation
bounds their entries, while random choices of the exponents defining
the incidence labels and expansion of the base graph control the
remaining contributions. Thus complex characters establish expansion,
while finite field characters supply the codewords. The two families
use the same character indices, with arithmetic in $\C$ and $F$,
respectively.

Golowich and Lin obtain a punctured Reed--Muller subcode through
affine line restrictions of low total degree polynomials
\cite[Lemmas~4.1 and~5.2]{GL}. Here we start from the full
multiplicative character basis and require low polynomial degree
only on local restrictions. This provides sufficiently many candidates
over a fixed field. \Cref{gp:theorem} combines the spectral and
dimension conclusions.

\subsection{The graph covers and their local codes}
\label{gp:construction}

The local evaluation sets lie in subfields $K_i\subset F$.
All polynomial coefficients and codeword entries lie in $F$.
We first specify the construction, then choose its parameters to
obtain expansion and a large space of codewords.

For a finite set $X$, write $F^X$ and $\C^X$ for the vector
spaces of functions on $X$ over the indicated field. For $T\subseteq X$,
$\one_T$ denotes its indicator, and $\one$ denotes the constant-one
function on the ambient set. For a condition $\mathcal E$,
$\one_{\mathcal E}$ is its indicator scalar. On complex coordinate
spaces use
$\langle x,y\rangle=\sum_{u\in X}x(u)\overline{y(u)}$ and
$\|x\|_2^2=\sum_{u\in X}|x(u)|^2$. Let $I$ denote the identity
operator on the space in question. Throughout, $U$ is a polynomial
indeterminate, and the zero polynomial has degree $-\infty$.

\paragraph{Fields and parameters.}
Fix a power $P\ge2$ of two and integers $d_0\ge3$ and $D\ge1$.
For an odd integer $r\ge3$, set
\begin{equation}\label{gp:construction-fields}
 \begin{gathered}
 m_i=r+i-1\quad(1\le i\le3),\qquad m=m_1m_2m_3,\\
 q=P^m,\qquad F=\F_q,\qquad K_i=\F_{P^{m_i}}\subset F,
 \qquad M=q-1,\\
 n=d_0\lfloor P^r/d_0\rfloor,\qquad b=n/d_0,
 \qquad k=\lfloor n/7\rfloor.
 \end{gathered}
\end{equation}
Take $r$ sufficiently large that $n\ge7$ and the roots of
\cref{gp:construction-slopes} exist
(\cref{gp:slope-roots-exist}). Each $K_i$ is the unique subfield
of $F$ of size $P^{m_i}$, with $[F:K_i]=m/m_i>1$.
The graph degree and local code length are both $n$, the local
dimension is $k$, and $D$ is the number of affine factors in
each incidence label. In particular,
$1\le k\le n\le |K_i|$. These parameters remain fixed as the
base graphs grow.

The identities $c^{P^{m_i}}=c$ for $c\in K_i$ reduce the degrees
of local polynomial representatives, while the consecutive degrees
$m_i$ keep the subfield sizes within a factor of $P^2$.
Moreover, the $m_i$ are pairwise coprime: consecutive integers
are coprime, and $\gcd(r,r+2)=1$ because $r$ is odd. This lets
the Chinese remainder theorem count digit positions satisfying
the degree conditions in all three subfields
(\cref{gp:exponent-set-size}).

\paragraph{The base graph and the group.}
Let $t\ge d_0$ be an integer and let $\Gamma_0$ be a finite
simple $d_0$-regular bipartite graph with vertex set
$V=V_-\sqcup V_+$ and $|V_-|=|V_+|=t$.
Write $\operatorname{side}(v)=\sigma$ when $v\in V_\sigma$.
Fix a decomposition of the edges
\begin{equation}
 E(\Gamma_0)=\mathcal M_1\sqcup\cdots\sqcup\mathcal M_{d_0}
\end{equation}
into perfect matchings. Such a decomposition exists for every
regular bipartite graph.
The group used for all three covers is
\begin{equation}
 H=(F^\times)^t,\qquad |H|=M^t.
\end{equation}
Multiplication and division in $H$ are coordinatewise, with identity
$1_H=(1,\ldots,1)$. The vertex copies $(v,z)$ have \emph{sheet label}
$z\in H$. Edges may join different sheets.
Fix a generator $\omega$ of $F^\times$ and identify
$(\Z/M\Z)^t$ with $H$ by
$(r_1,\ldots,r_t)\mapsto(\omega^{r_1},\ldots,\omega^{r_t})$.
We write elements of $H$ multiplicatively and exponent vectors
additively modulo $M$, which need not be prime. For
$\alpha,\beta\in(\Z/M\Z)^t$, set
$\alpha\cdot\beta=\sum_{j=1}^t\alpha_j\beta_j\in\Z/M\Z$.

\begin{definition}[Roots and slopes]
\label{gp:construction-slopes}
A choice of roots is an indexed family
\[
 A_{\mathrm{root}}
 =\bigl(a_{i,\sigma,\ell}\bigr)_
 {1\le i\le3,\ \sigma\in\{-,+\},\ 1\le\ell\le D}
 \quad\text{with }a_{i,\sigma,\ell}\in F
\]
such that, for each fixed $i$, its $2D$ elements have pairwise
distinct minimal polynomials over $K_i$, all of degree $m/m_i$.
The distinctness requirement is imposed separately for each $i$.

A choice of \emph{slopes} is a family
\[
 L_{i,v,\ell}=(L_{i,v,\ell,1},\ldots,L_{i,v,\ell,t})
 \in(\Z/M\Z)^t,
 \qquad 1\le i\le3,\quad v\in V,\quad 1\le\ell\le D.
\]
\end{definition}

Each root $a_{i,\sigma,\ell}$ generates $F$ over $K_i$, so it lies outside $K_i$
and $c-a_{i,\sigma,\ell}\ne0$ for every $c\in K_i$.
Distinct minimal polynomials will give the character sum cancellation
needed for expansion (\cref{gp:roots-and-interpolation}).
The roots are shared by vertices on the same side while the slopes may
vary with $v$. For $x\in F^\times$ and
$L=(L_1,\ldots,L_t)\in(\Z/M\Z)^t$, write
\begin{equation}
 x^L=(x^{L_1},\ldots,x^{L_t})\in H.
\end{equation}
This is independent of the integer representatives of the $L_j$
since $x^M=1$. Fix roots and slopes for the definitions that follow.

\begin{definition}[Colored base graph]
\label{gp:construction-colored-base}
For each $i$, choose pairwise disjoint sets
\[
 B_{i,1},\ldots,B_{i,d_0}\subset K_i,\qquad
 |B_{i,j}|=b,\qquad S_i=\bigcup_{j=1}^{d_0}B_{i,j}\subseteq K_i.
\]
Let $\overline\Gamma_0$ be the bipartite multigraph obtained
from $\Gamma_0$ by replacing each edge by $b$ distinct parallel
edges. More precisely, its vertex set is $V$ and its edge set is
$E(\Gamma_0)\times\{1,\ldots,b\}$, with each copy having the
same endpoints as the original edge.
For each $i$, choose a coloring $c_i:E(\overline\Gamma_0)\longrightarrow S_i$
whose restriction to the $b$ copies of any edge in $\mathcal M_j$
is a bijection onto $B_{i,j}$.
For $v\in V$ and $c\in S_i$, let $e_{i,v,c}$ denote the unique
edge of $\overline\Gamma_0$ incident to $v$ with color $c$.
\end{definition}

These choices are possible because $n\le|K_i|$.
Each matching contributes $b$ incident edges at each vertex,
one of each color in $B_{i,j}$. Thus every vertex is incident
to exactly one edge of every color in $S_i$.
The three colorings use the same multigraph $\overline\Gamma_0$.
Replacing each base edge by $b$ copies gives $n$ local evaluation
coordinates while retaining $d_0$ base neighbors. It preserves
normalized adjacency because $b/n=1/d_0$. In the cover, the
lifts of these copies can connect to different sheets, allowing
their character phases to cancel (\cref{gp:parallel-weight}).

For a function $f:H\to F$, give the lifted edge $(e,x)$
the value $f(x)$. An incidence label $g_{i,v}(c)\in H$
relates the sheet $z$ of a vertex to the sheet $x$ of its
incident color-$c$ edge by $x=zg_{i,v}(c)$.
The incident value is therefore $f(zg_{i,v}(c))$.
We choose the incidence labels as products of powers of affine
factors in $c$, so that character restrictions have the same form
up to a nonzero scalar.

\begin{definition}[Incidence labels, voltages, and graph covers]
\label{gp:construction-cover}
For $v\in V_\sigma$, define $g_{i,v}:S_i\to H$ coordinatewise by
\begin{equation}\label{gp:construction-labels}
 g_{i,v}(c)_j
 =\prod_{\ell=1}^D
   (c-a_{i,\sigma,\ell})^{L_{i,v,\ell,j}},
 \qquad 1\le j\le t.
\end{equation}
Equivalently,
$g_{i,v}(c)=\prod_{\ell=1}^D(c-a_{i,\sigma,\ell})^{L_{i,v,\ell}}$.

The graph $\Gamma_i$ has vertex set $V\times H$ and edge set
$E(\overline\Gamma_0)\times H$. The edge $(e,x)$ has color $c_i(e)$.
If $e$ has endpoints $u\in V_-$ and $v\in V_+$ and
$c=c_i(e)$, the endpoints of its lifted copy $(e,x)$ are
\begin{equation}\label{gp:construction-endpoints}
 \left(u,xg_{i,u}(c)^{-1}\right)
 \quad\text{and}\quad
 \left(v,xg_{i,v}(c)^{-1}\right).
\end{equation}
Define the projection $\pi_i:\Gamma_i\to\overline\Gamma_0$
by $\pi_i(v,z)=v$ and $\pi_i(e,x)=e$. Define the $H$-action on
vertices and edges by
\[
 h\cdot(v,z)=(v,hz),\qquad h\cdot(e,x)=(e,hx),\qquad h\in H.
\]

The \emph{voltage} of $e$, oriented from $u$ to $v$, is
$g_{i,u}(c)g_{i,v}(c)^{-1}\in H$. The reverse orientation has the
inverse voltage. Equivalently, traversing $e$ from
$(u,z)$ leads to $\left(v,z\,g_{i,u}(c)g_{i,v}(c)^{-1}\right)$.
\end{definition}

The endpoint formula solves $x=zg_{i,v}(c)$ for $z$.
At the two ends of a lifted edge,
$x=zg_{i,u}(c)=z'g_{i,v}(c)$, giving the stated voltage.

\begin{proposition}[Incidence and covering properties]
\label{gp:construction-cover-properties}
The map $\pi_i$ is an $|H|$-sheeted graph cover of
$\overline\Gamma_0$: it induces a bijection from the edges
incident to each lifted vertex to the edges incident to its
image. The graph $\Gamma_i$ is bipartite and $n$-regular, with
\[
 |V(\Gamma_i)|=2t|H|,\qquad |E(\Gamma_i)|=tn|H|.
\]
The unique color-$c$ edge incident to $(v,z)$ is
\begin{equation}\label{gp:construction-incidence}
 \bigl(e_{i,v,c},\,zg_{i,v}(c)\bigr).
\end{equation}
The $H$-action in \cref{gp:construction-cover} preserves
incidence and colors, and is free and
transitive on each vertex or edge fiber of $\pi_i$.
\end{proposition}
\begin{proof}
For a base edge $e$ incident to $v$, the endpoint equation
$xg_{i,v}(c_i(e))^{-1}=z$ has the unique solution
$x=zg_{i,v}(c_i(e))$.
This gives the local bijection and
\eqref{gp:construction-incidence}. The bipartition is
$(V_-\times H)\sqcup(V_+\times H)$, and the counts follow
from $|E(\overline\Gamma_0)|=tn$.
The action preserves \eqref{gp:construction-endpoints} and on each
fiber it is the multiplication action of $H$ on itself.
\end{proof}

The covers are defined for arbitrary slopes. However, we will require additional constraints for the graph to be simple and connected. The covering map is
to $\overline\Gamma_0$, since collapsing its parallel edges is not
a covering map when $b>1$.

\begin{definition}[Local codes and Tanner codes]
\label{gp:construction-tanner}
Define
\[
 \RS(S_i,k)
 =\bigl\{(p(c))_{c\in S_i}:p\in F[U],\ \deg p<k\bigr\}
 \subseteq F^{S_i}.
\]
Choose shifts $s_{i,v,\ell}\in\Z/M\Z$ for every $i,v,\ell$,
and define the nonzero multiplier functions
$\sigma_{i,v}:S_i\to F^\times$ by
\begin{equation}\label{gp:construction-rescaling}
 \sigma_{i,v}(c)
 =\prod_{\ell=1}^D
   (c-a_{i,\operatorname{side}(v),\ell})^{s_{i,v,\ell}},
 \qquad c\in S_i.
\end{equation}
The local code at every lifted copy of $v$ is the generalized
Reed--Solomon code
\[
 L_i(v):=\sigma_{i,v}\odot\RS(S_i,k)
 \subseteq F^{S_i}.
\]
Its column multipliers are the values $\sigma_{i,v}(c)$.
The \emph{Tanner code} $\mathcal T_i\subseteq F^{E(\Gamma_i)}$
consists of the edge assignments $w$ such that
\begin{equation}\label{gp:construction-tanner-condition}
 \left(
   \frac{w(e_{i,v,c},\,zg_{i,v}(c))}{\sigma_{i,v}(c)}
 \right)_{c\in S_i}
 \in\RS(S_i,k)
 \qquad\text{for every }v\in V,\ z\in H.
\end{equation}
\end{definition}

The incident values must therefore lie in $L_i(v)$. Their polynomial
may depend on the codeword and lifted vertex, but the multipliers are fixed for
the entire code. Changing the shifts changes the constraints without
changing the graph. The induced action on edge assignments is
\begin{equation}
 (h\cdot w)(e,x)=w(e,h^{-1}x),
 \qquad h\in H,\quad w\in F^{E(\Gamma_i)}.
\end{equation}

\begin{proposition}[Basic code properties]
\label{gp:construction-code-properties}
Each $L_i(v)$ is an $F$-linear code of length $n$, dimension
$k$, and minimum Hamming distance $n-k+1$.
Each $\mathcal T_i$ is an $F$-linear code of length $tn|H|$
and is invariant under this induced $H$-action.
\end{proposition}
\begin{proof}
A nonzero polynomial of degree less than $k$ has at most $k-1$
zeros in $S_i$. Thus evaluation of the $k$-dimensional polynomial
space is injective and the local
distance is at least $n-k+1$. A polynomial with $k-1$ distinct
roots in $S_i$ attains this bound. When $k=1$, take the constant
polynomial $1$. Nonzero column multipliers preserve dimension
and Hamming weight.
The Tanner constraints are linear. The local vector of $h\cdot w$
at $(v,z)$ is the local vector of $w$ at $(v,h^{-1}z)$, so the
constraints are preserved by $H$.
\end{proof}

For fixed $i,v$, the threefold Schur product of $L_i(v)$ is contained in
$\sigma_{i,v}^3\odot\RS(S_i,3k-2)$: multiplying three polynomials
of degree less than $k$ gives degree at most $3(k-1)$
(\cref{appA:schur}). The choice
$k=\lfloor n/7\rfloor$ ensures $3k-2<n/2$, as required
by the deleted edge rank argument in \cref{cg:rank}.

\begin{definition}[Edge assignments induced by functions]
\label{gp:construction-functions}
For $f\in F^H$ and $1\le i\le3$, define
\[
 w_{i,f}\in F^{E(\Gamma_i)},\qquad w_{i,f}(e,x)=f(x).
\]
When the direction $i$ is understood, we abbreviate $w_{i,f}$
to $w_f$.
\end{definition}

Every base edge receives the same sheet array $(f(x))_{x\in H}$.
By \eqref{gp:construction-incidence}, a multiplicative character has
incident values
\begin{equation}
 f(zg_{i,v}(c))=f(z)f(g_{i,v}(c)).
\end{equation}
The scalar $f(z)$ is nonzero. Since the
local code is linear, multiplication by $f(z)$ preserves
membership in that code. Thus the local constraint at every
lift of $v$ holds exactly when the function
$c\mapsto f(g_{i,v}(c))/\sigma_{i,v}(c)$ on $S_i$ is represented
by a polynomial of degree less than $k$.
We will choose characters satisfying this condition at every
base vertex in all three directions. First, we state the
local constraints for an arbitrary $F$-linear subspace
$\mathcal V\subseteq F^H$.

\begin{proposition}[Criterion for a common space of codewords]
\label{gp:construction-common-space}
We have $w_{i,f}\in\mathcal T_i$ for every $i$ and every
$f\in\mathcal V$ if and only if
\begin{equation}\label{gp:construction-local-target}
 \left(
   \frac{f(zg_{i,v}(c))}{\sigma_{i,v}(c)}
 \right)_{c\in S_i}
 \in\RS(S_i,k)
\end{equation}
for every $f\in\mathcal V$, $1\le i\le3$, $v\in V$, and $z\in H$.
When this holds, each map $f\mapsto w_{i,f}$ embeds
$\mathcal V$ linearly into $\mathcal T_i$. In particular,
$\dim_F\mathcal T_i\ge\dim_F\mathcal V$.
\end{proposition}
\begin{proof}
Substituting $w_{i,f}(e,x)=f(x)$ into
\eqref{gp:construction-tanner-condition} gives the equivalence.
The map is linear and injective: restricting $w_{i,f}$ to
$(e,x)$ for one fixed base edge $e$ and all $x\in H$ gives $f$.
\end{proof}

\subsection{Fourier decomposition and spectral parameters}
\label{gp:fixed-parameters}

Fourier decomposition separates the sheet coordinate from the base
vertex, reducing adjacency on the cover to matrices indexed by $V$.

\subsubsection{Characters and translations}
\label{gp:character-intuition}

By our construction, traversing an edge translates the sheet label. To study adjacency
on complex vertex functions, we therefore use a basis of $\C^H$
on which translations act by scalars.

\begin{definition}[Complex characters and translations]
\label{gp:character-definition}
A \emph{complex character} of $H$ is a group homomorphism
$H\to\C^\times$.
Set $\zeta_M=\exp(2\pi\mathrm i/M)\in\C$, where
$\mathrm i^2=-1$ is the imaginary unit, distinct from the
direction index $i$.
For each $\alpha\in(\Z/M\Z)^t$, define
$\widehat\chi_\alpha:H\to\C^\times$ using the generator
$\omega$ fixed in \cref{gp:construction} by
\begin{equation}\label{gp:complex-character}
 \widehat\chi_\alpha(\omega^{r_1},\ldots,\omega^{r_t})
 =\zeta_M^{\,\alpha\cdot\mathbf r},
 \qquad \mathbf r=(r_1,\ldots,r_t)\in(\Z/M\Z)^t.
\end{equation}
For $h\in H$, define the translation operator $T_h:\C^H\to\C^H$ by
\[
 (T_h\varphi)(z)=\varphi(zh).
\]
\end{definition}

The action induced on functions by $z\mapsto hz$ is $T_{h^{-1}}$,
consistently with \cref{gp:construction-code-properties}.

\begin{lemma}[Complex character basis]\label{gp:character-basis}
The functions $\widehat\chi_\alpha$, indexed by
$\alpha\in(\Z/M\Z)^t$, are precisely the complex characters of
$H$, each occurring once. They have absolute value one and satisfy
\[
 T_h\widehat\chi_\alpha
 =\widehat\chi_\alpha(h)\widehat\chi_\alpha
 \qquad(h\in H).
\]
Moreover, for every $\alpha,\beta\in(\Z/M\Z)^t$,
\begin{equation}\label{gp:complex-orthogonality}
 \sum_{z\in H}\widehat\chi_\alpha(z)
                  \overline{\widehat\chi_\beta(z)}
 =|H|\,\one_{\alpha=\beta}.
\end{equation}
Consequently, the $|H|$ functions
$|H|^{-1/2}\widehat\chi_\alpha$ form an orthonormal basis of
$\C^H$.
\end{lemma}
\begin{proof}
The exponent in \eqref{gp:complex-character} is well-defined
modulo $M$. Addition of exponents proves multiplicativity,
which also gives the translation identity.
A character is determined by its values on the $t$ coordinate
generators of $H$. Each generator has order $M$, so its image
is a unique power of $\zeta_M$. These powers determine a unique
index $\alpha$.
For orthogonality, the sum factors as
\begin{equation}
 \prod_{j=1}^t\left(
   \sum_{u=0}^{M-1}\zeta_M^{(\alpha_j-\beta_j)u}
 \right).
\end{equation}
Each factor is $M$ if $\alpha_j=\beta_j$ and zero otherwise,
by the geometric sum identity. This proves
\eqref{gp:complex-orthogonality}. The $M^t=|H|$ nonzero
orthogonal functions form a basis of the $|H|$-dimensional
space $\C^H$.
\end{proof}

\subsubsection{Fourier decomposition of vertex functions}
\label{gp:vertex-fourier}

For a vertex function $\varphi\in\C^{V\times H}$, we expand
$z\mapsto\varphi(v,z)$ in the character basis at each base vertex $v$.

\begin{definition}[Adjacency and character subspaces]
\label{gp:character-subspaces}
Let $\widetilde A_i$ be the normalized adjacency matrix of
$\Gamma_i$, acting on $\C^{V\times H}$: its entry at two
vertices is the number of edges joining them divided by $n$.
Equivalently,
\[
 (\widetilde A_i\varphi)(v,z)
 =\frac1n\sum_{(w,y)\sim(v,z)}\varphi(w,y),
\]
where the sum counts neighbors with edge multiplicity.
Extend translation to $\C^{V\times H}$ by
$(T_h\varphi)(v,z)=\varphi(v,zh)$.
For each character index $\alpha$, define
\[
 \mathcal W_\alpha
 =\left\{\varphi\in\C^{V\times H}:
   \varphi(v,z)=b_\alpha(v)\widehat\chi_\alpha(z)
   \text{ for some }b_\alpha\in\C^V\right\}.
\]
For any $\varphi\in\C^{V\times H}$, define its Fourier coefficient
function $b_\alpha\in\C^V$ by
\begin{equation}\label{gp:fourier-coefficients}
 b_\alpha(v)=|H|^{-1}\sum_{z\in H}
              \varphi(v,z)\overline{\widehat\chi_\alpha(z)}.
\end{equation}
\end{definition}

In $\mathcal W_\alpha$, the character fixes the sheet dependence
and the coefficient $b_\alpha(v)$ varies over base vertices.

\begin{proposition}[Fourier decomposition on the cover]
\label{gp:fourier-decomposition}
For the coefficient functions in \eqref{gp:fourier-coefficients},
every $\varphi\in\C^{V\times H}$ has the unique expansion
\begin{equation}\label{gp:partial-fourier}
 \varphi(v,z)=\sum_{\alpha\in(\Z/M\Z)^t}
                 b_\alpha(v)\widehat\chi_\alpha(z).
\end{equation}
With the coordinate inner product on $\C^{V\times H}$,
\begin{equation}\label{gp:character-space-decomposition}
 \C^{V\times H}
 =\bigoplus_{\alpha\in(\Z/M\Z)^t}\mathcal W_\alpha
 \qquad\text{orthogonally},\qquad
 \dim_\C\mathcal W_\alpha=2t.
\end{equation}
In addition,
\begin{equation}\label{gp:partial-parseval}
 \sum_{v\in V}\sum_{z\in H}|\varphi(v,z)|^2
 =|H|\sum_{\alpha\in(\Z/M\Z)^t}\sum_{v\in V}|b_\alpha(v)|^2.
\end{equation}
For each $i$, every $\mathcal W_\alpha$ is invariant under
$\widetilde A_i$.
\end{proposition}
\begin{proof}
Apply \cref{gp:character-basis} to the function
$z\mapsto\varphi(v,z)$ for each fixed $v$.
This gives the expansion, its uniqueness, and the orthogonal
decomposition. The coefficients $b_\alpha(v)$ can be chosen
independently for $v\in V$, so each
summand has dimension $|V|=2t$. Orthogonality also gives
\eqref{gp:partial-parseval}.

For invariance, multiplicativity gives
\begin{equation}
 \mathcal W_\alpha
 =\{\varphi:T_h\varphi=\widehat\chi_\alpha(h)\varphi
                  \text{ for every }h\in H\}.
\end{equation}
Indeed, evaluating the condition at $(v,1_H)$ gives the required
form of $\varphi(v,h)$. Translations are graph automorphisms by
\cref{gp:construction-cover-properties}, so adjacency commutes
with them and preserves this common eigenspace.
\end{proof}

\subsubsection{Character blocks and the spectral target}
\label{gp:adjacency-blocks}

Let $A\in\mathbb R^{V\times V}$ be the normalized adjacency
matrix of $\Gamma_0$: $A(u,v)=1/d_0$ for adjacent vertices
and is zero otherwise. It is also the normalized adjacency of
$\overline\Gamma_0$, since $b/n=1/d_0$.
Fix a direction $i$ and suppress it in the notation $A_\alpha$.
The following matrices describe adjacency acting on the Fourier
coefficient vectors $b_\alpha$.

\begin{definition}[Character blocks]
\label{gp:character-blocks}
For $\alpha\in(\Z/M\Z)^t$, define $A_\alpha\in\C^{V\times V}$ by
\begin{equation}\label{gp:block-entries}
 A_\alpha(u,v)=
 \begin{cases}
 \displaystyle\frac1n\sum_{c\in B_{i,j}}
   \widehat\chi_\alpha\!\left(g_{i,u}(c)g_{i,v}(c)^{-1}\right),
       &\{u,v\}\in\mathcal M_j,\\[6pt]
 0, &\{u,v\}\notin E(\Gamma_0).
 \end{cases}
\end{equation}
For a matrix $T$ on a coordinate space over $\C$, write
$\|T\|=\sup_{x\ne0}\|Tx\|_2/\|x\|_2$ for its operator norm.
\end{definition}

\begin{proposition}[Adjacency on a character subspace]
\label{gp:block-representation}
For every coefficient vector $b_\alpha\in\C^V$, set
$\varphi(v,z)=b_\alpha(v)\widehat\chi_\alpha(z)$. Then
\begin{equation}\label{gp:block-action}
 (\widetilde A_i\varphi)(u,z)
 =\widehat\chi_\alpha(z)(A_\alpha b_\alpha)(u).
\end{equation}
Each $A_\alpha$ is Hermitian, $A_0=A$, and
\begin{equation}\label{gp:block-decomposition}
 \widetilde A_i\cong
 \bigoplus_{\alpha\in(\Z/M\Z)^t}A_\alpha,
\end{equation}
where $\cong$ denotes unitary equivalence. In particular, the
spectrum of the cover is the union, with multiplicity, of the
spectra of these $2t\times2t$ matrices.
\end{proposition}
\begin{proof}
By \cref{gp:construction-cover}, an edge of color $c$ from
$(u,z)$ leads to $(v,z g_{i,u}(c)g_{i,v}(c)^{-1})$.
Multiplicativity therefore gives
\begin{equation}
 \varphi\!\left(v,z g_{i,u}(c)g_{i,v}(c)^{-1}\right)
 =\widehat\chi_\alpha(z)
  \widehat\chi_\alpha\!\left(g_{i,u}(c)g_{i,v}(c)^{-1}\right)
  b_\alpha(v).
\end{equation}
Sum over incident edges and divide by $n$ to obtain
\eqref{gp:block-action}.
Reversing an edge inverts its voltage and hence conjugates its
character value, so $A_\alpha(v,u)=\overline{A_\alpha(u,v)}$.
For $\alpha=0$, every summand is one, giving $A_0=A$.
Finally, \cref{gp:fourier-decomposition} and the normalized
characters $|H|^{-1/2}\widehat\chi_\alpha$ give the unitary
decomposition \eqref{gp:block-decomposition}.
\end{proof}

The trivial block $A_0=A$ acts on functions constant across sheets.
The nontrivial blocks describe variation between sheets, and their
norms depend on the voltages. Expansion requires a bound for every
$\alpha\ne0$, including characters outside the codeword space.
Since $A_\alpha$ is Hermitian, its norm is its largest absolute
eigenvalue.

\subsubsection{Choice of spectral parameters}
\label{gp:analysis-parameters}

For each cover, the normalized biadjacency matrix has rows indexed
by $V_-\times H$ and columns by $V_+\times H$, with each entry
equal to the number of joining edges divided by $n$.

Fix real constants $0<\lambda_0,\delta,\eta<1/2$ with
$\lambda_0+\delta+\eta<1$. Here $\lambda_0$ bounds the absolute
values of the base adjacency eigenvalues other than $\pm1$,
and $\eta$ bounds averages of nontrivial character products
over parallel edges. For each nontrivial character of $H$,
$\delta$ bounds the fraction of exceptional base vertices,
where all $D$ scalar characters are trivial
(\cref{gp:exceptional-set}). Together, these estimates will
bound the nontrivial normalized biadjacency singular values
of each cover by $\lambda_0+\delta+\eta$
(\cref{gp:spectrum}).

To obtain base graphs with the required spectral bound,
choose $d_0$ sufficiently large that
\begin{equation}\label{gp:base-degree-condition}
 \frac{2\sqrt{d_0-1}}{d_0}\le\lambda_0.
\end{equation}
For the spectral analysis, choose $\Gamma_0$ from an unbounded
sequence of connected simple $d_0$-regular bipartite graphs
whose normalized adjacency $A$ has the simple eigenvalues
$1,-1$ and all other eigenvalues in $[-\lambda_0,\lambda_0]$.
Such a sequence exists by \eqref{gp:base-degree-condition}
and the bipartite Ramanujan theorem~\cite[Theorem~5.5]{MSS}.

Next choose $D$ large enough to make the exceptional vertex
estimate hold simultaneously for all nontrivial characters.
Throughout, logarithms are natural unless a base is indicated.
Writing $h(x)=-x\log x-(1-x)\log(1-x)$ for $0<x<1$, we require
\begin{equation}\label{gp:occupancy-condition}
 (2D\delta-1)\log3>2h(\delta).
\end{equation}
This condition makes the union bound in
\cref{gp:uniform-exceptional-set} tend to zero as $t$ grows.

Choose $d_0$ and $D$ before $r$. With these constants fixed,
take $r$ sufficiently large for the character cancellation
and polynomial restriction estimates, then hold all local
parameters fixed as $t$ grows.

\subsection{Evaluation sets with uniform character cancellation}
\label{gp:character-cancellation}

We choose the color sets $B_{i,j}$ to make the entries of
\eqref{gp:block-entries} small whenever the associated tuple of
scalar characters is nontrivial. The choice depends on the
roots of \cref{gp:construction-slopes}, and will work for
every subsequent choice of slopes.

We adapt the classical character sum method for proving spectral
expansion of algebraic Cayley graphs. In those graphs,
multiplicative character sums over the generating set give
adjacency eigenvalues; the degree one construction uses an
affine line over a subfield
\cite[Sections~1 and~4]{LuWanWangZhang2013}.
Here products of characters evaluated on the affine factors
$c-a_{i,\sigma,\ell}$ occur in the entries of the Fourier blocks.
The proof of \cref{gp:palettes} bounds these sums using Wan's
finite algebra character sum theorem.

\subsubsection{Character weights of parallel edges}
\label{gp:parallel-character-weights}

We first express each character block entry as an average of
products of scalar characters.

\begin{definition}[Characters of $F^\times$]
\label{gp:scalar-characters}
Using $\omega$ and $\zeta_M$ from
\cref{gp:construction,gp:character-definition}, define
\[
 \psi_\beta:F^\times\longrightarrow\C^\times,
 \qquad \psi_\beta(\omega^s)=\zeta_M^{\beta s},
 \qquad \beta,s\in\Z/M\Z.
\]
These are the $M$ multiplicative characters of $F^\times$ and
$\psi_0$ is the trivial character.
A tuple $\boldsymbol\psi=(\psi_{\sigma,\ell})_{
 \sigma\in\{-,+\},\,1\le\ell\le D}$ of homomorphisms
$\psi_{\sigma,\ell}:F^\times\to\C^\times$ is \emph{nontrivial}
if at least one of its $2D$ component characters is nontrivial.
\end{definition}

\begin{definition}[Projected exponents]
\label{gp:complex-projections}
For a fixed direction $i\in\{1,2,3\}$ and index
$\alpha\in(\Z/M\Z)^t$, write
\[
 \beta_{v,\ell}=\alpha\cdot L_{i,v,\ell}\in\Z/M\Z
 \qquad(v\in V,\ 1\le\ell\le D).
\]
The notation suppresses $i$ and $\alpha$ and both are held fixed
whenever these projected exponents are used.
\end{definition}

For $x\in F^\times$ and $L\in(\Z/M\Z)^t$,
\begin{equation}
 \widehat\chi_\alpha(x^L)
 =\prod_{j=1}^t\psi_{\alpha_j}(x^{L_j})
 =\psi_{\alpha\cdot L}(x).
\end{equation}
Thus the slope determines which scalar character is applied to
each factor. There are only $2D$ factors per edge, independently
of $t$.

For an edge with endpoints $u\in V_-$ and $v\in V_+$, its
associated tuple is
$\psi_{-,\ell}=\psi_{\beta_{u,\ell}}$ and
$\psi_{+,\ell}=\psi_{-\beta_{v,\ell}}$, for $1\le\ell\le D$.
The minus sign arises because the character is evaluated on
$g_{i,v}(c)^{-1}$ in the voltage $g_{i,u}(c)g_{i,v}(c)^{-1}$.

\begin{proposition}[Character weights of parallel edges]
\label{gp:parallel-weight}
Fix $i$, a character index $\alpha\in(\Z/M\Z)^t$, and an
edge $e_0\in\mathcal M_j$ with endpoints $u\in V_-$ and
$v\in V_+$, and use its associated tuple defined above.
For every $c\in B_{i,j}$, the character value of the voltage
on the color-$c$ copy of $e_0$, oriented from $u$ to $v$, is
\begin{equation}\label{gp:voltage-character-product}
 \widehat\chi_\alpha\bigl(g_{i,u}(c)g_{i,v}(c)^{-1}\bigr)
 =\prod_{\sigma\in\{-,+\}}\prod_{\ell=1}^D
     \psi_{\sigma,\ell}(c-a_{i,\sigma,\ell}).
\end{equation}
The $b$ parallel edges over $e_0$ give the following entry
$A_\alpha(u,v)$ of the character block defined in
\cref{gp:character-blocks}:
\begin{equation}\label{gp:parallel-average}
 \frac1{d_0}\left(
   \frac1b\sum_{c\in B_{i,j}}
     \prod_{\sigma,\ell}
       \psi_{\sigma,\ell}(c-a_{i,\sigma,\ell})
 \right).
\end{equation}
The tuple $\boldsymbol\psi$ is nontrivial exactly
when at least one of the $2D$ projected exponents
$\beta_{u,\ell},\beta_{v,\ell}$ is nonzero.
\end{proposition}
\begin{proof}
The coordinate formula for $\widehat\chi_\alpha$ and
\eqref{gp:construction-labels} give, for $w\in\{u,v\}$,
\begin{equation}\label{gp:complex-label-factorization}
 \widehat\chi_\alpha(g_{i,w}(c))
 =\prod_{\ell=1}^D
   \psi_{\alpha\cdot L_{i,w,\ell}}
      (c-a_{i,\operatorname{side}(w),\ell}).
\end{equation}
Taking the ratio of the two endpoint values proves
\eqref{gp:voltage-character-product}, including the minus sign
in $\psi_{+,\ell}$.
For $\varphi(w,z)=b_\alpha(w)\widehat\chi_\alpha(z)$,
the edges over $e_0$ contribute
\begin{equation}
 \frac{\widehat\chi_\alpha(z)b_\alpha(v)}{n}
 \sum_{c\in B_{i,j}}
   \widehat\chi_\alpha\bigl(g_{i,u}(c)g_{i,v}(c)^{-1}\bigr)
\end{equation}
to $(\widetilde A_i\varphi)(u,z)$, by
\cref{gp:construction-cover}.
Since $n=d_0b$, this gives \eqref{gp:parallel-average}.
Finally, $\psi_\beta$ is trivial if and only if $\beta=0$.
\end{proof}

\subsubsection{Roots and polynomial interpolation}
\label{gp:roots-and-interpolation}

\begin{lemma}[Existence of roots]
\label{gp:slope-roots-exist}
For fixed $P$ and $D$, the fields in
\eqref{gp:construction-fields} admit a choice of roots for
every sufficiently large odd $r$.
\end{lemma}
\begin{proof}
Fix $i$ and write $s=[F:K_i]=m/m_i$.
An element of degree less than $s$ over $K_i$ lies in a proper
subfield of $F$ containing $K_i$.
There are at most $s$ such subfields. Each has size at most
$|K_i|^{s/2}=q^{1/2}$, because its degree over $K_i$ is a proper
divisor of $s$.
Consequently, at least $q-sq^{1/2}$ elements have degree $s$.
Each minimal polynomial of such an element has exactly $s$
distinct roots in $F$, so the number of available minimal
polynomials is at least $(q-sq^{1/2})/s$.
This tends to infinity with $r$, since $s=O(r^2)$ and
$q=P^{\Theta(r^3)}$. Choose one root from each of $2D$
distinct minimal polynomials. The argument applies to all
three directions, and the roots may be chosen separately for each.
\end{proof}

To apply Wan's character sum bound below, we need polynomial
evaluation over $K_i$ to attain arbitrary prescribed values at
all $2D$ roots simultaneously. Each root has degree $[F:K_i]$
over $K_i$, so evaluation at that root can attain every value
in $F$. Since the roots have distinct minimal polynomials,
the Chinese remainder theorem lets us attain all the prescribed
values with a single polynomial.

Write
$\mathbf a_i=(a_{i,\sigma,\ell})_{\sigma,\ell}\in F^{2D}$,
with coordinates indexed by $\{-,+\}\times\{1,\ldots,D\}$.
Addition and multiplication in $F^{2D}$ are coordinatewise,
and a scalar $c\in K_i$ is identified with the constant tuple
$(c,\ldots,c)$. This makes $F^{2D}$ a commutative $K_i$-algebra.
The subalgebra generated by $\mathbf a_i$ consists of its
polynomial evaluations:
\begin{equation}
 K_i[\mathbf a_i]=\{p(\mathbf a_i):p\in K_i[U]\},
 \qquad
 p(\mathbf a_i)=(p(a_{i,\sigma,\ell}))_{\sigma,\ell}.
\end{equation}

\begin{lemma}[Polynomial evaluation at the roots]
\label{gp:root-interpolation}
For every choice of roots as in \cref{gp:construction-slopes}
and every direction $i$, the evaluation map
\[
 K_i[U]\longrightarrow F^{2D},\qquad
 p\longmapsto\bigl(p(a_{i,\sigma,\ell})\bigr)_{\sigma,\ell}
\]
is surjective. Equivalently, the $K_i$-algebra generated by
the single tuple $\mathbf a_i$ satisfies
\begin{equation}\label{gp:generated-algebra}
 K_i[\mathbf a_i]=F^{2D},
 \qquad \dim_{K_i}F^{2D}=2D\,m/m_i.
\end{equation}
\end{lemma}
\begin{proof}
Let $p_{\sigma,\ell}\in K_i[U]$ be the minimal polynomial
of $a_{i,\sigma,\ell}$.
The polynomials $p_{\sigma,\ell}$ are distinct and irreducible,
hence pairwise coprime.
Each evaluation map induces an isomorphism
$K_i[U]/(p_{\sigma,\ell})\cong F$, because the root has degree
$[F:K_i]$.
The Chinese remainder theorem therefore gives
\begin{equation}
 K_i[U]\Big/\Big(\prod_{\sigma,\ell}p_{\sigma,\ell}\Big)
 \cong\prod_{\sigma,\ell}K_i[U]/(p_{\sigma,\ell})
 \cong F^{2D},
\end{equation}
with the displayed evaluation map as the quotient map.
The dimension is the sum of the $2D$ component dimensions.
\end{proof}

Distinct roots can have the same minimal polynomial over $K_i$.
Suppose two chosen roots are $a$ and its conjugate
$a'=a^{|K_i|}\ne a$. Choose a nontrivial character
$\psi:F^\times\to\C^\times$, assign the characters
$\psi^{|K_i|}$ and $\psi^{-1}$ to these two roots, and take
all other component characters to be trivial.
For $c\in K_i$, we have $c-a'=(c-a)^{|K_i|}$, so
$\psi(c-a)^{|K_i|}\psi(c-a')^{-1}=1$.
Thus the character tuple is nontrivial, but its product is
identically one and its sum over $K_i$ equals $|K_i|$.

\subsubsection{Character cancellation and the choice of colors}
\label{gp:choosing-colors}

\begin{lemma}[Character cancellation]\label{gp:palettes}
Fix $P,d_0,D$, and $\eta$ as in
\cref{gp:construction,gp:analysis-parameters}.
For every sufficiently large odd $r$ and every choice of roots
as in \cref{gp:construction-slopes}, there are sets $B_{i,j}$
satisfying \cref{gp:construction-colored-base} such that, for
every $1\le i\le3$, every $1\le j\le d_0$, and every
nontrivial tuple $\boldsymbol\psi$,
\begin{equation}\label{gp:palette-bound}
 \left|\frac1b\sum_{c\in B_{i,j}}
   \prod_{\sigma,\ell}
      \psi_{\sigma,\ell}(c-a_{i,\sigma,\ell})\right|
 \le\eta.
\end{equation}
The threshold for $r$ is uniform over the choice of roots.
\end{lemma}
\begin{proof}
We first bound the mean of each character product over $K_i$,
then sample color sets whose averages are close to that mean.
For a direction $i$ and a nontrivial tuple $\boldsymbol\psi$,
define $Z:K_i\to\C$ and $\mu_Z\in\C$ by
\begin{equation}
 Z(c)=\prod_{\sigma,\ell}
      \psi_{\sigma,\ell}(c-a_{i,\sigma,\ell}),
 \qquad
 \mu_Z=\frac1{|K_i|}\sum_{c\in K_i}Z(c).
\end{equation}
By \cref{gp:root-interpolation}, $K_i[\mathbf a_i]=F^{2D}$,
an algebra of dimension $2Dm/m_i$ over $K_i$.
The tuple defines a character $\Psi:(F^\times)^{2D}\to\C^\times$
of its unit group by
\begin{equation}
 \Psi\bigl((x_{\sigma,\ell})_{\sigma,\ell}\bigr)
 =\prod_{\sigma,\ell}\psi_{\sigma,\ell}(x_{\sigma,\ell}).
\end{equation}
Since at least one component is nontrivial, $\Psi$ is nontrivial
on the unit group of $K_i[\mathbf a_i]=F^{2D}$, as required by
Wan's bound.
Writing $\mathbf 1=(1,\ldots,1)$, each $c\mathbf 1-\mathbf a_i$
with $c\in K_i$ is a unit because the roots lie outside $K_i$,
and $\Psi(c\mathbf 1-\mathbf a_i)=Z(c)$. Wan's finite algebra
bound~\cite[Corollary~2.2]{Wan}, applied to $F^{2D}$ over $K_i$
with generating element $\mathbf a_i$ and character $\Psi$
extended by zero on nonunits, therefore gives
\begin{equation}\label{gp:full-field-character-bound}
 |\mu_Z|\le (2Dm/m_i-1)|K_i|^{-1/2}
 =O(D r^2P^{-r/2}).
\end{equation}
For sufficiently large $r$, this is at most $\eta/2$, uniformly
over the directions, roots, and nontrivial character tuples.

For each $i$, sample an ordered list of $n$ distinct elements
of $K_i$ uniformly and partition it into $d_0$ consecutive
blocks of size $b$. Their underlying sets are $B_{i,1},\ldots,
B_{i,d_0}$. They are disjoint by construction, and each
$B_{i,j}$ is a uniform $b$-subset of $K_i$.

For fixed $i,j$ and a nontrivial tuple, $|Z(c)|=1$.
Hoeffding's inequality for sampling without replacement
\cite[Theorems~2 and~4]{Hoeffding} applies to the real and
imaginary parts of $Z$, which lie in $[-1,1]$. A complex deviation
greater than $\eta/2$ forces one part to deviate by more than
$\eta/(2\sqrt2)$, so
\begin{equation}\label{gp:color-sampling-tail}
 \Pr\!\left[
   \left|\frac1b\sum_{c\in B_{i,j}}Z(c)-\mu_Z\right|>
       \frac\eta2\right]
 \le 4\exp(-b\eta^2/16).
\end{equation}
There are at most $M^{2D}$ nontrivial tuples. A union bound over
these tuples, the three directions, and the $d_0$ blocks bounds
the probability of any violation by
\begin{equation}
 12d_0M^{2D}\exp(-b\eta^2/16)=o_r(1).
\end{equation}
The limit follows from $\log M=\Theta(r^3)$ and
$b=\Theta(P^r)$, with $P,d_0,D,\eta$ fixed.
Thus, for sufficiently large $r$, there is a choice of disjoint
color sets for which every required average satisfies
\begin{equation}
 \begin{aligned}
 \left|\frac1b\sum_{c\in B_{i,j}}Z(c)\right|
 &\le |\mu_Z|+
   \left|\frac1b\sum_{c\in B_{i,j}}Z(c)-\mu_Z\right|\\
 &\le\frac\eta2+\frac\eta2=\eta.
 \end{aligned}
\end{equation}
This proves \eqref{gp:palette-bound} with a threshold for $r$
independent of the choice of roots.
\end{proof}

The estimate covers every nontrivial scalar character tuple,
so the color sets can be fixed independently of $t$ and of the slopes.

\subsection{Random slopes and the spectrum}
\label{gp:random-slopes-spectrum}

Fix the local parameters and roots as above, and color sets from
\cref{gp:palettes}. For any choice of slopes, their cancellation
bound controls each character block entry unless all $D$ exponents
vanish at both endpoints. In that case, the entry equals its base
adjacency weight. We use expansion of the base graph to bound the
contribution of these remaining entries. This will reduce the spectral
estimate to choosing slopes for which few vertices have all $D$
exponents zero, simultaneously for every nontrivial character.

\subsubsection{Exceptional vertices and the block norm}
\label{gp:exceptional-vertices}

Fix a direction $i$. We suppress $i$ in $A_\alpha$ and
$Z_\alpha$ below; each direction has its own matrices and sets.

\begin{definition}[Exceptional vertices]
\label{gp:exceptional-set}
For the fixed direction $i$ and index $\alpha\in(\Z/M\Z)^t$,
use the exponents $\beta_{v,\ell}$ of \cref{gp:complex-projections}
and define
\[
 Z_\alpha=\{v\in V:\beta_{v,\ell}=0
                         \text{ for every }1\le\ell\le D\}.
\]
We call the vertices in $Z_\alpha$ exceptional for this character.
\end{definition}

For $Z\subseteq V$, let $A_Z\in\mathbb R^{V\times V}$ be the
base adjacency restricted to edges with both endpoints in $Z$:
\begin{equation}
 A_Z(u,v)=\one_Z(u)A(u,v)\one_Z(v).
\end{equation}
For a vector $x$ or matrix $T$, write $|x|$ or $|T|$ for
entrywise absolute values; $T^*$ denotes conjugate transpose.

\begin{proposition}[Norm bound for the character blocks]
\label{gp:block-norm-bound}
Under the base spectral hypothesis of \cref{gp:analysis-parameters},
for color sets satisfying \eqref{gp:palette-bound}, any choice
of slopes, and any $\alpha\in(\Z/M\Z)^t\setminus\{0\}$,
\begin{equation}\label{gp:block-domination}
 |A_\alpha|\le A_{Z_\alpha}+\eta A
 \qquad\text{entrywise},
\end{equation}
and
\begin{equation}\label{gp:block-norm}
 \|A_\alpha\|
 \le\lambda_0+(1-\lambda_0)\frac{|Z_\alpha|}{2t}+\eta.
\end{equation}
In particular, if $|Z_\alpha|\le2t\delta$, then
$\|A_\alpha\|\le\lambda_0+\delta+\eta$.
\end{proposition}
\begin{proof}
By \cref{gp:parallel-weight,gp:palettes}, a base edge $\{u,v\}$
has $A_\alpha(u,v)=A(u,v)$ when $u,v\in Z_\alpha$, and
$|A_\alpha(u,v)|\le\eta A(u,v)$ otherwise. Nonedges have zero
entries, so \eqref{gp:block-domination} follows.

To bound the contribution within $Z_\alpha$, use the base spectral
hypothesis. For a real vector $x$ supported on $Z_\alpha$, write
$x=a\one+y$, where $a=(2t)^{-1}\sum_v x(v)$ and $y\perp\one$.
Every eigenvalue of $A$ on $\one^\perp$ is at most $\lambda_0$,
including the bipartite eigenvalue $-1$. Hence
\begin{equation}
 \begin{aligned}
 x^{\mathsf T}Ax
 &\le\lambda_0\|x\|_2^2+(1-\lambda_0)\,2t a^2\\
 &\le\left(\lambda_0+(1-\lambda_0)\frac{|Z_\alpha|}{2t}\right)
       \|x\|_2^2,
 \end{aligned}
\end{equation}
where the second inequality is Cauchy--Schwarz on $Z_\alpha$.
Since $A_{Z_\alpha}$ is entrywise nonnegative,
$|x^{\mathsf T}A_{Z_\alpha}x|\le |x|^{\mathsf T}A_{Z_\alpha}|x|$.
Applying the preceding estimate to $|x|$ bounds every absolute
Rayleigh quotient on $Z_\alpha$. As $A_{Z_\alpha}$ vanishes
outside this set, we obtain
\begin{equation}\label{gp:small-set-norm}
 \|A_{Z_\alpha}\|
 \le\lambda_0+(1-\lambda_0)\frac{|Z_\alpha|}{2t}.
\end{equation}
Finally, for every $x\in\C^V$, \eqref{gp:block-domination} gives
\begin{equation}
 |x^*A_\alpha x|
 \le |x|^{\mathsf T}(A_{Z_\alpha}+\eta A)|x|
 \le(\|A_{Z_\alpha}\|+\eta)\|x\|_2^2,
\end{equation}
using $\|A\|=1$. Since $A_\alpha$ is Hermitian, taking the
supremum over unit vectors proves \eqref{gp:block-norm}.
\end{proof}

It remains to make $|Z_\alpha|\le2t\delta$ for every nontrivial
character. Independently for every $i,v,\ell$, choose
$L_{i,v,\ell}$ uniformly from $(\Z/M\Z)^t$.
Probabilities below refer only to these slopes. All local data remain
fixed as $t$ grows along the base graph sequence of
\cref{gp:analysis-parameters}; $o_t(1)$ denotes a quantity tending
to zero, uniformly over base graphs with the stated spectral bound.
We group characters by order in the union bound so that $D$
can be chosen independently of $M$.

\begin{lemma}[Uniform bound on exceptional vertices]
\label{gp:uniform-exceptional-set}
In the random slope model of \cref{gp:random-slopes-spectrum},
with probability $1-o_t(1)$, every direction $i\in\{1,2,3\}$
and every $\alpha\in(\Z/M\Z)^t\setminus\{0\}$ satisfy $|Z_\alpha|\le2t\delta$.
\end{lemma}
\begin{proof}
Fix $i$ and an index $\alpha$ of additive order $s>1$.
The map $L\mapsto\alpha\cdot L$ is a homomorphism from
$(\Z/M\Z)^t$ to $\Z/M\Z$ with image of size $s$.
Indeed, if $d=\gcd(M,\alpha_1,\ldots,\alpha_t)$, its image
is $d\Z/M\Z$, of size $M/d=s$.
The fibers have equal size, so $\alpha\cdot L$ is uniformly
distributed over this image when $L$ is uniform. Independence
of the $D$ slopes at each vertex gives $\Pr[v\in Z_\alpha]=s^{-D}$,
independently over the $2t$ vertices.

For each divisor $s\mid M$, at most $s^t$ indices have order
exactly $s$: they satisfy $s\alpha=0$, which allows $s$ choices
for each coordinate. Set $R_\delta=\lceil2t\delta\rceil$.
If $|Z_\alpha|\ge R_\delta$, there are $R_\delta$ vertices at
each of which all $D$ exponents vanish. A union bound gives
\begin{equation}\label{gp:order-union}
 \Pr[\exists\alpha\ne0:\ |Z_\alpha|\ge R_\delta]
 \le\sum_{\substack{s\mid M\\s>1}}
       s^t\binom{2t}{R_\delta}s^{-DR_\delta}.
\end{equation}
Using $\binom{2t}{R_\delta}\le
\exp\bigl(2t h(R_\delta/(2t))\bigr)$, the logarithm of each
summand is at most
\begin{equation}
 t\bigl[(1-2D\delta)\log s+2h(\delta)\bigr]+O(1).
\end{equation}
Since $M$ is odd, $s\ge3$. Condition
\eqref{gp:occupancy-condition} makes the coefficient of $t$
strictly negative for every such $s$.
The number of divisors of $M$ is fixed, so the sum tends to zero.
The same holds after a union bound over the three directions.
On the complementary event,
$|Z_\alpha|<\lceil2t\delta\rceil$, hence
$|Z_\alpha|\le2t\delta$.
\end{proof}

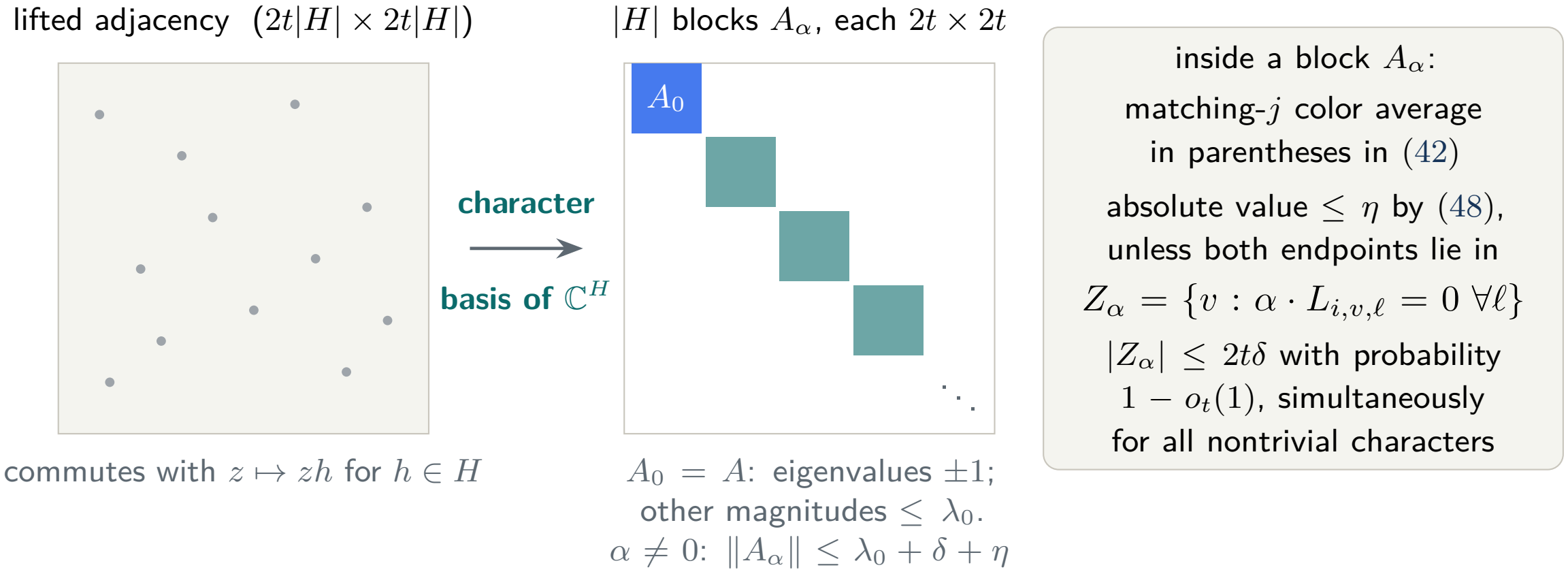
\begin{figure}[htbp]
\centering
\begin{tikzpicture}[x=1cm,y=1cm]
\draw[linec,fill=panelc] (0,0) rectangle (3.6,3.6);
\foreach \p in {(0.4,3.1),(1.2,2.7),(2.3,3.2),(3.0,2.2),(0.8,1.6),(1.9,1.2),(2.8,0.6),(0.5,0.5),(1.5,2.1),(2.5,1.7),(3.2,1.1),(1.0,0.9)}
  \fill[mutedc!60] \p circle(1.3pt);
\node[smk,above=3pt] at (1.8,3.6) {lifted adjacency \ ($2t|H|\times 2t|H|$)};
\node[sml,below=3pt] at (1.8,0) {commutes with $z\mapsto zh$ for $h\in H$};
\draw[flow] (4.0,1.8) -- (5.1,1.8);
\node[sml,text=accentc] at (4.55,2.25) {\textbf{character}};
\node[sml,text=accentc] at (4.55,1.35) {\textbf{basis of $\C^H$}};
\draw[linec] (5.5,0) rectangle (9.1,3.6);
\fill[dira!85] (5.57,2.92) rectangle (6.25,3.6);
\fill[accentc!60] (6.29,2.20) rectangle (6.97,2.88);
\fill[accentc!60] (7.01,1.48) rectangle (7.69,2.16);
\fill[accentc!60] (7.73,0.76) rectangle (8.41,1.44);
\node[lab,text=mutedc] at (8.75,0.45) {$\ddots$};
\node[font=\footnotesize\sffamily,text=white] at (5.91,3.26) {$A_0$};
\node[smk,above=3pt] at (7.3,3.6) {$|H|$ blocks $A_\alpha$, each $2t\times 2t$};
\node[sml,below=3pt,text width=4.3cm,align=center] at (7.3,0) {$A_0=A$: eigenvalues $\pm1$;\\ other magnitudes $\le\lambda_0$.\\ $\alpha\ne0$: $\|A_\alpha\|\le\lambda_0+\delta+\eta$};
\node[fbox,text width=4.7cm] at (12.1,1.8)
 {{\footnotesize\sffamily inside a block $A_\alpha$:}\\[2pt]
  {\footnotesize matching-$j$ color average in parentheses in
   \eqref{gp:parallel-average}}\\[3pt]
  {\footnotesize absolute value $\le\eta$ by \eqref{gp:palette-bound}, unless both endpoints lie in}\\[2pt]
  $Z_\alpha=\{v:\alpha\cdot L_{i,v,\ell}=0\ \forall\ell\}$\\[3pt]
  {\footnotesize $|Z_\alpha|\le 2t\delta$ with probability $1-o_t(1)$,
  simultaneously for all nontrivial characters}};
\end{tikzpicture}
\caption{The spectral argument for a fixed direction $i$.
The Fourier decomposition gives one block for each character.
For a nontrivial character, color cancellation bounds every
entry outside $Z_\alpha\times Z_\alpha$ by $\eta$ times
its base adjacency weight. Random slopes make $Z_\alpha$
small, and base expansion bounds adjacency within that set.
\Cref{gp:block-norm-bound} combines these estimates.}
\label{fig:fourier}
\end{figure}

\Cref{fig:fourier} summarizes this argument.

\subsubsection{Simple expanding covers}
\label{gp:simple-expanding-covers}

To prove simplicity, we bound the probability that two parallel
base edges have the same voltage.

\begin{lemma}[Simplicity of the covers]
\label{gp:simple-covers}
In the random slope model of \cref{gp:random-slopes-spectrum},
the probability that at least one of the three covers is not
simple is at most
\[
 3d_0t\binom b2\,3^{-t}=o_t(1).
\]
\end{lemma}
\begin{proof}
Fix $i$, an edge of $\Gamma_0$ with endpoints $u\in V_-$,
$v\in V_+$, and two distinct colors $c,c'$ on its parallel
copies. Condition on all slopes except $L_{i,u,1}$.
By \eqref{gp:construction-labels}, the two voltages agree exactly when
\begin{equation}
 \rho^{L_{i,u,1}}=h,
 \qquad
 \rho=\frac{c-a_{i,-,1}}{c'-a_{i,-,1}}
 \in F^\times.
\end{equation}
Here $h\in H$ is fixed by the conditioned slopes.
Since $c\ne c'$, we have $\rho\ne1$. Its order is an odd divisor
of $M$, hence at least three. The $t$ coordinates of
$\rho^{L_{i,u,1}}$ are independent and uniform in the subgroup
generated by $\rho$, so this equality has probability at most
$3^{-t}$, also after removing the conditioning.

At any sheet above $u$, the two lifted edges have the same other
endpoint exactly when their voltages agree; no union over sheets
is needed. Since $\Gamma_0$ is simple and the covers are bipartite,
only parallel copies of one base edge can create a multiple edge,
and loops are excluded. A union bound over the $d_0t\binom b2$
pairs in each of the three directions proves the claim.
\end{proof}

\begin{lemma}[Simple expanding covers]\label{gp:spectrum}
Under the parameter and base graph assumptions of
\cref{gp:analysis-parameters} and the fixed color sets and random
slopes of \cref{gp:random-slopes-spectrum},
with probability $1-o_t(1)$ all three covers are simple and
connected. In each cover, normalized
biadjacency has the simple singular value $1$ associated with
the constant vectors, and every other singular value is at most
$\lambda_0+\delta+\eta<1$.
\end{lemma}
\begin{proof}
By \cref{gp:uniform-exceptional-set,gp:block-norm-bound,gp:simple-covers},
with probability $1-o_t(1)$ all three covers are simple and every
nontrivial character block has norm at most $\lambda_0+\delta+\eta$.
Together with the base spectral hypothesis for $A_0=A$,
\cref{gp:block-representation} shows that each $\widetilde A_i$
has the simple eigenvalues $1,-1$ and all others of absolute value
at most $\lambda_0+\delta+\eta$. The covers are connected because
the multiplicity of the eigenvalue $1$ equals the number of
connected components.

For each $i$, let $B$ be the normalized biadjacency matrix of
$\Gamma_i$. In bipartition order,
$\widetilde A_i=\left(\begin{smallmatrix}0&B\\B^*&0\end{smallmatrix}\right)$,
whose eigenvalues are $\pm$ the singular values of $B$.
This proves the singular value claim.
\end{proof}

\subsection{Low degree restrictions and a large character space}
\label{gp:polynomial-character-space}

Fix the data of \cref{gp:construction} and any family of slopes.
We now choose the shifts $s_{i,v,\ell}$ defining the local code
multipliers $\sigma_{i,v}$ and a space $\mathcal V\subseteq F^H$
satisfying \eqref{gp:construction-local-target}. The shifts affect
only the local codes, so the covers retain their spectral properties.
All coefficients in this subsection lie in $F$.

A count of local constraints explains why we need to construct
codewords explicitly.

\begin{lemma}[Constraint count]\label{gp:constraint-count}
For the Tanner codes of \cref{gp:construction-tanner},
\begin{equation}\label{gp:constraint-dimension-bound}
 \dim_F\mathcal T_i
 \ge tn|H|-2t|H|(n-k)
 =t|H|(2k-n),\qquad 1\le i\le3.
\end{equation}
\end{lemma}
\begin{proof}
Each of the $2t|H|$ vertices imposes $n-k$ linear constraints
on the $tn|H|$ edge coordinates, by
\cref{gp:construction-cover-properties,gp:construction-code-properties}.
Rank--nullity gives the bound.
\end{proof}

For $k=\lfloor n/7\rfloor$, this bound is negative. We obtain a
positive bound by finding many functions whose induced assignments
belong to all three Tanner codes (\cref{gp:construction-common-space}).

\subsubsection{Finite field characters and local restrictions}
\label{gp:finite-field-restrictions}

\begin{definition}[Finite field characters]
\label{gp:field-characters}
For $\alpha\in(\Z/M\Z)^t$, define
\begin{equation}\label{gp:field-character}
 \chi_\alpha:H\longrightarrow F^\times,
 \qquad \chi_\alpha(z)=\prod_{j=1}^t z_j^{\alpha_j}.
\end{equation}
The exponents are well-defined modulo $M$ because $z_j^M=1$.
\end{definition}

At $z=(\omega^{r_1},\ldots,\omega^{r_t})$, the finite field
character has value $\omega^{\alpha\cdot\mathbf r}$; the complex
character of \eqref{gp:complex-character} has value
$\zeta_M^{\alpha\cdot\mathbf r}$. The common index records the same
exponent, with arithmetic performed in $F$ and $\C$, respectively.

\begin{lemma}[The finite field character basis]
\label{gp:field-character-basis}
The functions $\chi_\alpha$ form an $F$-basis of $F^H$.
For every $\alpha,\beta\in(\Z/M\Z)^t$,
\begin{equation}\label{gp:field-orthogonality}
 \sum_{z\in H}\chi_\alpha(z)\chi_{-\beta}(z)
 =|H|\,\one_{\alpha=\beta},
\end{equation}
where the integer $|H|$ is interpreted as an element of $F$.
\end{lemma}
\begin{proof}
The coordinatewise geometric sums of \cref{gp:character-basis}
apply over $F$ with $\omega$ in place of $\zeta_M$, giving
\eqref{gp:field-orthogonality}. Since $|H|$ is odd, it is nonzero
in $F$, so the $|H|$ characters are independent and form a basis
(see also \cref{pre:fourier}(ii)).
\end{proof}

For fixed $k$, polynomials of total degree less than $k$ in $t$
variables span only $\binom{t+k-1}{k-1}=t^{O(k)}$ dimensions.
The character basis provides $M^t$ candidates; we impose degree
bounds only on their local restrictions. This uses the multiplicative
group: the additive group of an $F$-vector space has no nontrivial
characters valued in $F^\times$.

\begin{proposition}[Local restrictions of a character]
\label{gp:character-local-restriction}
For every $i\in\{1,2,3\}$, $v\in V$,
$\alpha\in(\Z/M\Z)^t$, and $c\in S_i$,
\begin{equation}\label{gp:projected}
 \chi_\alpha(g_{i,v}(c))
 =\prod_{\ell=1}^D
   (c-a_{i,\operatorname{side}(v),\ell})^{\alpha\cdot L_{i,v,\ell}}.
\end{equation}
For any fixed shifts from \cref{gp:construction-tanner}, the local
vector at a sheet $z\in H$, after division by $\sigma_{i,v}$, is
\begin{equation}\label{gp:character-sheet-factorization}
 \left(\frac{\chi_\alpha(zg_{i,v}(c))}{\sigma_{i,v}(c)}\right)_{c\in S_i}
 =\chi_\alpha(z)
  \left(\frac{\chi_\alpha(g_{i,v}(c))}{\sigma_{i,v}(c)}\right)_{c\in S_i}.
\end{equation}
Consequently, the induced assignment $w_{i,\chi_\alpha}$
satisfies the local constraint at every copy of $v$ if and only
if the right-hand vector in \eqref{gp:character-sheet-factorization}
without its scalar factor lies in $\RS(S_i,k)$.
\end{proposition}
\begin{proof}
Substitute \eqref{gp:construction-labels} into
\eqref{gp:field-character} and collect the exponent of each
factor $c-a_{i,\operatorname{side}(v),\ell}$.
Multiplicativity gives
\eqref{gp:character-sheet-factorization}; its scalar
$\chi_\alpha(z)$ is nonzero and independent of $c$.
The last assertion follows from linearity of $\RS(S_i,k)$.
\end{proof}

Each local restriction has only $D$ projected exponents, independent
of $t$. Dividing by the multiplier shifts them:
\begin{equation}
 \frac{\chi_\alpha(g_{i,v}(c))}{\sigma_{i,v}(c)}
 =\prod_{\ell=1}^D
   (c-a_{i,\operatorname{side}(v),\ell})^{
      \alpha\cdot L_{i,v,\ell}-s_{i,v,\ell}}
 \qquad(c\in S_i).
\end{equation}
It remains to choose many exponents for which these products have
low degree polynomial representatives on all three subfields.

\subsubsection{Exponent digits and polynomial degree}
\label{gp:digit-restrictions}

Fix $i\in\{1,2,3\}$, $a\in F$, and an integer
$e\in\{0,\ldots,M-1\}$. Write its base-$P$ expansion as
$e=\sum_{j=0}^{m-1}e_jP^j$, where each $e_j\in\{0,\ldots,P-1\}$.
For $c\in K_i$, the Frobenius identity and $c^{P^{m_i}}=c$ give
\begin{equation}\label{gp:frobenius-intuition}
 (c-a)^e
 =\prod_{j=0}^{m-1}(c^{P^j}-a^{P^j})^{e_j}
 =\prod_{j=0}^{m-1}(c^{P^{j\bmod m_i}}-a^{P^j})^{e_j}.
\end{equation}
Replacing $c$ by $U$ gives a polynomial representative on $K_i$
of degree $\sum_j e_jP^{j\bmod m_i}$. The coefficient $a^{P^j}$
remains unchanged: only the iterate index on $c$ is reduced modulo
$m_i$. Thus a digit's position modulo $m_i$ determines its local
degree contribution.

To bound the degree in all three subfields, we require digits at
positions with residues near $m_i-1$ to be zero. The cutoff must allow
for repeated residues and the product of $D$ factors.

\begin{definition}[Allowed exponent digits]
\label{gp:allowed-exponents}
Define the positive integer
$\ell_0=\lceil\log_P(100DP^3m)\rceil$.
Whenever $\ell_0<r$, define
\begin{equation}\label{gp:digits}
 \begin{gathered}
 \mathcal J=\{j\in\{0,\ldots,m-1\}:\ j\bmod m_i<m_i-\ell_0
                                  \text{ for }i=1,2,3\},\\
 E=\left\{\sum_{j\in\mathcal J}e_jP^j:
                   e_j\in\{0,\ldots,P-1\}\right\}.
 \end{gathered}
\end{equation}
Thus digits in $\mathcal J$ are free, and all other digits
are zero. Regard $E$ as a subset of $\Z/M\Z$ using the
integer representatives justified below, and define the real
number $\epsilon_E=1-\log_M|E|$.
\end{definition}

The cutoff reduces the degree bound $DmP^{r+2}$ by a factor
$P^{\ell_0}\ge100DP^3m$, putting it below the local allowance.
Since $m=r(r+1)(r+2)$, this requires only $O_{P,D}(\log r)$
excluded residues per modulus.

\begin{lemma}[Size of the exponent set]
\label{gp:exponent-set-size}
Fix a power $P\ge2$ of two and an integer $D\ge1$.
As $r$ tends to infinity through odd integers,
\begin{equation}\label{gp:cutoff-growth}
 \ell_0=3\log_P r+O_{P,D}(1)=O_{P,D}(\log r).
\end{equation}
In particular, $\ell_0<r$ eventually, and
$E\subseteq\{0,\ldots,M-1\}$. Moreover,
\begin{equation}\label{gp:digit-density}
 \begin{gathered}
 |\mathcal J|=\prod_{i=1}^3(m_i-\ell_0),
 \qquad |E|=P^{|\mathcal J|},\\
 0\le\epsilon_E
 \le1-\frac{|\mathcal J|}{m}
 \le\ell_0\sum_{i=1}^3\frac1{m_i}
 \le\frac{3\ell_0}{r}
 =O_{P,D}\!\left(\frac{\log r}{r}\right).
 \end{gathered}
\end{equation}
\end{lemma}
\begin{proof}
Since $r^3\le m\le(r+2)^3$, the definition of $\ell_0$ gives
\eqref{gp:cutoff-growth}, and hence $\ell_0<r$ for large $r$.
The position $m-1$ is excluded from $\mathcal J$, because
its residue modulo every $m_i$ is $m_i-1$.
Thus every $e\in E$ is at most $P^{m-1}-1<M$.

By pairwise coprimality, the Chinese remainder map
$j\mapsto(j\bmod m_1,j\bmod m_2,j\bmod m_3)$ is a bijection
from $\{0,\ldots,m-1\}$ to $\prod_i\{0,\ldots,m_i-1\}$.
There are $\prod_i(m_i-\ell_0)$ allowed residue tuples, giving
$|\mathcal J|$; the $P$ choices per allowed digit give
$|E|=P^{|\mathcal J|}$.

The inclusion $E\subseteq\{0,\ldots,M-1\}$ gives
$\epsilon_E\ge0$. Since $\log M<m\log P$,
\begin{equation}
 \epsilon_E
 =1-\frac{|\mathcal J|\log P}{\log M}
 \le1-\frac{|\mathcal J|}{m}
 =1-\prod_{i=1}^3\left(1-\frac{\ell_0}{m_i}\right)
 \le\ell_0\sum_{i=1}^3\frac1{m_i}.
\end{equation}
The last inequality is $1-\prod_i(1-x_i)\le\sum_i x_i$
for $0\le x_i\le1$. Now use $m_i\ge r$ and
\eqref{gp:cutoff-growth}.
\end{proof}

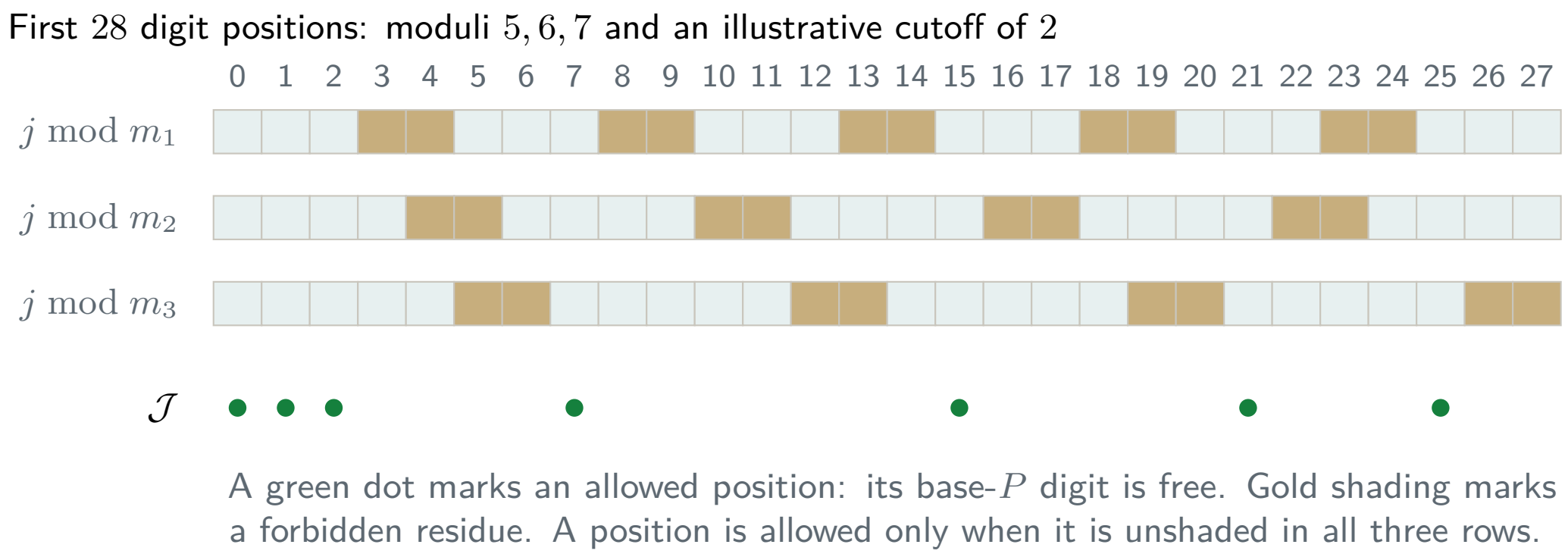
\begin{figure}[htbp]
\centering
\begin{tikzpicture}[x=0.42cm,y=1cm]
\node[smk,anchor=west] at (-4.6,1.2) {First $28$ digit positions: moduli $5,6,7$ and an illustrative cutoff of $2$};
\node[sml,anchor=east] at (-0.4,0.3) {$j\bmod m_1$};
\node[sml,anchor=east] at (-0.4,-0.45) {$j\bmod m_2$};
\node[sml,anchor=east] at (-0.4,-1.2) {$j\bmod m_3$};
\node[smk,anchor=east] at (-0.4,-2.1) {$\mathcal J$};
\foreach \j in {0,...,27}{
  \node[font=\scriptsize\sffamily,text=mutedc] at (\j+0.5,0.8) {\j};
  \pgfmathtruncatemacro{\ra}{mod(\j,5)}
  \pgfmathtruncatemacro{\rb}{mod(\j,6)}
  \pgfmathtruncatemacro{\rc}{mod(\j,7)}
  \ifnum\ra>2 \fill[goldc!55] (\j,0.12) rectangle ++(1,0.38); \else \fill[accentc!10] (\j,0.12) rectangle ++(1,0.38); \fi
  \draw[linec] (\j,0.12) rectangle ++(1,0.38);
  \ifnum\rb>3 \fill[goldc!55] (\j,-0.63) rectangle ++(1,0.38); \else \fill[accentc!10] (\j,-0.63) rectangle ++(1,0.38); \fi
  \draw[linec] (\j,-0.63) rectangle ++(1,0.38);
  \ifnum\rc>4 \fill[goldc!55] (\j,-1.38) rectangle ++(1,0.38); \else \fill[accentc!10] (\j,-1.38) rectangle ++(1,0.38); \fi
  \draw[linec] (\j,-1.38) rectangle ++(1,0.38);
  \pgfmathtruncatemacro{\ok}{(\ra<3)&&(\rb<4)&&(\rc<5) ? 1 : 0}
  \ifnum\ok=1 \fill[goodc] (\j+0.5,-2.1) circle(2.2pt); \fi
}
\node[sml,anchor=west,text width=11.6cm] at (0,-3.0)
 {A green dot marks an allowed position: its base-$P$ digit is free.
  Gold shading marks a forbidden residue. A position is allowed only
  when it is unshaded in all three rows.};
\end{tikzpicture}
\caption{The simultaneous digit test for moduli $5,6,7$ and an
illustrative cutoff of $2$. A digit at position $j$ contributes
degree $e_jP^{j\bmod m_i}$ on $K_i$; it is allowed only if its
residue avoids the excluded range for every $i$. The construction
uses the cutoff of \cref{gp:allowed-exponents} on all $m$ positions.}
\label{fig:digits}
\end{figure}

Thus $|E|=M^{1-\epsilon_E}$ with $\epsilon_E\to0$, although
the fraction $|E|/M$ may be small. \Cref{fig:digits} illustrates
the simultaneous residue test.

\begin{lemma}[Local polynomial representatives]\label{gp:frobenius}
Fix a power $P\ge2$ of two and integers $d_0\ge3$, $D\ge1$.
For all sufficiently large odd $r$, for each $i\in\{1,2,3\}$
and each choice of $e_1,\ldots,e_D\in E$ and
$a_1,\ldots,a_D\in F$, there is a polynomial $f\in F[U]$ such that
\[
 f(c)=\prod_{\ell=1}^D(c-a_\ell)^{e_\ell}
 \quad(c\in K_i),\qquad \deg f<n/100<k.
\]
The bound is uniform over the exponents and the $a_\ell$.
\end{lemma}
\begin{proof}
Write $e_\ell=\sum_{j\in\mathcal J}e_{\ell,j}P^j$.
By \eqref{gp:frobenius-intuition}, an explicit representative is
\begin{equation}\label{gp:explicit-polynomial}
 f(U)=\prod_{\ell=1}^D\prod_{j\in\mathcal J}
       (U^{P^{j\bmod m_i}}-a_\ell^{P^j})^{e_{\ell,j}}.
\end{equation}
Every residue modulo $m_i$ occurs exactly $m/m_i$ times
among all $m$ positions, and at most that often in
$\mathcal J$. Only residues $0,\ldots,m_i-\ell_0-1$
are allowed. Since $e_{\ell,j}\le P-1$,
\begin{align}
 \deg f
 &\le D\frac m{m_i}(P-1)
             \sum_{d=0}^{m_i-\ell_0-1}P^d\notag\\
 &=D\frac m{m_i}(P^{m_i-\ell_0}-1)
 <DmP^{r+2-\ell_0}
 \le\frac{P^r}{100P}.
 \label{gp:local-degree-bound}
\end{align}
Here $m_i\le r+2$, and the final inequality is the defining
choice $P^{\ell_0}\ge100DP^3m$.
For large $r$, $P^r\ge2d_0$ and hence
$n=d_0\lfloor P^r/d_0\rfloor\ge P^r/2$.
Together with $P\ge2$, this gives $\deg f<n/100$.
Finally, $n\ge7$ implies $k=\lfloor n/7\rfloor\ge n/14>n/100$.
\end{proof}

\subsubsection{Common multipliers and character selection}
\label{gp:common-character-selection}

The projected exponents need not lie in $E$. Random shifts give
each fixed character the same probability of satisfying every local
condition, and averaging yields a single set of shifts accepting
many characters.

\begin{definition}[Accepted characters and their span]
\label{gp:accepted-characters}
Take $r$ large enough for \cref{gp:frobenius}, and fix shifts
$s_{i,v,\ell}\in\Z/M\Z$. Define
\begin{equation}\label{gp:accepted}
 \mathcal A=
 \left\{\alpha\in(\Z/M\Z)^t:
   \alpha\cdot L_{i,v,\ell}-s_{i,v,\ell}\in E
                   \text{ for every }i,v,\ell\right\},
\end{equation}
where $i\in\{1,2,3\}$, $v\in V$, and $1\le\ell\le D$;
membership in $E$ uses the representative in
$\{0,\ldots,M-1\}$. Let
\begin{equation}\label{gp:function-space}
 \mathcal V=\Span_F\{\chi_\alpha:\alpha\in\mathcal A\}
 =\left\{\sum_{\alpha\in\mathcal A}u_\alpha\chi_\alpha:
                                           u_\alpha\in F\right\}.
\end{equation}
For $\alpha\in\mathcal A$, let
$e_{i,v,\alpha,\ell}\in E$ be the integer representative of
$\alpha\cdot L_{i,v,\ell}-s_{i,v,\ell}$.
Use a polynomial indeterminate $U_i$ for each direction $i$.
Define $f_{i,v,\alpha}\in F[U_i]$ by
\eqref{gp:explicit-polynomial} with $U=U_i$,
$a_\ell=a_{i,\operatorname{side}(v),\ell}$ and
$e_\ell=e_{i,v,\alpha,\ell}$.
By \cref{gp:frobenius}, $\deg f_{i,v,\alpha}<n/100<k$.
\end{definition}

\begin{lemma}[Character selection]\label{gp:selection}
Take $r$ large enough for \cref{gp:frobenius}.
For every fixed choice of slopes there are shifts such that
\begin{equation}\label{gp:selection-size}
 \dim_F\mathcal V=|\mathcal A|
 \ge M^t(|E|/M)^{6tD}.
\end{equation}
For every choice of shifts, every $\alpha\in\mathcal A$,
and all $i\in\{1,2,3\}$ and $v\in V$, the polynomials
of \cref{gp:accepted-characters} satisfy
\begin{equation}\label{gp:restriction-identity}
 \chi_\alpha(g_{i,v}(c))
 =\sigma_{i,v}(c)f_{i,v,\alpha}(c)
 \qquad(c\in S_i).
\end{equation}
This identity determines $f_{i,v,\alpha}$ uniquely among
polynomials of degree less than $k$.
\end{lemma}
\begin{proof}
Choose the $6tD$ shifts independently and uniformly in $\Z/M\Z$.
For fixed $\alpha$, the shifted exponents are independent and
uniform, so each lies in $E$ with probability $|E|/M$. Hence
\begin{equation}
 \mathbb E|\mathcal A|
 =\sum_{\alpha\in(\Z/M\Z)^t}\Pr[\alpha\in\mathcal A]
 =M^t(|E|/M)^{6tD}.
\end{equation}
Some choice attains at least this mean. Independence between
characters is unnecessary, and \cref{gp:field-character-basis}
gives the dimension equality.

For any shifts and an accepted $\alpha$, equations~\eqref{gp:projected} and
\eqref{gp:construction-rescaling} give
\begin{equation}
 \frac{\chi_\alpha(g_{i,v}(c))}{\sigma_{i,v}(c)}
 =\prod_{\ell=1}^D
   (c-a_{i,\operatorname{side}(v),\ell})^{e_{i,v,\alpha,\ell}}.
\end{equation}
All bases are nonzero, so reducing exponents modulo $M$
preserves their values. The definition of $f_{i,v,\alpha}$
and \cref{gp:frobenius} give \eqref{gp:restriction-identity}.
Since $|S_i|=n\ge k$, this is also the unique polynomial of
degree less than $k$ with these values after division by
$\sigma_{i,v}$.
\end{proof}

\begin{figure}[htbp]
\centering
\begin{tikzpicture}[x=1cm,y=1cm]
\draw[linec] (0,0) rectangle (0.55,2.9);
\foreach \y in {0.42,0.84,1.26,1.68,2.10,2.52} \draw[linec] (0,\y) -- (0.55,\y);
\node[smk,below=2pt] at (0.27,0) {$\alpha$};
\node[sml,below=14pt] at (0.27,0) {$t$ entries (grows)};
\draw[flow] (0.62,2.3) -- (1.9,2.55);
\draw[flow] (0.62,1.45) -- (1.9,1.55);
\draw[flow] (0.62,0.6) -- (1.9,0.55);
\node[fbox,minimum width=2.6cm] (p1) at (3.25,2.55) {$\alpha\cdot L_{i,v,1}$};
\node[fbox,minimum width=2.6cm] (p2) at (3.25,1.55) {$\alpha\cdot L_{i,v,2}$};
\node[fbox,minimum width=2.6cm] (p3) at (3.25,0.55) {$\cdots\ \alpha\cdot L_{i,v,D}$};
\node[sml,anchor=west] at (0.9,-0.55) {$D$ projected exponents per (direction, vertex)};
\draw[flow] (4.65,1.55) -- (6.0,1.55);
\node[sml,text=accentc] at (5.3,2.35) {\textbf{$-\,s_{i,v,\ell}$}};
\node[sml] at (5.3,2.0) {(defines $\sigma_{i,v}$)};
\draw[mutedc,line width=1pt] (6.2,1.55) -- (13.4,1.55);
\draw[mutedc] (6.2,1.42) -- (6.2,1.68);
\draw[mutedc] (13.4,1.42) -- (13.4,1.68);
\node[sml,below=2pt] at (6.2,1.45) {$0$};
\node[sml,below=2pt] at (13.4,1.45) {$M-1$};
\foreach \xx in {6.38,6.52,6.66, 7.05,7.19, 7.67,7.81,7.95, 8.5,8.64, 9.2,9.34,9.48, 9.95,10.09, 10.7,10.84,10.98, 11.5,11.64, 12.25,12.39, 12.9,13.04}
  \draw[goodc,line width=1.5pt] (\xx,1.55) -- (\xx,1.82);
\node[smk,text=goodc] at (9.8,2.2) {\textbf{digit set $E\subset\Z/M\Z$}};
\node[sml,text width=7.6cm,align=center] at (9.8,0.05)
 {accept $\alpha\iff$ all $6tD$ shifted exponents belong to $E$\\
  $\Pr$ per condition $=|E|/M$ \;$\Rightarrow$\;
  $\mathbb{E}|\mathcal A|=|H|\,(|E|/M)^{6tD}$};
\end{tikzpicture}
\caption{The $D$ projected exponents for one direction and base
vertex $(i,v)$. Independent uniform shifts make a fixed character
pass all $6tD$ tests with probability $(|E|/M)^{6tD}$, yielding
\cref{gp:selection}. The shifts change only the local code multipliers.}
\label{fig:funnel}
\end{figure}
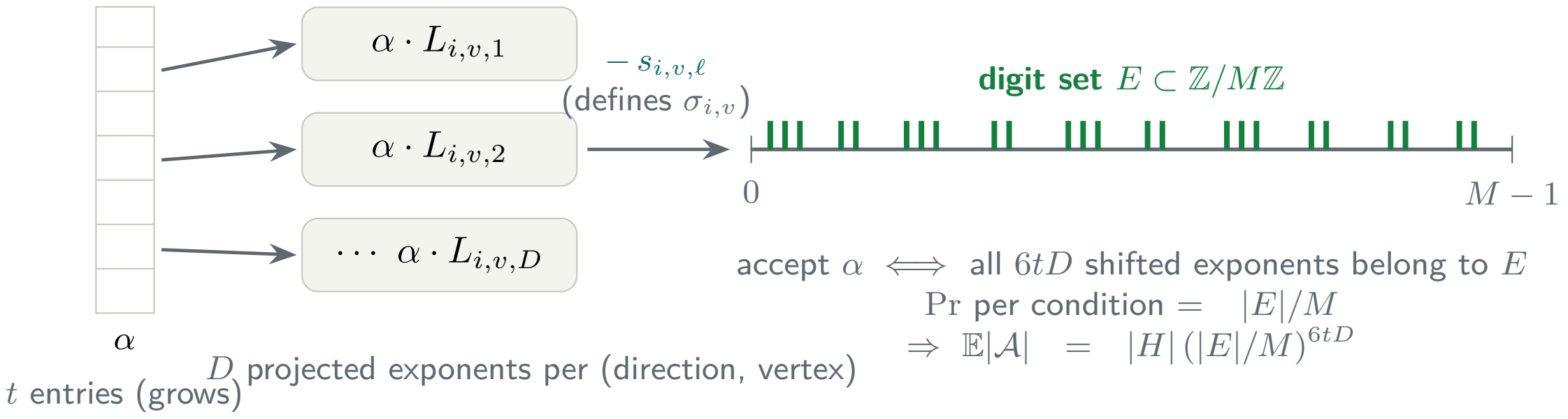

\Cref{fig:funnel} shows how the acceptance conditions select many
characters with one common family of local multipliers.

\begin{proposition}[Local constraints for the character span]
\label{gp:character-codewords}
For any shifts and the space $\mathcal V$ of
\cref{gp:accepted-characters}, each map $f\mapsto w_{i,f}$ is an injective linear map from
$\mathcal V$ into $\mathcal T_i$. In particular,
\[
 \dim_F\mathcal T_i\ge\dim_F\mathcal V=|\mathcal A|
 \qquad(i=1,2,3).
\]
\end{proposition}
\begin{proof}
For $f=\sum_{\alpha\in\mathcal A}u_\alpha\chi_\alpha$,
multiplicativity and \eqref{gp:restriction-identity} give
\begin{equation}\label{gp:whole-space-local}
 \frac{f(zg_{i,v}(c))}{\sigma_{i,v}(c)}
 =\sum_{\alpha\in\mathcal A}
      u_\alpha\chi_\alpha(z)f_{i,v,\alpha}(c).
\end{equation}
The right side evaluates a polynomial in $F[U_i]$ of degree
less than $k$, since each summand has that property.
This verifies \eqref{gp:construction-local-target} for every
$f,i,v,z$. The conclusion now follows from
\cref{gp:construction-common-space,gp:field-character-basis}.
\end{proof}

\subsection{Dimension bound and compatible parameter choices}
\label{gp:conclusion}

Choose shifts attaining the bound in \cref{gp:selection}.
The deficit $\epsilon_E$ from \cref{gp:allowed-exponents}
satisfies $|E|/M=M^{-\epsilon_E}$.
Substituting this identity into \eqref{gp:selection-size} gives
\begin{equation}\label{gp:dimension-exponent}
 \dim_F\mathcal V=|\mathcal A|
 \ge M^{t-6tD\epsilon_E}
 =|H|^{1-6D\epsilon_E}.
\end{equation}
Since $D$ is fixed and $\epsilon_E\to0$ by
\cref{gp:exponent-set-size}, for any $\zeta>0$ we can fix $r$
with $6D\epsilon_E<\zeta$. Then
\begin{equation}\label{gp:near-full-exponent}
 \dim_F\mathcal V\ge M^{(1-\zeta)t}
 =|H|^{1-\zeta}.
\end{equation}
This choice of $r$ is independent of $t$.

\begin{theorem}[Expanding covers with a common space of codewords]
\label{gp:theorem}
Fix a power $P\ge2$ of two, $0<\xi<1$, and $\zeta>0$.
There are fixed integers $d_0\ge3$ and $D\ge1$ such that,
for every sufficiently large odd $r$, the fields and parameters in
\eqref{gp:construction-fields} admit a choice of roots as in
\cref{gp:construction-slopes} and color sets as in
\cref{gp:construction-colored-base} with the following property.

For each $t$ in an unbounded sequence of positive integers,
there is a common simple
$d_0$-regular bipartite base graph $\Gamma_0$ with $t$ vertices
on each side, and a choice of slopes as in
\cref{gp:construction-slopes} for which the covers
$\Gamma_i\to\overline\Gamma_0$ of \cref{gp:construction-cover}
are simple, connected, and $n$-regular, with group $H=(F^\times)^t$.
In each cover, normalized biadjacency has the simple singular
value $1$ associated with the constant vectors, and all other
singular values are at most $\xi$.
There are shifts defining $\sigma_{i,v}$ as in
\eqref{gp:construction-rescaling} whose accepted index set
$\mathcal A$ from \cref{gp:accepted-characters} satisfies
$|\mathcal A|\ge|H|^{1-\zeta}$. For every
$\alpha\in\mathcal A$, $i\in\{1,2,3\}$, and $v\in V$,
\[
 \chi_\alpha(g_{i,v}(c))
 =\sigma_{i,v}(c)f_{i,v,\alpha}(c)
 \quad(c\in S_i),\qquad
 f_{i,v,\alpha}\in F[U_i],\quad \deg f_{i,v,\alpha}<n/100<k.
\]
The character span $\mathcal V$ of
\cref{gp:accepted-characters} has dimension $|\mathcal A|$ and
embeds by $f\mapsto w_{i,f}$ into each Tanner code $\mathcal T_i$.
All fields, roots, color sets, and parameters $d_0,D,n,k,r$
are independent of $t$.
\end{theorem}
\begin{proof}
Choose $0<\lambda_0,\delta,\eta<1/2$ with
$\lambda_0+\delta+\eta<\xi$.
Choose $d_0$ satisfying \eqref{gp:base-degree-condition}
and $D$ satisfying \eqref{gp:occupancy-condition}.
With $P,d_0,D$ fixed,
\cref{gp:slope-roots-exist,gp:palettes,gp:exponent-set-size,gp:frobenius}
hold for every sufficiently large odd $r$. By \eqref{gp:digit-density},
we may also require $6D\epsilon_E<\zeta$. Fix any such $r$.
Fix a choice of roots and color sets supplied by
\cref{gp:palettes}; all local data are now fixed.

Take the unbounded sequence of base graphs from
\cref{gp:analysis-parameters}.
For each sufficiently large $t$ in this sequence,
\cref{gp:spectrum} has positive success probability, so
there is a choice of slopes giving three simple connected
covers with spectral bound less than $\xi$.
For these fixed slopes, \cref{gp:selection} supplies shifts
and polynomial representatives.
Equation~\eqref{gp:dimension-exponent} gives
$|\mathcal A|\ge|H|^{1-\zeta}$, and
\cref{gp:character-codewords} gives the common subspace of
Tanner codewords. The shifts affect only the local codes,
so the spectral conclusions remain valid.
\end{proof}

Compatibility rests on \eqref{gp:order-union}: it allows $D$
to be fixed before increasing $r$ to meet the local degree and
dimension requirements.

To express the dimension bound in terms of code length, let $N_t$ be a
positive integer for each $t$ in the sequence under consideration,
and let $C>0$ be a real constant independent of $t$.

\begin{corollary}[Polynomial factors in the block length]
\label{gp:polynomial-length}
Fix $0<\zeta<1$ and take a family supplied by
\cref{gp:theorem} with deficit target $\zeta/2$.
If the lengths above satisfy $N_t\le C t^3|H|$,
then, for all sufficiently large $t$ in the resulting sequence,
\[
 \dim_F\mathcal V\ge N_t^{1-\zeta}.
\]
\end{corollary}
\begin{proof}
The theorem gives $\dim_F\mathcal V\ge M^{(1-\zeta/2)t}$.
Since $M>1$ is fixed,
\begin{equation}
 \frac{M^{(1-\zeta/2)t}}{(Ct^3M^t)^{1-\zeta}}
 =\frac{M^{\zeta t/2}}{C^{1-\zeta}t^{3(1-\zeta)}}
 \longrightarrow\infty.
\end{equation}
The stated upper bound on $N_t$ proves the claim.
\end{proof}

The same covers also give classical LDPC codes with a multiplication
property. The local Reed--Solomon multiplication rule determines
the target code, and graph expansion gives linear distance for
both codes.

\begin{corollary}[Classical LDPC codes with a multiplication property]
\label{gp:classical-multiplication}
For every $0<\eps<1$, the covers of \cref{gp:theorem} can be
chosen over a fixed finite field $F$ of characteristic two, with
code lengths $N_t=tn|H|\to\infty$, so that the following holds.
For each $i\in\{1,2,3\}$, there is an $F$-linear code
$\mathcal T_i'\subseteq F^{E(\Gamma_i)}$ such that
\begin{equation}
 \dim_F\mathcal T_i\ge N_t^{1-\eps},\qquad
 \mathcal T_i\odot\mathcal T_i\odot\mathcal T_i
 \subseteq\mathcal T_i'.
\end{equation}
Both codes have minimum Hamming distance $\Omega_\eps(N_t)$
and admit parity-check matrices whose row and column weights
are $O_\eps(1)$.
\end{corollary}
\begin{proof}
Apply \cref{gp:theorem} with $P=2$, $\xi=1/4$, and
$\zeta=\eps/2$. Each $\mathcal T_i$ has length $N_t=tn|H|$.
Applying \cref{gp:polynomial-length} with $\zeta=\eps$ and
$C=n$ gives the dimension bound for all sufficiently large $t$.
Define $\mathcal T_i'$ on the same graph by imposing the local
code $\sigma_{i,v}^3\odot\RS(S_i,3k-2)$ at every lift of $v$.
The local Schur product inclusion in \cref{appA:schur}, applied
at every vertex, gives the stated global inclusion.

The local codes defining $\mathcal T_i$ and $\mathcal T_i'$
have distances $n-k+1$ and $n-3k+3$, respectively. Both
relative distances exceed $4/7$, since $k=\lfloor n/7\rfloor$.
The Sipser--Spielman distance bound
\cite[Lemma~3.31]{GL}, with $\xi=1/4$, therefore gives minimum
distance at least $\frac47(\frac47-\frac14)N_t=9N_t/49$
for both Tanner codes.
Finally, imposing a parity-check matrix at each vertex gives
row weight at most $n$ and column weight at most $2n$,
since every edge meets two vertices. These bounds are constant
because $n$ is fixed.
\end{proof}

%% file: sections/trilinear.tex
\section{CSS codes and addressable logical non-Clifford gates}\label{cg:section}

We construct three CSS code blocks from the graph covers of
\cref{gp:section}. On the Cartesian product of the three graphs,
a 1-cocycle can be built as the tensor product of one graph
1-cocycle and two graph 0-cocycles. The common character space
$\mathcal V$ supplies such cocycles
that are invariant under the group action defining the balanced
product. This lets us compute their cup products one graph factor
at a time and then descend to the smaller balanced product complex.

Placing the three edge factors over marked base edges makes their
cup product supported over one base cube. Its sheet values are the
pointwise product of the three input functions. To distinguish these
cup products in cohomology, we construct a linear map that vanishes on
coboundaries and sends a cochain supported above the marked cube
to its sheet array. Choosing an information set for $\mathcal V$
then makes the logical pairing diagonal. Varying the evaluation
functional selects any subset of the resulting logical triples.

\subsection{Graph cocycles and their tensor products}\label{cg:graph-sec}

\paragraph{Fixed data and notation.}
Retain the fields, local parameters, covers $\Gamma_i$, incidence labels
$g_{i,v}$, and multipliers $\sigma_{i,v}$ of \cref{gp:section}.
Set $Q:=|H|=M^t$.
Thus each of the three copies of the colored base graph
$\overline\Gamma_0$ has $2t$ vertices and $tn$ edges, counting
parallel copies. Choose the base spectral bound to satisfy
$\lambda_0\le1/14$. We take the local parameter $r$ sufficiently
large for \cref{gp:theorem,cg:dual-existence}, then hold the field
and local parameters fixed as $t$ grows.

Throughout this section, $i\in\{1,2,3\}$ indexes a geometric
direction, $h\in\{1,2,3\}$ indexes a code block, and $j\in\{0,1,2,3\}$
is a cochain degree. An edge color in direction $i$ is denoted by
$c\in S_i$, and a sheet by $x\in H$. When the direction is fixed,
we abbreviate the edge color map $c_i(e)$ of \cref{gp:section}
by $c(e)$. Retain $U_i$ for the polynomial indeterminate in direction $i$.

To keep track of all sheet values, we use the group algebra
$R=F[H]$. For $x\in H$, write $\gb x$ for its group basis vector;
square brackets remain reserved for (co)homology classes. The
multiplication $\gb x\,\gb y=\gb{xy}$ extends $F$-bilinearly:
\begin{equation}\label{pre:convolution}
 \left(\sum_xa_x\gb x\right)\left(\sum_yb_y\gb y\right)
 =\sum_{z\in H}\left(\sum_{xy=z}a_xb_y\right)\gb z.
\end{equation}
For a function $G:H\to F$, its coefficient array is
$\gb G=\sum_{x\in H}G(x)\gb x$. We use $\star$ for convolution
and $\odot$ for pointwise multiplication:
\begin{equation}\label{pre:sheet-products}
 \begin{aligned}
 (G_1\star G_2)(z)&=\sum_{xy=z}G_1(x)G_2(y),\\
 (G_1\odot G_2)(z)&=G_1(z)G_2(z).
 \end{aligned}
\end{equation}
Thus $\gb{G_1\star G_2}=\gb{G_1}\gb{G_2}$, while $\odot$
is the coordinatewise operation used for local codes in
\cref{pre:sheafsec}. The identity is $1_R=\gb{1_H}$, supported
at the identity sheet. Multiplication by $\gb g$ translates arrays:
\begin{equation}\label{pre:group-translation}
 \gb g\left(\sum_xa_x\gb x\right)
 =\sum_xa_x\gb{gx}
 =\sum_xa_{g^{-1}x}\gb x.
\end{equation}

\paragraph{Choosing the local codes.}
A 1-cell in a Cartesian product of three graphs has one edge factor
and two vertex factors. To construct logical operators in block
$h$, we will take tensor products of a graph 1-cocycle in
direction $h$ and graph 0-cocycles in the other two directions.
We retain the local codes of \cref{gp:section} in those two
directions so that the accepted characters supply the required
0-cocycles. In direction $h$, every graph 1-cochain is a cocycle,
regardless of the vertex-to-edge restriction maps, because a graph
has no 2-cochains. We can therefore modify these maps
while preserving the tensor products as cocycles.

Modify the local code $L_i(v)$ of \cref{gp:construction-tanner}
at each direction-$i$ base vertex $v$ by setting
\begin{equation}
 P_i(v):=\gamma_{i,v} \odot L_i(v),
 \qquad \gamma_{i,v}:S_i\longrightarrow F^\times.
\end{equation}
Powers of multiplier vectors are coordinatewise. For now the
$\gamma_{i,v}$ are arbitrary vectors with nonzero entries.
Block $h$ uses $P_h$ in its edge direction and $L_i$ in the
other directions:

\begin{equation}
\begin{array}{c|ccc}
 &\text{direction }1&\text{direction }2&\text{direction }3\\ \hline
 \text{block }1&P_1&L_2&L_3\\
 \text{block }2&L_1&P_2&L_3\\
 \text{block }3&L_1&L_2&P_3
\end{array}
\end{equation}
Each \emph{base halfedge} $(v,e)$ of color $c$ receives its own
factor $\gamma_{i,v}(c)$, independently of the other endpoint.
These rescalings preserve local dimension and distance while
changing the agreement equations across edges. Using the same
factor on every lift makes the restriction maps commute with sheet
translations. We will choose
these factors so that the resulting cup products can be detected
in cohomology (\cref{cg:dual-existence}).

\paragraph{The graph coefficient systems.}
Write $\mathcal L_i$ and $\mathcal P_i$ for the coefficient systems
on $\Gamma_i$ with local codes $L_i(v)$ and $P_i(v)$,
respectively. Both have vertex stalk $F[U_i]_{<k}$ and edge stalk
$F$. For block $h$, denote the system in direction $i$ by
$\shf_{i,\{h\}}$: it is $\mathcal P_i$ when $i=h$ and
$\mathcal L_i$ otherwise. At a base incidence $(v,e)$ of color
$c=c(e)$, its restriction map is
\begin{equation}
 (\shf_{i,\{h\}})_{v\to e}:F[U_i]_{<k}\longrightarrow F,
 \qquad p\longmapsto
 \gamma_{i,v}(c)^{\one_{i=h}}\sigma_{i,v}(c)p(c).
\end{equation}
The same map is used at every lift. Since the lifted edge
$(e,x)$ meets $(v,x/g_{i,v}(c))$, the associated cochain complex is
\begin{equation}\label{cg:directional}
\begin{gathered}
 D_{i,\{h\}}^\bullet:=C^\bullet(\Gamma_i;\shf_{i,\{h\}}),
 \qquad
 D_{i,\{h\}}^0=R^{2tk}
 \xrightarrow{\ \delta^0_{i,\{h\}}\ }
 D_{i,\{h\}}^1=R^{tn},\\
 (\delta^0_{i,\{h\}}f)(e,x)
 =\sum_{v\in\partial e}
   (\shf_{i,\{h\}})_{v\to e}
   \bigl(f(v,x/g_{i,v}(c(e)))\bigr).
\end{gathered}
\end{equation}
Here $f(v,z)$ is the polynomial at the lifted vertex $(v,z)$.
At each vertex, we use the monomial basis
$1,U_i,\ldots,U_i^{k-1}$; at each edge, we use the basis
$\{1\}$ of its scalar stalk $F$. For each basis element at a
base vertex or edge, we collect its coefficients on all sheets
into one element of $R$. This gives the free $R$-module
coordinates above. The restriction maps commute with sheet
translations, so the differential is $R$-linear.

\begin{definition}[Marked edge cochain]\label{cg:marked-edge}
Fix one base edge $b_i$ in each direction. For every block $h$,
the graph complex $D_{i,\{h\}}^\bullet$ has 1-cochain space
$D_{i,\{h\}}^1=R^{tn}$. In this common space, let $\bar b_i$
be the vector with coordinate $1_R$ at $b_i$ and zero at
every other base edge.
Equivalently, it is the indicator of the single lifted edge
$(b_i,1_H)$:
\begin{equation}\label{cg:marked-coordinates}
 \bar b_i(e,x)=\one_{e=b_i}\one_{x=1_H},\qquad
 \bar b_i(e):=\one_{e=b_i},\qquad
 (\bar b_i a)(e,x)=\bar b_i(e)\,a_x
 \quad\left(a=\sum_{x\in H}a_x\gb x\in R\right).
\end{equation}
The notation $\bar b_i(e)$ therefore records only the base edge,
whereas $\bar b_i(e,x)$ also specifies the sheet. The $R$-span of
$\bar b_i$ consists of all cochains supported over $b_i$.
\end{definition}

\paragraph{Accepted characters and local polynomials.}
Use the accepted index set $\mathcal A$, its character span
$\mathcal V$, and the local polynomials
$f_{i,v,\alpha}\in F[U_i]_{<k}$ from \cref{gp:accepted-characters}.
These polynomials satisfy \eqref{gp:restriction-identity}.

\begin{lemma}[Graph cocycles from accepted characters]\label{cg:constituents}
For $\alpha\in\mathcal A$ and $i\in\{1,2,3\}$, define
\begin{equation}
 \zeta^{(0)}_{i,\alpha}(v,z)
   =\chi_\alpha(z)f_{i,v,\alpha}
       \in F[U_i]_{<k},
 \qquad
 \xi^{(1)}_{i,\alpha}(e,x)
   =\bar b_i(e)\chi_\alpha(x).
\end{equation}
Then $\zeta^{(0)}_{i,\alpha}$ is a 0-cocycle for
$\mathcal L_i$. Restricting its polynomial from either endpoint
to the lifted edge $(e,x)$ gives $\chi_\alpha(x)$. The 1-cochain
$\xi^{(1)}_{i,\alpha}$ is a cocycle for $\mathcal P_i$ (and
for any of the coefficient systems on $\Gamma_i$).
\end{lemma}
\begin{proof}
At the endpoint $(v,x/g_{i,v}(c))$ of a color-$c$ edge,
\eqref{gp:restriction-identity} gives
\begin{equation}\label{cg:compensation}
 \sigma_{i,v}(c)f_{i,v,\alpha}(c)
       \chi_\alpha\!\left(x/g_{i,v}(c)\right)
 =\chi_\alpha(g_{i,v}(c))
       \frac{\chi_\alpha(x)}{\chi_\alpha(g_{i,v}(c))}
 =\chi_\alpha(x).
\end{equation}
Thus the polynomial evaluation cancels the change of sheet at the
endpoint. Both endpoints give the same value, so their sum is
zero in characteristic two, proving that $\zeta^{(0)}_{i,\alpha}$
is a cocycle. The cochain $\xi^{(1)}_{i,\alpha}$ is a cocycle
because a graph has no 2-cochains.
\end{proof}

\paragraph{Cocycles on the Cartesian product.}\label{pre:crosssec}
A cell in a Cartesian product of graphs is a product of vertices and
edges; its dimension is the number of edge factors. A cochain assigns
a vector in the corresponding coefficient space to each cell of its
degree. To combine graph cocycles, we take tensor products of their
coefficients. Throughout this section, $\otimes$ without a subscript
denotes tensor product over $F$.

For sheaves $\shf,\shf'$ on graphs $X,Y$, their \emph{external
product} $\shf\boxtimes\shf'$ is a coefficient system on the
Cartesian product $X\times Y$. For cells $s\subseteq X$ and
$s'\subseteq Y$, its stalk is
\begin{equation}\label{pre:external-stalk}
 (\shf\boxtimes\shf')_{(s,s')}=\shf_s\otimes\shf'_{s'}.
\end{equation}
Each restriction map applies the graph's restriction map in the
coordinate that changes and the identity in the other coordinate.

For $f\in C^i(X;\shf)$ and $g\in C^j(Y;\shf')$, their
\emph{cross product} $f\times g$ is the $(i+j)$-cochain on
$X\times Y$ obtained by taking the tensor product of their values:
\begin{equation}\label{pre:cross-coordinates}
 (f\times g)_{(s, s')}=f_s\otimes g_{s'}
 \qquad(\dim s=i,\ \dim s'=j).
\end{equation}
It is zero on cells of other bidegrees.
Extending $f\otimes g\mapsto f\times g$
linearly gives the natural isomorphism
\begin{equation}\label{pre:external-cochains}
 C^\bullet(X;\shf)\otimes C^\bullet(Y;\shf')
 \cong C^\bullet(X\times Y;\shf\boxtimes\shf').
\end{equation}
Under this identification, the cross product $f\times g$
corresponds to the tensor $f\otimes g$.
On the left, degree $j$ is the direct sum of
$C^i(X;\shf)\otimes C^{j-i}(Y;\shf')$ over $i$.
The differential acts one direction at a time:
\begin{equation}\label{pre:external-differential}
 \delta(f\times g)=\delta f\times g+f\times\delta g.
\end{equation}
The usual signs disappear in characteristic two. In the product of
three graphs, a $j$-cell has $j$ edge factors and $3-j$ vertex factors.

Form the tensor product cochain complex over $F$:
\begin{equation}
 T_{\{h\}}^\bullet
 :=D_{1,\{h\}}^\bullet\otimes_F D_{2,\{h\}}^\bullet
                            \otimes_F D_{3,\{h\}}^\bullet,
 \qquad \widetilde B:=\Gamma_1\times\Gamma_2\times\Gamma_3.
\end{equation}
By \eqref{pre:external-cochains}, this is the cochain complex on
$\widetilde B$ with the external product of the graph coefficient
systems. Its $j$-cochain space is the direct sum over choices of
$j$ edge directions: those factors contribute $D_{i,\{h\}}^1$
and the remaining factors contribute $D_{i,\{h\}}^0$.
This is the three factor version of the hypergraph product (HGP)
construction, whose tensor products of graph cocycles underlie the
cup product calculations in \cite{LiEtAl2026Cohomological}.

By the differential formula \eqref{pre:external-differential},
the tensor product of one graph 1-cocycle and two graph
0-cocycles is a 1-cocycle on $\widetilde B$.
Thus, for $\alpha\in\mathcal A$,
\cref{cg:constituents} gives the following 1-cocycles:
\begin{equation}\label{cg:pullback-representatives}
 \begin{array}{c|l}
 \text{block}&\text{1-cocycle on }\widetilde B\\ \hline
 1&\xi^{(1)}_{1,\alpha}\otimes_F\zeta^{(0)}_{2,\alpha}
                         \otimes_F\zeta^{(0)}_{3,\alpha}\\
 2&\zeta^{(0)}_{1,\alpha}\otimes_F\xi^{(1)}_{2,\alpha}
                         \otimes_F\zeta^{(0)}_{3,\alpha}\\
 3&\zeta^{(0)}_{1,\alpha}\otimes_F\zeta^{(0)}_{2,\alpha}
                         \otimes_F\xi^{(1)}_{3,\alpha}
 \end{array}
\end{equation}
Their dependence on the three sheet coordinates is through
$\chi_\alpha(x_1)\chi_\alpha(x_2)\chi_\alpha(x_3)
=\chi_\alpha(x_1x_2x_3)$. We can therefore identify sheet triples
with the same product while retaining these cocycles.

\subsection{Balanced product codes and descended cocycles}
\label{cg:balanced}\label{cg:embed-sec}\label{pre:balsec}

Covering maps transport logical pairings between HGP and
larger complexes in \cite[Section~5]{LLL}. Here the Cartesian
product $\widetilde B$ is the covering complex. To form the
balanced product, we express the identification of sheet triples
as a quotient by a group action that preserves the local restriction
maps. This gives two equivalent descriptions of cochains on the quotient.

\begin{definition}[Free actions; equivariant sheaves]\label{pre:equivar}
A finite group $G$ acts freely on a directed graph $\Gamma$ if it
preserves directed incidences and no nonidentity element fixes a vertex.
A system $\shf$ is equivariant when, under the fixed identifications
of stalks along each orbit,
$\shf_{gv\to ge}=\shf_{v\to e}$ for every $g,v,e$.
The induced action on a cochain $c$ is
$(gc)_s=c_{g^{-1}s}$, using these stalk identifications;
it commutes with $\delta$.
\end{definition}

For complexes with coordinate bases indexed by cells, freeness
refers to the action on those indices. The cochain spaces themselves
are vector spaces on which the action is linear.

\begin{definition}[Balanced product; cf.~\cite{BE}]\label{pre:balanced}
Let $G$ be a finite abelian group, and let $A^\bullet,B^\bullet$ be
cochain complexes with specified coordinate bases permuted freely by
$G$. Assume that the actions commute with the differentials. On their
tensor product over $F$, let $g\in G$ act by
\begin{equation}\label{pre:balanced-action}
 g\cdot (a\otimes b)=ga\otimes g^{-1}b.
\end{equation}
The \emph{balanced product} in the invariant model is the subcomplex
\begin{equation}\label{pre:balanced-invariants}
 (A^\bullet\otimes B^\bullet)^G
 =\{c\in A^\bullet\otimes B^\bullet:
       g\cdot c=c\text{ for every }g\in G\}.
\end{equation}
A basis for this space consists of orbit sums of coordinate basis
pairs under $(a,b)\sim(ga,g^{-1}b)$. The residual action
$a\otimes b\mapsto ga\otimes b$ commutes with the balanced action
and acts freely on the orbit sum basis. Consequently, each
cochain group is a free $F[G]$-module for this residual action;
the basis can have several $G$-orbits.
\end{definition}

\begin{remark}[Invariants versus coinvariants]\label{pre:invcoinv}
Tensoring over the commutative group algebra $F[G]$ gives the quotient
\begin{equation}\label{pre:balanced-coinvariants}
 A^\bullet\otimes_{F[G]}B^\bullet
 =\frac{A^\bullet\otimes B^\bullet}
 {\Span_F\{ga\otimes b-a\otimes gb:
              g\in G,\ a\in A^\bullet,\ b\in B^\bullet\}}.
\end{equation}
This is the space of \emph{coinvariants} for the action in
\eqref{pre:balanced-action}. Assume now that $|G|$ is odd. Since
$|G|\cdot1_F=1$, the averaging map to the space of invariants is
\begin{equation}\label{pre:balanced-transfer}
 \begin{aligned}
 \mathrm{avg}:A^\bullet\otimes_{F[G]}B^\bullet
    &\longrightarrow(A^\bullet\otimes B^\bullet)^G,\\
 a\otimes_{F[G]}b&\longmapsto\sum_{g\in G}g a\otimes g^{-1}b.
 \end{aligned}
\end{equation}
The balancing relation preserves this sum by reindexing $g$, so the
map is well defined. It commutes with the differential and is inverse to the quotient
projection restricted to invariants: either composite is the sum of
$|G|$ identical terms. Thus the two spaces are isomorphic complexes.

\end{remark}

For the three graph covers, let
\begin{equation}
 K:=\{(h_1,h_2,h_3)\in H^3:h_1h_2h_3=1_H\}\cong H^2
\end{equation}
act on sheets by
$(x_1,x_2,x_3)\mapsto(x_1h_1,x_2h_2,x_3h_3)$.
The orbit of a sheet triple is determined by $x=x_1x_2x_3$.
Thus the quotient replaces the $Q^3$ sheet triples above each
triple of base cells by $Q$ balanced sheets, with $Q^2$ lifts of
each quotient cell. Write $B=\widetilde B/K$, let
$\pi:\widetilde B\to B$ be the quotient map, and let $B(j)$ be
its $j$-cells. For each block $h$, the external product of its
graph coefficient systems descends to $\shf_h=\shf_{\{h\}}$ on $B$.

\begin{definition}[Balanced product complexes and code blocks]
\label{cg:complexes}
For $h\in\{1,2,3\}$, define
\begin{equation}
 \mathcal C_h^\bullet=\mathcal C_{\{h\}}^\bullet
 :=D_{1,\{h\}}^\bullet\otimes_R D_{2,\{h\}}^\bullet
                              \otimes_R D_{3,\{h\}}^\bullet.
\end{equation}
The degree summands are the same as in $T_{\{h\}}^\bullet$, with
tensor products now taken over $R$. Write
$\delta_h^j:\mathcal C_h^j\to\mathcal C_h^{j+1}$ for the sum of
differentials acting in the three tensor factors.
\end{definition}

The balancing relations for $\otimes_R$ identify a sheet
translation in one graph factor with the same translation in
another. These are precisely the coinvariant relations for the
$K$-action (\cref{pre:balanced}). Consequently,
\begin{equation}\label{cg:quotient-models}
 \mathcal C_h^\bullet=C^\bullet(B;\shf_h)
 \cong(T_{\{h\}}^\bullet)_K
 \cong(T_{\{h\}}^\bullet)^K.
\end{equation}
The identification with invariant cochains is realized by
pullback along $\pi:\widetilde B\to B$. Together with the orbit
projection, this gives the maps
\begin{equation}\label{cg:quotient-maps}
 \pi^*:\mathcal C_h^\bullet\longrightarrow(T_{\{h\}}^\bullet)^K,
 \qquad
 \tau:T_{\{h\}}^\bullet\longrightarrow\mathcal C_h^\bullet.
\end{equation}
Pullback copies a quotient cell's coefficient to every lift;
the orbit projection sums coefficients over those lifts.
Each fiber contains $|K|=Q^2$ cells. Since $F$ has characteristic
two and $Q^2$ is odd, summing a common coefficient over a fiber
returns that coefficient. Hence $\tau\pi^*=\mathrm{id}$ and
$\pi^*\tau=\mathrm{id}$ on invariant cochains.
Thus $\pi^*$ and the restriction of $\tau$ to invariant
cochains are inverse, giving the averaging identification of
\cref{pre:invcoinv}.

\begin{definition}[Descended cochains]\label{cg:phidef}
For $\alpha\in\mathcal A$, the block-$h$ cocycle in
\eqref{cg:pullback-representatives} is $K$-invariant, since its
sheet dependence is $\chi_\alpha(x_1x_2x_3)$. Define
$\Phi_h(\chi_\alpha)\in\mathcal C_h^1$ uniquely by requiring
its pullback to equal this cocycle.
The descended cochain vanishes on edges in directions other than
$h$, and on a direction-$h$ cell it is
\begin{equation}\label{cg:embedding}
 \Phi_h(\chi_\alpha)(e_h;(v_i)_{i\ne h};x)
 =\bar b_h(e_h)\chi_\alpha(x)
      \prod_{i\ne h}f_{i,v_i,\alpha}(U_i).
\end{equation}
Here $x$ is the product of the sheet labels of the edge factor
and the two vertex factors in any lift of this 1-cell to
$\widetilde B$. The polynomial product denotes the coefficient
tensor in the two vertex directions.
Extend this assignment on the character basis $F$-linearly to a map
$\Phi_h:\mathcal V\to\mathcal C_h^1$: for
$G=\sum_{\alpha\in\mathcal A}u_\alpha\chi_\alpha$ with
$u_\alpha\in F$, set
$\Phi_h(G)=\sum_{\alpha\in\mathcal A}u_\alpha\Phi_h(\chi_\alpha)$.
\end{definition}

\begin{corollary}[The descended cochains are cocycles]
\label{cg:cocycle}
For every $G\in\mathcal V$, $\Phi_h(G)\in\ker\delta_h^1$.
For $\alpha\in\mathcal A$, consider a cube with base edges
$(e_1,e_2,e_3)$ and balanced sheet $x$. The coefficient of
$\Phi_h(\chi_\alpha)$ on any direction-$h$ edge of this cube maps to
$\bar b_h(e_h)\chi_\alpha(x)$ under the edge-to-cube restriction map.
\end{corollary}
\begin{proof}
The tensor products in \eqref{cg:pullback-representatives} are cocycles,
so $\delta\pi^*\Phi_h(\chi_\alpha)=0$ for $\alpha\in\mathcal A$.
Since $\pi^*$ is injective
and commutes with coboundaries, $\delta_h^1\Phi_h(\chi_\alpha)=0$.
Linearity gives the claim for every $G\in\mathcal V$.

For the restriction formula, lift the cube to the Cartesian
product, with graph edges $(e_i,x_i)$ and balanced sheet
$x=x_1x_2x_3$. Each direction-$h$ edge of this cube uses a vertex
factor in the other two directions. By \eqref{cg:compensation},
the polynomial assigned by $\zeta^{(0)}_{i,\alpha}$ to either
endpoint of $(e_i,x_i)$ maps to $\chi_\alpha(x_i)$ under the
corresponding vertex-to-edge map.
The direction-$h$ edge factor contributes
$\bar b_h(e_h)\chi_\alpha(x_h)$. Applying the edge-to-cube map
therefore gives
\begin{equation}
 \bar b_h(e_h)\chi_\alpha(x_h)
       \prod_{i\ne h}\chi_\alpha(x_i)
 =\bar b_h(e_h)\chi_\alpha(x).
\end{equation}
\end{proof}

The modified direction $h$ contributes only its edge stalk $F$,
so no $\gamma$ multiplier occurs in \eqref{cg:embedding}.

\paragraph{CSS checks and physical qudits.}
\begin{proposition}[CSS length and sparsity]
\label{cg:css-parameters}
For $0\le j\le3$, the number of $j$-cells and the length
of each of the three CSS blocks are
\begin{equation}\label{cg:length}
 |B(j)|=\binom3j2^{3-j}t^3n^jQ,\qquad
 N_0:=\dim_F\mathcal C_h^1=12t^3nk^2Q.
\end{equation}
In monomial coefficient bases, take
\begin{equation}
 H_{X,h}:=(\delta_h^0)^{\mathsf T},\qquad
 H_{Z,h}:=\delta_h^1.
\end{equation}
These matrices define commuting X- and Z-type stabilizer checks.
The encoded $F$-dimension of block $h$ is
$\calK_{0,h}=\dim_F H^1(\mathcal C_h)$.
The column and row weights of $\delta_h^0$ are at most $3n$ and
$2k$, respectively; those of $\delta_h^1$ are at most $2n$ and
$4k$. In particular, X- and Z-check weights are at most $3n$
and $4k$, and a physical qudit is incident to at most
$2k+2n$ checks in total.
\end{proposition}
\begin{proof}
Choose the $j$ edge directions. Each contributes $tn$ base
edges, while each remaining direction contributes $2t$ base
vertices. There are $Q$ balanced sheets above each base cell,
giving the stated cell count. The stalk on a $j$-cell has
dimension $k^{3-j}$, so $N_0=|B(1)|k^2$.

For $j=0,1$, each restriction map evaluates one polynomial variable
and leaves the other exponents unchanged. A coefficient on a
$j$-cell therefore contributes to at most one coefficient on
each of its $(3-j)n$ containing $(j+1)$-cells. Thus the column
weight of $\delta_h^j$ is at most $(3-j)n$.
Conversely, a $(j+1)$-cell has $2(j+1)$ boundary $j$-cells.
For a fixed output coefficient, the evaluated variable allows
at most $k$ input coefficients from each boundary cell.
Hence the row weight is at most $2(j+1)k$.

Since $\delta_h^1\delta_h^0=0$, the check matrices satisfy
$H_{X,h}H_{Z,h}^{\mathsf T}=0$, and the encoded dimension is
$N_0-\operatorname{rank}\delta_h^0-
\operatorname{rank}\delta_h^1=\dim_F H^1(\mathcal C_h)$.
Transposing $\delta_h^0$ gives the stated check weights;
each physical qudit occurs in at most $2k$ X-checks
and $2n$ Z-checks.
\end{proof}

Each physical qudit is indexed by a polynomial coefficient on a
1-cell. With the local parameters fixed, the check weights are
bounded independently of $t$.

\subsection{Cup products on the marked cube}\label{cg:cup-sec}

We first construct a trilinear map from the computational basis label spaces
to top cochains. When two inputs are cocycles and the third is a
coboundary, its output must be a coboundary. The cup product has this
property by the Leibniz rule. We define it using polynomial
multiplication in the local coefficient spaces, then evaluate it on the
cocycles already constructed.

\subsubsection{Compatible coefficient pairings}\label{pre:cupsec}

To multiply cochains from different code blocks, we need bilinear
pairings of their coefficient spaces on each cell. These pairings must commute with
restriction: pairing two coefficients and then restricting must give
the same result as restricting each coefficient and then pairing
the results.

\begin{definition}[Sheaf pairing]\label{pre:pairing}
A \emph{pairing} $m$ of cellular sheaves $\shf,\shf'$ on $\Gamma$
with values in $\shf''$ consists of $F$-bilinear maps
\[
 m_v:\shf_v\times\shf'_v\longrightarrow\shf''_v,
 \qquad
 m_e:\shf_e\times\shf'_e\longrightarrow\shf''_e
\]
such that, for every incidence $v\in\partial e$ and every
$a\in\shf_v$, $b\in\shf'_v$,
\begin{equation}\label{pre:pairing-compatibility}
 m_e\big(\shf_{v\to e}(a),\shf'_{v\to e}(b)\big)
 =\shf''_{v\to e}\big(m_v(a,b)\big).
\end{equation}
\end{definition}

The maps in \cref{pre:pairing} pair coefficients on the same vertex
or the same edge. For the cup product of a 0-cochain and a 1-cochain,
first restrict the vertex coefficient to the edge, then apply the
edge pairing. Both inputs to that pairing lie in edge stalks.

For two polynomial stalks with degree bounds $k$ and $k'$ satisfying
$k+k'-1\le |S|$, evaluation remains injective in the target
coefficient system. The vertex
pairing is polynomial multiplication:
\begin{equation}\label{pre:polynomial-pairing}
 m_v:F[U]_{<k}\times F[U]_{<k'}
       \longrightarrow F[U]_{<k+k'-1},
 \qquad (f,g)\longmapsto fg.
\end{equation}
Thus $m_v(U^a,U^b)=U^{a+b}$. It factors uniquely through a linear map
from the $kk'$-dimensional tensor product
$F[U]_{<k}\otimes F[U]_{<k'}$. The target
space has dimension $k+k'-1$. When $k=k'=2$, the distinct tensors
$U\otimes1$ and $1\otimes U$ both map to $U$.
On edges, $m_e:F\times F\to F$ is ordinary multiplication in $F$. These maps commute with restriction because
\begin{equation}\label{pre:restriction-pairing}
 \sigma(c)f(c)\,\sigma'(c)g(c)
   =(\sigma(c)\sigma'(c))(fg)(c).
\end{equation}
Evaluating polynomial products gives the coordinatewise products
defined before \cref{pre:vrow}.
In particular,
\begin{equation}\label{pre:schurrs}
\big(\sigma \odot \RS(S,k)\big) \odot \big(\sigma' \odot \RS(S,k')\big)
\;\subseteq\;(\sigma \odot \sigma') \odot \RS(S,\,k+k'-1),
\end{equation}
because evaluations multiply coordinatewise and polynomial degrees add
(\cref{app:classical}).

In the cup product formulas below, edge pairings are ordinary
multiplication in $F$, as in the polynomial systems above.

For the three code blocks, the following target systems accommodate
all products of their local polynomials.

\begin{definition}[Coefficient systems on the graphs]
\label{cg:coefficient-systems}
For nonempty $J\subseteq\{1,2,3\}$, put
\begin{equation}\label{cg:productcodes}
 r_J:=|J|(k-1)+1,\qquad
 C_{i,J}(v):=\gamma_{i,v}^{\one_{i\in J}} \odot
          \sigma_{i,v}^{|J|} \odot \RS(S_i,r_J).
\end{equation}
For the target of the triple cup product, put
$u:=r_{\{1,2,3\}}=3k-2<n/2$,
using $k=\lfloor n/7\rfloor$.
For $|J|\ge2$, define $\shf_{i,J}$ on $\Gamma_i$ to have vertex
stalk $F[U_i]_{<r_J}$ and edge stalk $F$. At a base incidence
$(v,e)$ of color $c=c(e)$, set
\begin{equation}
 (\shf_{i,J})_{v\to e}:F[U_i]_{<r_J}\longrightarrow F,
 \qquad p\longmapsto
 \gamma_{i,v}(c)^{\one_{i\in J}}
 \sigma_{i,v}(c)^{|J|}p(c).
\end{equation}
The same map is used at every lift. For $J=\{h\}$, these formulas
agree with the input systems: $C_{i,\{h\}}(v)$ is $P_i(v)$ if
$i=h$ and $L_i(v)$ otherwise. The $n$ maps at each vertex
together identify its stalk with $C_{i,J}(v)\subseteq F^{S_i}$,
since evaluation at the distinct colors is injective for $r_J\le u<n$.
\end{definition}

For disjoint nonempty sets $J,J'\subseteq\{1,2,3\}$,
$r_{J\cup J'}=r_J+r_{J'}-1$, and the restriction multipliers
multiply to those of $J\cup J'$. By \eqref{pre:restriction-pairing},
polynomial multiplication at vertices and scalar multiplication
at edges therefore give the sheaf pairing of \cref{pre:pairing},
with target $\shf_{i,J\cup J'}$. On evaluation codes, this is
the Schur product containment \eqref{pre:schurrs}.

For $|J|\ge2$, let
$D_{i,J}^\bullet=C^\bullet(\Gamma_i;\shf_{i,J})$.
As for the input systems, form the graph and product complexes
\begin{equation}\label{cg:product-complexes}
 \begin{aligned}
 D_{i,J}^0=R^{2tr_J}
 &\xrightarrow{\ \delta^0_{i,J}\ }D_{i,J}^1=R^{tn},\\
 T_J^\bullet
 &=D_{1,J}^\bullet\otimes_F D_{2,J}^\bullet\otimes_F D_{3,J}^\bullet,\\
 \mathcal C_J^\bullet
 &=D_{1,J}^\bullet\otimes_R D_{2,J}^\bullet\otimes_R D_{3,J}^\bullet.
 \end{aligned}
\end{equation}
The graph differential is given by \eqref{cg:directional} with
$\{h\}$ replaced by $J$. The external product of the graph
systems descends to $\shf_J$ on $B$, with the same quotient
identifications and maps as in
\eqref{cg:quotient-models}--\eqref{cg:quotient-maps}. Write
$\delta_J^j:\mathcal C_J^j\to\mathcal C_J^{j+1}$ for the
tensor product differential, agreeing with $\delta_h^j$ when
$J=\{h\}$. For the target of the triple cup product, write
$\shf_W:=\shf_{\{1,2,3\}}$,
$\mathcal C_W^\bullet:=\mathcal C_{\{1,2,3\}}^\bullet$, and
$\delta_W^j:=\delta_{\{1,2,3\}}^j$.

\begin{figure}[htbp]
\centering
\begin{tabular}{@{}cclc@{}}
\toprule
cell type & degree $j$ & stalk in $\mathcal C_J$ & dimension\\
\midrule
VVV & $0$ & $F[U_1]_{<r_J}\otimes F[U_2]_{<r_J}
                   \otimes F[U_3]_{<r_J}$ & $r_J^3$\\
EVV & $1$ & $F[U_2]_{<r_J}\otimes F[U_3]_{<r_J}$ & $r_J^2$\\
EEV & $2$ & $F[U_3]_{<r_J}$ & $r_J$\\
EEE & $3$ & $F$ & $1$\\
\bottomrule
\end{tabular}
\caption{Coefficient spaces in the balanced product. Here V and E
 denote vertex and edge factors, $J\subseteq\{1,2,3\}$ indexes the
multiplied code blocks, and $r_J=|J|(k-1)+1$, where $k$ is the input
local code dimension. The EVV and EEV rows show representative orientations; permuting directions
permutes the remaining polynomial variables. A coboundary replaces
one vertex factor by an edge factor, evaluating that variable and
applying its incidence multiplier. The coefficient label $J$ determines
the polynomial degree bounds, while the cell degree $j$ determines
which polynomial variables remain.}
\label{fig:sheaf}
\end{figure}

\subsubsection{Cup products on graphs and cubes}

\begin{definition}[Cup product on a graph]\label{pre:cup}
Given a pairing $m$ as above, the bilinear cup product
$\smile:C^i(\Gamma,\shf)\times C^j(\Gamma,\shf')\to
C^{i+j}(\Gamma,\shf'')$ is given, for $e=(v,v')$, by
\begin{align*}
(i,j)=(0,0):&\quad (c\smile c')_v=m_v(c_v,c'_v),\\
(i,j)=(0,1):&\quad (c\smile c')_e
       =\shf_{v\to e}(c_v)\cdot c'_e
       \qquad\text{(tail)},\\
(i,j)=(1,0):&\quad (c\smile c')_e
       =c_e\cdot\shf'_{v'\to e}(c'_{v'})
       \qquad\text{(head)}.
\end{align*}
The cup product of two 1-cochains vanishes because the graph has
no $2$-cells.
\end{definition}

\begin{lemma}[Leibniz rule and cup product on cohomology]\label{pre:leibniz}
In characteristic two,
$\delta(c\smile c')=\delta c\smile c'+c\smile\delta c'$.
Consequently the cochain cup product induces a well-defined bilinear map
$H^i(\Gamma,\shf)\times H^j(\Gamma,\shf')\to
H^{i+j}(\Gamma,\shf'')$.
\end{lemma}
\begin{proof}
Only $(i,j)=(0,0)$ requires a calculation on a graph. At one edge,
write $c_\pm=\shf_{v_\pm\to e}(c_{v_\pm})$, and use the same
convention for $c'_\pm$. Then
\begin{equation}\label{pre:leibniz-edge}
 \underbrace{c_+c'_+-c_-c'_-}_{\delta(c\smile c')}
 =\underbrace{(c_+-c_-)}_{\delta c}\,c'_+
   +c_-\,\underbrace{(c'_+-c'_-)}_{\delta c'}.
\end{equation}
This is the Leibniz rule with the endpoint choices of \cref{pre:cup};
addition and subtraction agree in characteristic two. If both inputs
are cocycles, the rule shows that their product is a cocycle.
If $c=\delta a$ and $\delta c'=0$, the rule gives
$c\smile c'=\delta(a\smile c')$. The same argument applies to the
other slot, proving that the product depends only on cohomology classes.
\end{proof}

\begin{lemma}[The cup product is a Schur product on cocycles]\label{pre:cupschur}
Identify a 0-cocycle $z\in Z^0(\Gamma,\shf)$ with its
evaluated Tanner codeword $c(z)$ of \eqref{pre:evaluated-tanner}.
For a $1$-cochain $x\in C^1(\Gamma,\shf')=F^{E(\Gamma)}$ and a
$0$-cocycle $z'$,
\[
 z\smile x=c(z) \odot x,\qquad
 c(z\smile z')=c(z) \odot c(z')
      \in\Tan(\Gamma,\shf'').
\]
In particular, for the indicator $x=\one_e$ of a single edge,
$z\smile\one_e=c(z)_e\,\one_e$.
\end{lemma}
\begin{proof}
In degrees $(0,1)$, \cref{pre:cup} gives
$(z\smile x)_e=\shf_{v\to e}(z_v)x_e=c(z)_ex_e$.
The cocycle condition makes the chosen endpoint immaterial.
In degrees $(0,0)$, restrict the vertex pairing to an edge and apply
the compatibility in \cref{pre:pairing}.
\end{proof}

The constructions of~\cite{GL,GTZ} use this identity to select
coordinates. Codewords that restrict to coordinate indicators are
cupped with an edge indicator; their Schur product tests whether the
selected coordinates agree. \Cref{cg:keyidentity} uses the same
calculation with sheet labels as the coordinates.

\begin{figure}[htbp]
\centering
\begin{tikzpicture}[x=1cm,y=1cm]
\node[smk] at (2.6,2.95) {\textbf{degrees $(0,1)$: $c\smile c'$}};
\draw[mutedc,line width=1.2pt] (0.8,1.3) -- (4.4,1.3);
\fill[dira] (0.8,1.3) circle(2.6pt);
\draw[dira,line width=1pt] (4.4,1.3) circle(2.6pt);
\node[sml,below=4pt] at (0.8,1.3) {$v$ (tail, $-$)};
\node[sml,below=4pt] at (4.4,1.3) {$v'$ (head, $+$)};
\node[draw=accentc,fill=accentc!8,rounded corners=2.5mm,font=\footnotesize\sffamily,text=accentc] (chipA) at (0.9,2.25) {$c_v\in\shf_v$};
\draw[-{Stealth[length=2.2mm]},accentc] (chipA.south east) -- (1.9,1.42);
\node[sml] at (3.1,1.95) {restriction $\shf_{v\to e}$};
\node[smk] at (2.6,0.45) {$(c\smile c')_e=\shf_{v\to e}(c_v)\cdot c'_e$};
\node[sml] at (2.6,0.0) {0-cochain factor before the edge $\to$ front face};
\draw[linec,dashed] (6.4,-0.2) -- (6.4,3.1);
\node[smk] at (10.1,2.95) {\textbf{degrees $(1,0)$: $c\smile c'$}};
\draw[mutedc,line width=1.2pt] (8.6,1.3) -- (12.2,1.3);
\draw[dirb,line width=1pt] (8.6,1.3) circle(2.6pt);
\fill[dirb] (12.2,1.3) circle(2.6pt);
\node[sml,below=4pt] at (8.6,1.3) {$v$};
\node[sml,below=4pt] at (12.2,1.3) {$v'$};
\node[draw=goldc,fill=goldc!10,rounded corners=2.5mm,font=\footnotesize\sffamily,text=goldc] (chipB) at (12.1,2.25) {$c'_{v'}\in\shf'_{v'}$};
\draw[-{Stealth[length=2.2mm]},goldc] (chipB.south west) -- (11.1,1.42);
\node[smk] at (10.4,0.45) {$(c\smile c')_e=c_e\cdot \shf'_{v'\to e}(c'_{v'})$};
\node[sml] at (10.4,0.0) {0-cochain factor after the edge $\to$ back face};
\end{tikzpicture}
\caption{The cup product of cellular sheaves on a graph
(\cref{pre:cup}). The position of a 0-cochain factor relative to the
1-cochain factor in the ordered product determines which vertex-to-edge
map is applied: use the tail (front face) before the edge and the head
(back face) after it. With the polynomial coefficient pairings used here,
this convention gives associativity and the
Leibniz rule, and iterating it over three directions produces the cubical
diagonal approximation of \cref{fig:cube}.}
\label{fig:cuprule}
\end{figure}
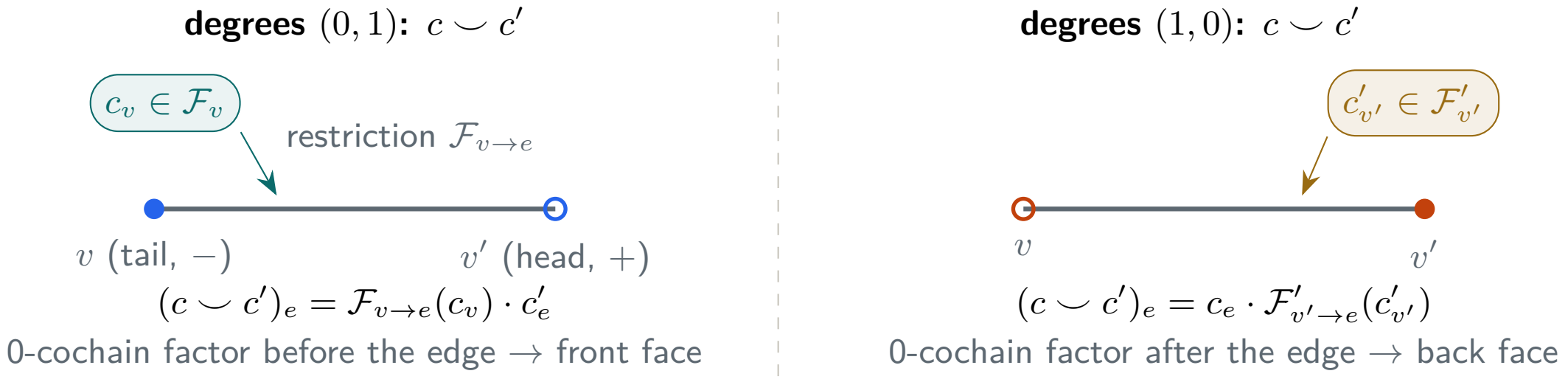

For the polynomial multiplication pairings used here, iterated cup
products are associative. At a vertex this is associativity of
polynomial multiplication. On an edge, every 0-cochain factor
before the unique 1-cochain factor is restricted at the tail, and
every factor after it is restricted at the head, regardless of
parenthesization. Thus the endpoint convention and associative
coefficient multiplication give associativity
(\cref{fig:cuprule}).

\begin{lemma}[Cup product on a Cartesian product]\label{pre:tensorcup}
Suppose the graph coefficient systems have compatible associative
pairings, as supplied by polynomial multiplication. Under the
identification in \eqref{pre:external-cochains}, their graph cup
products define a cup product on the Cartesian product by the rule
\begin{equation}\label{pre:tensorcup-rule}
 (a\times b)\smile(a'\times b')
   =(a\smile a')\times(b\smile b'),
\end{equation}
where $a,a'$ are cochains on the first graph and $b,b'$ are cochains
on the second. Extend this rule bilinearly to arbitrary cochains.
The target coefficient system is the external product of the two
factor target systems. The resulting cup product is associative
and satisfies the Leibniz rule.
\end{lemma}
\begin{proof}
The right-hand side of \eqref{pre:tensorcup-rule} is $F$-multilinear
in $a,b,a',b'$, so it induces a well-defined bilinear map on the
tensor products of cochain spaces. This is the usual cup product
on tensor products of complexes; the graded sign is one in
characteristic two~\cite[Lemma~3.6]{GTZ}.
Associativity follows from associativity in each graph factor.
Substituting \eqref{pre:external-differential} and applying the
factorwise Leibniz rules proves the Leibniz rule on the product.
\end{proof}

This construction gives the cubical cup product used in~\cite{GL,GTZ}.
For a splitting of a cube's directions into complementary sets
$A\sqcup B$, the first cochain is evaluated on the front $A$-face:
the directions in $B$ are fixed at their tails. The second is
evaluated on the back $B$-face: the directions in $A$ are fixed at
their heads. Restrict the coefficient vectors to the cube, apply
the specified coefficient pairings, and sum over splittings. This
rule for front and back faces is the cubical diagonal approximation.

\begin{lemma}[Descent of cup products]\label{pre:descent}
Suppose a finite group acts on cochain complexes by cochain
automorphisms and preserves their cup product pairings:
$g(a\smile a')=(ga)\smile(ga')$ for every $g$ and every pair of
inputs. Then the cup product restricts to invariant cochains and
satisfies the Leibniz rule there. Under the pullback identification
of quotient cochains with invariant cochains, it defines a cup
product on the quotient, and pullback preserves this product.
\end{lemma}
\begin{proof}
For invariant $c,c'$, compatibility with the pairing gives
\begin{equation}\label{pre:descent-equivariance}
 g(c\smile c')=(gc)\smile(gc')=c\smile c'.
\end{equation}
The differentials preserve invariants, so the Leibniz identity
restricts to them. Transporting the product along the pullback
isomorphism gives the quotient product and its compatibility with
pullback.
\end{proof}

\paragraph{The cup product on the balanced product complex.}
For disjoint nonempty block sets $J,J'\subseteq\{1,2,3\}$ and
nonnegative integer degrees $j_1,j_2$ with $j_1+j_2\le3$, the
coefficient pairings above give $F$-bilinear maps
\begin{equation}
 \smile:\mathcal C_J^{j_1}\times\mathcal C_{J'}^{j_2}
       \longrightarrow\mathcal C_{J\cup J'}^{j_1+j_2}.
\end{equation}
Use polynomial multiplication at vertices. On edges, apply the
vertex-to-edge maps at the tail or head as prescribed by
\cref{pre:cup}.

The $H$-action on each graph cover changes only the sheet
labels and preserves edge colors and orientations.
Since the restriction maps and coefficient pairings are
independent of the sheet labels, this action preserves
the graph cup product of \cref{pre:cup}.
By \eqref{pre:tensorcup-rule},
the cup product on $\widetilde B$ is therefore $H^3$-equivariant.
Since $K\subseteq H^3$, the cup product of two $K$-invariant
cochains is $K$-invariant. The pullback identification
\eqref{cg:quotient-models} then defines a cup product on $B$.
Associativity and the Leibniz rule follow from \cref{pre:tensorcup}.
Because we work in characteristic two, all graded signs are one.
\Cref{app:avgx} gives the equivalent coinvariants calculation.

Iterate \cref{pre:tensorcup} over three graph factors. To evaluate
the cup product of $f_h\in\mathcal C_h^1$, $h=1,2,3$, on a cube $s$,
expand each input into its three edge direction components. A
contributing term must choose a different direction for each input:
a graph factor has no 2-cochains, so it cannot supply two edge
factors. Thus the possible assignments are the six permutations
$\pi_0\in\mathfrak S_3$. The input order determines the order in which those
directions are traversed~\cite[Section~6.2]{GL}.

For the assignment $\pi_0=(1,2,3)$, $f_1$ uses a direction-1 edge
with the other directions at their tails. The next input $f_2$
uses a direction-2 edge with direction 1 at its head and direction 3
at its tail. Finally, $f_3$ uses a direction-3 edge with the first
two directions at their heads. This is the staircase path from
the all-minus to the all-plus corner. The front/back rule gives
the same description for every permutation.

Starting at the all-minus corner, traverse the directions in order
$\pi_0(1),\pi_0(2),\pi_0(3)$. Let $e_s^h(\pi_0)$ be the $h$-th edge of this
path; it runs in direction $\pi_0(h)$. The notation
$(f_h|_s)_{e_s^h(\pi_0)}$ means the component of $f_h$ on that edge,
with its remaining vertex coefficient factors restricted to the
edges of $s$. It is a scalar in $F$. Directions already traversed
are fixed at their heads; directions still to be traversed are fixed
at their tails. With these conventions,
\begin{equation}\label{cg:cup}
(f_1\smile f_2\smile f_3)(s)
=\sum_{\pi_0\in\mathfrak S_3}\ \prod_{h=1}^{3}\big(f_h|_s\big)_{e^h_s(\pi_0)}.
\end{equation}
There are six terms, one for each path in \cref{fig:cube}. The
index $h$ specifies the input's position in the ordered product;
$\pi_0(h)$ specifies the geometric direction of its chosen edge.
For example, the term $\pi_0=(2,1,3)$ evaluates $f_1$ on the
edge in direction $2$ from the all-minus corner, $f_2$ on the next
edge in direction $1$, and $f_3$ on the final edge in direction $3$.

The products in \eqref{cg:cup} are ordinary products in $F$.
The binary cup product is $F$-bilinear; the sheet action is equivariant
for simultaneous translation of both inputs, rather than bilinear
for convolution in $R$.

\newcommand{\mskel}{%
\draw[linec,line width=0.6pt]
 (0,0)--(0.95,0) (0.4,0.28)--(1.35,0.28) (0,0.8)--(0.95,0.8) (0.4,1.08)--(1.35,1.08)
 (0,0)--(0.4,0.28) (0.95,0)--(1.35,0.28) (0,0.8)--(0.4,1.08) (0.95,0.8)--(1.35,1.08)
 (0,0)--(0,0.8) (0.95,0)--(0.95,0.8) (0.4,0.28)--(0.4,1.08) (1.35,0.28)--(1.35,1.08);
\fill (0,0) circle(1.5pt);
\draw (1.35,1.08) circle(1.5pt);}

\begin{figure}[htbp]
\centering
\begin{tikzpicture}[x=1cm,y=1cm,line cap=round]
\begin{scope}[shift={(0.4,1.7)}]
\begin{scope}[line width=1.5pt]
\draw[dira] (0,0)--(2.2,0);   \draw[dira] (0.95,0.66)--(3.15,0.66);
\draw[dira] (0,1.9)--(2.2,1.9); \draw[dira] (0.95,2.56)--(3.15,2.56);
\draw[dirb] (0,0)--(0.95,0.66); \draw[dirb] (2.2,0)--(3.15,0.66);
\draw[dirb] (0,1.9)--(0.95,2.56); \draw[dirb] (2.2,1.9)--(3.15,2.56);
\draw[dirc] (0,0)--(0,1.9); \draw[dirc] (2.2,0)--(2.2,1.9);
\draw[dirc] (0.95,0.66)--(0.95,2.56); \draw[dirc] (3.15,0.66)--(3.15,2.56);
\end{scope}
\fill (0,0) circle(2.4pt);
\draw[line width=1pt] (3.15,2.56) circle(2.4pt);
\node[sml,below=4pt] at (0.3,-0.05) {\textbf{all-minus corner}};
\node[sml,above=4pt] at (3.15,2.6) {\textbf{all-plus corner}};
\end{scope}
\begin{scope}[shift={(6.3,1.2)}]
\node[smk,anchor=west] at (0,3.3) {\textbf{a cube cell $s=(e_1,e_2,e_3;x)$:}};
\node[sml,anchor=west] at (0,2.85) {one edge per direction, one shared sheet $x\in H$};
\draw[dira,line width=2pt] (0,2.4) -- (0.6,2.4);
\node[sml,anchor=west] at (0.75,2.4) {direction 1 (color $c_1\in S_1$)};
\draw[dirb,line width=2pt] (0,1.95) -- (0.6,1.95);
\node[sml,anchor=west] at (0.75,1.95) {direction 2 (color $c_2\in S_2$)};
\draw[dirc,line width=2pt] (0,1.5) -- (0.6,1.5);
\node[sml,anchor=west] at (0.75,1.5) {direction 3 (color $c_3\in S_3$)};
\node[sml,anchor=west,text width=7.2cm] at (0,0.8)
 {$j$-cells: $j$ edge directions, $3-j$ vertex directions; block-$h$ coefficient dim $=k^{3-j}$};
\node[smk,anchor=west] at (0,0.1) {\textbf{slot order $=$ evaluation order:}};
\node[sml,anchor=west,text width=7.2cm] at (0,-0.55)
 {$f_1$ on the path's 1st edge, $f_2$ on its 2nd, $f_3$ on its 3rd; the $h$-th edge runs in direction $\pi_0(h)$};
\end{scope}
\node[sml,anchor=west] at (0,0.5) {six staircase paths $\downarrow$ (one per $\pi_0\in\mathfrak S_3$)};
\begin{scope}[shift={(0,-1.6)},line width=1.6pt]
\begin{scope}[shift={(0,0)}]\mskel
 \draw[dira] (0,0)--(0.95,0); \draw[dirb] (0.95,0)--(1.35,0.28); \draw[dirc] (1.35,0.28)--(1.35,1.08);
 \node[sml] at (0.67,-0.45) {$\pi_0=(1,2,3)$};
\end{scope}
\begin{scope}[shift={(2.45,0)}]\mskel
 \draw[dira] (0,0)--(0.95,0); \draw[dirc] (0.95,0)--(0.95,0.8); \draw[dirb] (0.95,0.8)--(1.35,1.08);
 \node[sml] at (0.67,-0.45) {$\pi_0=(1,3,2)$};
\end{scope}
\begin{scope}[shift={(4.9,0)}]\mskel
 \draw[dirb] (0,0)--(0.4,0.28); \draw[dira] (0.4,0.28)--(1.35,0.28); \draw[dirc] (1.35,0.28)--(1.35,1.08);
 \node[sml] at (0.67,-0.45) {$\pi_0=(2,1,3)$};
\end{scope}
\begin{scope}[shift={(7.35,0)}]\mskel
 \draw[dirb] (0,0)--(0.4,0.28); \draw[dirc] (0.4,0.28)--(0.4,1.08); \draw[dira] (0.4,1.08)--(1.35,1.08);
 \node[sml] at (0.67,-0.45) {$\pi_0=(2,3,1)$};
\end{scope}
\begin{scope}[shift={(9.8,0)}]\mskel
 \draw[dirc] (0,0)--(0,0.8); \draw[dira] (0,0.8)--(0.95,0.8); \draw[dirb] (0.95,0.8)--(1.35,1.08);
 \node[sml] at (0.67,-0.45) {$\pi_0=(3,1,2)$};
\end{scope}
\begin{scope}[shift={(12.25,0)}]\mskel
 \draw[dirc] (0,0)--(0,0.8); \draw[dirb] (0,0.8)--(0.4,1.08); \draw[dira] (0.4,1.08)--(1.35,1.08);
 \node[sml] at (0.67,-0.45) {$\pi_0=(3,2,1)$};
\end{scope}
\end{scope}
\end{tikzpicture}
\caption{The cubical diagonal approximation for three 1-cochain inputs,
\eqref{cg:cup}. Each permutation $\pi_0$ is a staircase path from the
all-minus corner to the all-plus corner --- equivalently an iterated choice
of complementary front/back faces. In the term for $\pi_0$, factor $f_h$
contributes its value on the path's $h$-th edge, whose direction is
$\pi_0(h)$, with its remaining vertex factors restricted at the corner the
path occupies there. In the small cubes, the first, second, and third steps
of each path are colored by the direction they run in. In characteristic
two all six terms carry sign $+1$.}
\label{fig:cube}
\end{figure}
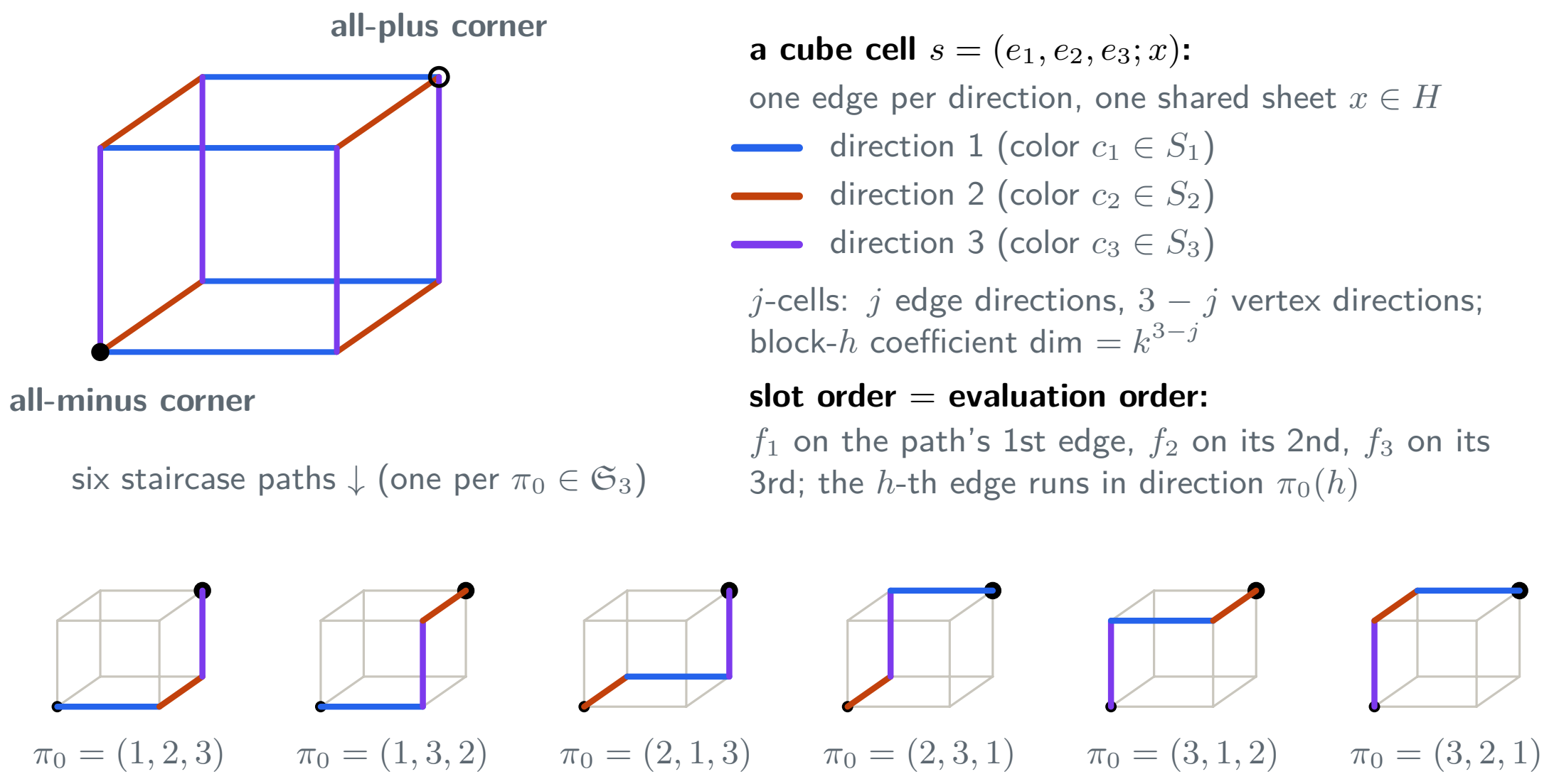

\begin{figure}[htbp]
\centering
\begin{tikzpicture}[x=1cm,y=1cm,line cap=round]
\begin{scope}[shift={(0.6,1.2)}]
\draw[linec,line width=0.8pt]
 (0,0)--(0,1.9) (0,1.9)--(2.2,1.9) (0,1.9)--(0.95,2.56)
 (2.2,1.9)--(3.15,2.56) (0.95,2.56)--(3.15,2.56) (0.95,0.66)--(0.95,2.56)
 (2.2,0)--(2.2,1.9) (0,0)--(0.95,0.66) (0.95,0.66)--(3.15,0.66);
\begin{scope}[line width=2.6pt]
\draw[dira] (0,0)--(2.2,0);
\draw[dirb] (2.2,0)--(3.15,0.66);
\draw[dirc] (3.15,0.66)--(3.15,2.56);
\end{scope}
\fill (0,0) circle(2.4pt);
\draw[line width=1pt] (3.15,2.56) circle(2.4pt);
\node[smk] at (1.7,3.15) {\textbf{$\pi_0=(1,2,3)$: contributes}};
\node[sml,text=dira] at (1.3,-0.45) {\textbf{step 1: $\Phi_1$ on a direction-1 edge \checkmark}};
\node[sml,text=dirb,anchor=west] at (3.45,0.35) {\textbf{step 2: $\Phi_2$ \checkmark}};
\node[sml,text=dirc,anchor=west] at (3.45,1.7) {\textbf{step 3: $\Phi_3$ \checkmark}};
\end{scope}
\begin{scope}[shift={(9.0,1.2)}]
\draw[linec,line width=0.8pt]
 (0,0)--(2.2,0) (0,1.9)--(2.2,1.9) (0.95,2.56)--(3.15,2.56)
 (2.2,1.9)--(3.15,2.56) (0,0)--(0,1.9) (0.95,0.66)--(0.95,2.56)
 (2.2,0)--(2.2,1.9) (0,1.9)--(0.95,2.56) (2.2,0)--(3.15,0.66);
\begin{scope}[line width=2.6pt,opacity=0.45]
\draw[dirb] (0,0)--(0.95,0.66);
\draw[dira] (0.95,0.66)--(3.15,0.66);
\draw[dirc] (3.15,0.66)--(3.15,2.56);
\end{scope}
\fill[mutedc] (0,0) circle(2.4pt);
\draw[mutedc,line width=1pt] (3.15,2.56) circle(2.4pt);
\node[smk] at (1.7,3.15) {\textbf{$\pi_0=(2,1,3)$: vanishes}};
\node[sml,text=badc,anchor=west] at (-0.3,-0.45) {\textbf{step 1 is a direction-2 edge,}};
\node[sml,text=badc,anchor=west] at (-0.3,-0.9) {\textbf{but $\Phi_1$ vanishes there $\Rightarrow$ term $=0$}};
\end{scope}
\end{tikzpicture}
\caption{Only the order $(1,2,3)$ contributes to the cup product of
the cocycles $\Phi_h(G_h)$, $G_h\in\mathcal V$: input $h$ has an edge component only in
direction $h$. Their marked edge supports then restrict the output
to the base cube $(b_1,b_2,b_3)$, as in \eqref{cg:collapse}.}
\label{fig:survive}
\end{figure}
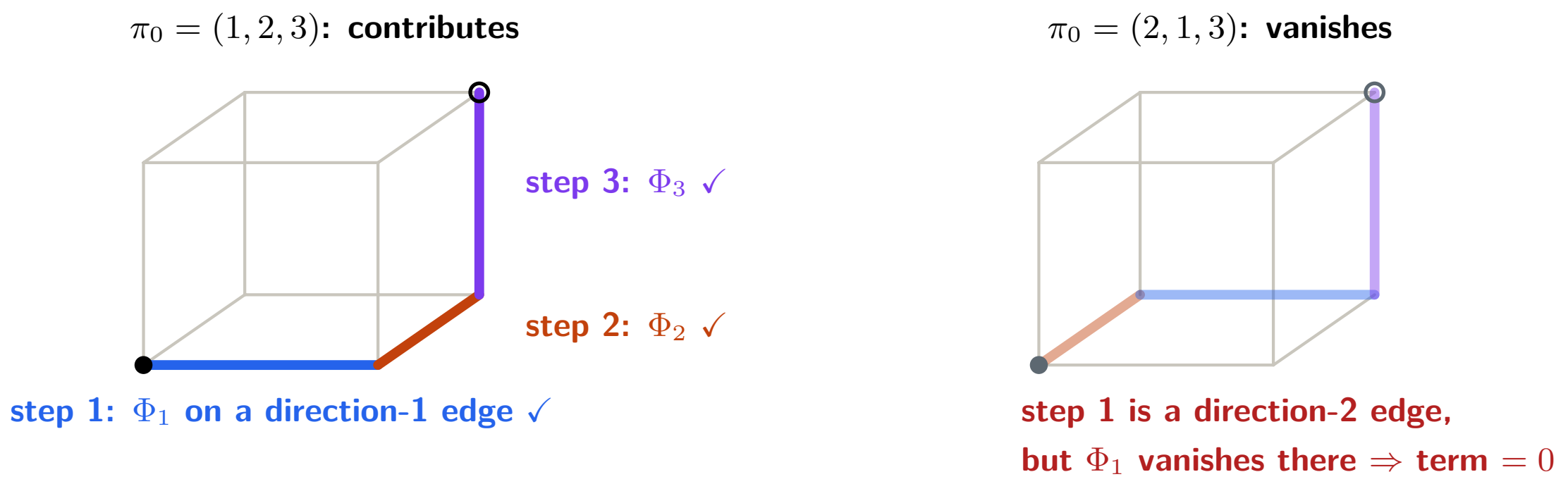

For the cocycles $\Phi_h(G_h)$, input $h$ has an edge component
only in direction $h$. Thus only the staircase order $(1,2,3)$
can contribute in \eqref{cg:cup} (\cref{fig:survive}).

In each graph direction $i$, the three blocks use local codes
$P_i(v)$, $L_i(v)$, and $L_i(v)$. Their Schur product lies in
the target local code
\begin{equation}\label{cg:target-system}
 W_i(v):=C_{i,\{1,2,3\}}(v)
 =\gamma_{i,v} \odot \sigma_{i,v}^{3} \odot \RS(S_i,u),
 \qquad u=3k-2.
\end{equation}
Write $W_i=\shf_{i,\{1,2,3\}}$ for the associated coefficient
system on $\Gamma_i$, and
$D_{i,W}^\bullet=D_{i,\{1,2,3\}}^\bullet$ for its cochain complex.
For the cocycles in \eqref{cg:pullback-representatives},
grouping the cup product by graph direction therefore gives a
1-cochain with coefficients in $W_i$ in each factor.

\begin{lemma}[Cup product on the marked cube]\label{cg:marked-product}
For $G_1,G_2,G_3\in\mathcal V$, the cup product of the cocycles
$\Phi_h(G_h)$ is supported over the base cube $(b_1,b_2,b_3)$ and satisfies
\begin{equation}\label{cg:collapse}
 \begin{aligned}
 &\big(\Phi_1(G_1)\smile\Phi_2(G_2)\smile\Phi_3(G_3)\big)
          (e_1,e_2,e_3;x)\\
 &\qquad=\bar b_1(e_1)\bar b_2(e_2)\bar b_3(e_3)
                  G_1(x)G_2(x)G_3(x).
 \end{aligned}
\end{equation}
Equivalently, as a top cochain in the free $R$-module,
\begin{equation}\label{cg:marked-cube-product}
 \Phi_1(G_1)\smile\Phi_2(G_2)\smile\Phi_3(G_3)
 =(\bar b_1\otimes_R\bar b_2\otimes_R\bar b_3)\,
                 \gb{G_1\odot G_2\odot G_3}.
\end{equation}
\end{lemma}
\begin{proof}
Both sides are $F$-trilinear, so it suffices to take
$G_h=\chi_{\alpha_h}$ for $\alpha_h\in\mathcal A$. Put
$\alpha=\alpha_1+\alpha_2+\alpha_3\in(\mathbb Z/M\mathbb Z)^t$.
Iterating \eqref{pre:tensorcup-rule} on the tensor products
in \eqref{cg:pullback-representatives} groups the factors by graph
direction while preserving the input order within each graph.
Since pullback preserves cup products
(\cref{pre:descent}), we obtain
\begin{equation}\label{cg:factorize}
 \begin{aligned}
 &\pi^*\big(\Phi_1(\chi_{\alpha_1})\smile
            \Phi_2(\chi_{\alpha_2})\smile
            \Phi_3(\chi_{\alpha_3})\big)\\
 &\quad=
   (\xi^{(1)}_{1,\alpha_1}\smile\zeta^{(0)}_{1,\alpha_2}
                              \smile\zeta^{(0)}_{1,\alpha_3})
       \otimes_F
   (\zeta^{(0)}_{2,\alpha_1}\smile\xi^{(1)}_{2,\alpha_2}
                              \smile\zeta^{(0)}_{2,\alpha_3})\\
 &\hspace{4em}\otimes_F
   (\zeta^{(0)}_{3,\alpha_1}\smile\zeta^{(0)}_{3,\alpha_2}
                              \smile\xi^{(1)}_{3,\alpha_3}).
 \end{aligned}
\end{equation}
Write $A_i\in D_{i,W}^1$ for the graph cup product in direction
$i$. The graph cup rule of \cref{pre:cup} applies the
vertex-to-edge maps to the two 0-cocycle values at the
appropriate endpoints.
Equation~\eqref{cg:compensation} therefore gives
\begin{equation}
 A_i(e_i,x_i)=\bar b_i(e_i)\prod_{h=1}^3\chi_{\alpha_h}(x_i)
            =\bar b_i(e_i)\chi_\alpha(x_i).
\end{equation}
Thus, on a lifted cube with graph edges $(e_i,x_i)$, the tensor product
in \eqref{cg:factorize} has value
\begin{equation}
 \prod_{i=1}^3\bar b_i(e_i)\chi_\alpha(x_i)
 =\bar b_1(e_1)\bar b_2(e_2)\bar b_3(e_3)
      \chi_\alpha(x_1x_2x_3).
\end{equation}
This depends only on the balanced sheet $x=x_1x_2x_3$.
By \cref{pre:transferid}, the orbit projection $\tau$ gives the
cochain on $B$ with this same value, proving \eqref{cg:collapse}
for character inputs. Trilinearity extends the identity to all
$G_h\in\mathcal V$.
The tensor product $\bar b_1\otimes_R\bar b_2\otimes_R\bar b_3$
has coefficient $1_R$ at the marked base cube and zero elsewhere.
Multiplying it by $\gb{G_1\odot G_2\odot G_3}$ therefore gives
\eqref{cg:marked-cube-product}.
\end{proof}

\subsection{Normalized functionals and the logical pairing}
\label{cg:lambda-sec}\label{cg:logical-sec}

Equation~\eqref{cg:marked-cube-product} shows that the cup product
vanishes above every base cube except $(b_1,b_2,b_3)$.
Above this marked cube, its values form the sheet array
$(G_1(x)G_2(x)G_3(x))_{x\in H}$.
This cochain identity alone does not show that its cohomology
class is nonzero. To distinguish classes by these values, we
seek an $R$-linear map $\Lambda:\mathcal C_W^3\to R$ such that,
for every $a\in R$ and $g\in\mathcal C_W^2$,
\begin{equation}\label{cg:recovery-goal}
 \Lambda\bigl((\bar b_1\otimes_R\bar b_2\otimes_R\bar b_3)a
                    +\delta_W^2g\bigr)=a.
\end{equation}
We construct $\Lambda$ as a tensor product of graph functionals
with the analogous property on the marked edges. The following
proposition guarantees such functionals after a suitable choice of
the halfedge multipliers.

\begin{proposition}[Existence of normalized graph functionals]
\label{cg:dual-existence}
Choose the base graph parameters as in
\cref{gp:analysis-parameters}, with $\lambda_0\le1/14$.
The local parameter $r$ can be chosen sufficiently large,
independently of $t$, so that the following holds for all
sufficiently large $t$ in the chosen base graph sequence.
After fixing the covers $\Gamma_i$, the multipliers
$\sigma_{i,v}$, and the marked edges $b_i$, there are nonzero
halfedge multipliers $\gamma_{i,v}(c)$ for which the resulting
target complexes admit $R$-linear maps
\begin{equation}\label{cg:lambda}
 \lambda_i:D_{i,W}^1=R^{tn}\longrightarrow R,
 \qquad \lambda_i\delta^0_{i,W}=0,
 \qquad \lambda_i(\bar b_i)=1_R
 \qquad(i=1,2,3).
\end{equation}
\end{proposition}

The proof is given in \cref{cg:dual-existence-sec}. Fix parameters,
multipliers, and maps as in this proposition for the rest of the
construction. By $R$-linearity, the two identities give
$\lambda_i(\bar b_i a+\delta^0_{i,W}v)=a$ for every $a\in R$
and $v\in D_{i,W}^0$.

\paragraph{Dual chains and evaluation.}\label{pre:kronsec}
For a cell $s$, write
$\shf_s^\vee=\operatorname{Hom}_F(\shf_s,F)$ for its dual stalk,
the space of linear functionals on $\shf_s$.
The notation $\shf^\vee$ records these dual spaces and the
transposes of the restriction maps. This differs from
$\shf^\perp$, which uses orthogonal complements of local edge
codes as in \cref{pre:sheaf}.

The chain complex $C_\bullet(X;\shf^\vee)$ is the linear dual
of $C^\bullet(X;\shf)$. A $j$-chain $\lambda$ assigns a functional
$\lambda_s\in\shf_s^\vee$ to each $j$-cell $s$, and the boundary
$\partial$ is the transpose of the coboundary $\delta$.
Elements of $\ker\partial$ are \emph{cycles}, and elements of
$\im\partial$ are \emph{boundaries}. Write
$Z_j(X;\shf^\vee)=\ker\partial_j$ for the cycle space and
$H_j(X;\shf^\vee)=Z_j(X;\shf^\vee)/\im\partial_{j+1}$
for its homology, where $\partial_j:C_j\to C_{j-1}$.

\begin{definition}[Kronecker pairing]\label{pre:kron}
For $\lambda\in C_j(X;\shf^\vee)$ and
$f\in C^j(X;\shf)$, define
\begin{equation}\label{pre:kronecker-evaluation}
 \langle \lambda,f\rangle
   =\sum_{j\text{-cells }s}\lambda_s(f_s).
\end{equation}
Here $\lambda_s$ is a linear functional and $f_s\in\shf_s$ is
its input \footnote{Note that the convention here is opposite to the usual textbook convention such as in \cite{Hatcher:2001ut}, where a cochain is used as a functional and chain as its input. Either convention is fine since the pairing is symmetric.}. This pairing evaluates a cochain against a dual chain
and returns a scalar. The coefficient pairing $m$ of
\cref{pre:pairing} takes two coefficients and returns a coefficient
in a target sheaf. The Kronecker pairing is canonical and requires
no choice of bases.
It induces a bilinear pairing
$\langle\cdot,\cdot\rangle:
H_j(X;\shf^\vee)\times H^j(X;\shf)\to F$,
with the chain written on the left.
\end{definition}

The identity that makes the pairing well defined on classes
comes directly from the definition of transpose:
\begin{equation}\label{pre:kronecker-adjoint}
 \langle\lambda,\delta a\rangle
   =\langle\partial\lambda,a\rangle,
\end{equation}
which is the cochain form of Stokes' theorem.
If $\lambda$ is a cycle, the right side is zero, so adding a
coboundary to a cocycle does not change its pairing with $\lambda$.
Similarly, if $z$ is a cocycle, then
$\langle \partial\mu,z\rangle
=\langle\mu,\delta z\rangle=0$; adding a boundary to $\lambda$
does not change the pairing. This proves independence of both
representatives using the cochain and chain differentials already
defined.

On a graph, there are no $2$-cells. Hence every $1$-cochain is
a cocycle and there are no 1-boundaries, so each
$1$-cycle is the unique representative of its homology class.
By \cref{pre:dualtanner}, these 1-cycles are exactly
the dual Tanner codewords.
We will use dual 3-cycles in the target system $\shf_W$ to evaluate
the cup products. Their vanishing on coboundaries will make the
resulting scalar forms well defined on cohomology classes.

\paragraph{Scalar cycles and the normalization.}
The maps $\lambda_i$ take values in $R$. To obtain $F$-valued
functionals, define the coefficient extraction map
\begin{equation}\label{cg:coefficient-extraction}
 \ev_w:R\longrightarrow F,\qquad
 \ev_w\!\left(\sum_{x\in H}a_x\gb x\right)=a_w
 \qquad(w\in H).
\end{equation}
For $z\in D_{i,W}^1$, the value $(\ev_w\circ\lambda_i)(z)$ is
the $w$-sheet coefficient of $\lambda_i(z)$. This scalar functional
vanishes on coboundaries and hence represents a dual 1-cycle.

\begin{lemma}[Cycles dual to the marked edge classes]\label{cg:eq19hom}
For every $w\in H$, the scalar functional
$\ev_w\circ\lambda_i$ is a cycle in
$Z_1(\Gamma_i;W_i^\vee)$, i.e.,
\begin{equation}
    \partial_1 \lambda_i =0,
\end{equation}
 and its Kronecker pairing satisfies
\begin{equation}\label{cg:scalar-normalization}
 \big\langle\ev_w\circ\lambda_i,[\bar b_i\gb x]\big\rangle
 =\one_{w=x}\qquad(x,w\in H).
\end{equation}
Consequently, for $z\in D_{i,W}^1$, the $R$-valued pairing
$\langle\lambda_i,[z]\rangle:=\lambda_i(z)$ is also
well defined on $H^1(D_{i,W}^\bullet)$ and obeys
$\langle\lambda_i,[\bar b_i a]\rangle=a$ for every $a\in R$.
\end{lemma}
\begin{proof}
Equation~\eqref{cg:lambda} gives
$(\ev_w\circ\lambda_i)\delta^0_{i,W}=0$.
By \cref{pre:dualtanner}, the functional is a dual Tanner word
and a 1-cycle. By $R$-linearity and $\lambda_i(\bar b_i)=1_R$,
$(\ev_w\circ\lambda_i)(\bar b_i\gb x)
=\ev_w(\gb x)=\one_{w=x}$.
Annihilation of coboundaries also makes the $R$-valued pairing
well defined.
\end{proof}

\paragraph{The tensor product of the graph functionals.}
Taking the tensor product of the maps $\lambda_i$ gives the desired
map on the balanced complex.
Define the $R$-linear functional
$\Lambda=\lambda_1\otimes_R\lambda_2\otimes_R\lambda_3$
on $\mathcal C_W^3$, using the identification
$R\otimes_R R\otimes_R R\cong R$ by multiplication.
We use functional notation and pairing notation interchangeably:
\begin{equation}\label{cg:Lambda}
 \langle\Lambda,z\rangle:=\Lambda(z)\in R,
 \qquad \langle\Lambda,\delta_W^2g\rangle=0
       \quad(g\in\mathcal C_W^2),
 \qquad \Lambda_w:=\ev_w\circ\Lambda
       \in Z_3(B;\shf_W^\vee).
\end{equation}
Each summand of $\mathcal C_W^2$ has one vertex factor and two
edge factors. Its top coboundary applies $\delta^0_{i,W}$ in the
vertex factor, so evaluation by $\Lambda$ contains the zero factor
$\lambda_i\delta^0_{i,W}$. Hence $\Lambda\delta_W^2=0$, and
each $\Lambda_w$ is a scalar 3-cycle.

More generally, every $F$-linear functional $\ell:R\to F$ gives
\begin{equation}\label{cg:Lambdaell}
 \Lambda_\ell:=\ell\circ\Lambda
 =\sum_{w\in H}\ell(\gb w)\,\Lambda_w
 \in Z_3(B;\shf_W^\vee),
 \qquad
 \langle\Lambda_\ell,[z]\rangle
 =\ell\big(\langle\Lambda,[z]\rangle\big).
\end{equation}
There are no 4-chains in $B$, so these cycles are identified
with their classes in $H_3(B;\shf_W^\vee)$. On the marked cube, the normalization in \eqref{cg:lambda} gives
\begin{equation}\label{cg:top-normalization}
 \Lambda\big((\bar b_1\otimes_R\bar b_2\otimes_R\bar b_3)a\big)=a
 \qquad(a\in R).
\end{equation}
Thus $\Lambda$ satisfies \eqref{cg:recovery-goal}.

\paragraph{The physical and logical pairing.}
\begin{definition}[Physical and logical trilinear forms]
\label{cg:trilinear-definition}
For an $F$-linear functional $\ell:R\to F$, define
\begin{equation}\label{eq:cupform-preview}
 T_\ell:\mathcal C_1^1\times\mathcal C_2^1\times\mathcal C_3^1
       \longrightarrow F,
 \qquad
 T_\ell(f_1,f_2,f_3)
 =\langle\Lambda_\ell,f_1\smile f_2\smile f_3\rangle.
\end{equation}
The arguments are the physical coefficient vectors of the three
blocks. Its restriction to cocycles will define the logical
trilinear form.
\end{definition}

\begin{proposition}[Coboundary invariance and local support]
\label{cg:invariance}
The form $T_\ell$ descends to an $F$-trilinear form on
$H^1(\mathcal C_1)\times H^1(\mathcal C_2)\times H^1(\mathcal C_3)$.
In its expansion in scalar entries, each entry
occurs in at most $6n^2k^4$ cubic monomial occurrences, before
repeated monomials are combined.
\end{proposition}
\begin{proof}
For cocycles $f_2,f_3$ and $v\in\mathcal C_1^0$, the Leibniz
rule and \eqref{cg:Lambdaell} give
$T_\ell(\delta_1^0v,f_2,f_3)
=\langle\Lambda_\ell,\delta_W^2(v\smile f_2\smile f_3)\rangle=0$.
The same argument in the other inputs, together with trilinearity,
shows that $T_\ell$ is well defined on cohomology.

A physical qudit is associated with a 1-cell incident to
$n^2$ cubes. Each cube contributes six staircase terms, with at
most $k^2$ choices of scalar entry in each of the other two inputs.
Thus the entry labeling this qudit occurs at most $n^2\cdot6\cdot(k^2)^2=6n^2k^4$
times. The linear functional $\Lambda_\ell$ only weights these
monomials.
\end{proof}

In homological notation the induced form is the Kronecker pairing
\begin{equation}\label{cg:kronform}
 T_\ell(f_1,f_2,f_3)
 =\big\langle[\Lambda_\ell],\,
 [f_1]\smile[f_2]\smile[f_3]\big\rangle,
 \qquad [\Lambda_\ell]\in H_3(B;\shf_W^\vee),
\end{equation}
for cocycle inputs.

The pairing can also be computed on the Cartesian product $\widetilde B$.
\paragraph{Evaluation across the quotient.}
To compare a logical pairing on $B$ with its calculation on the
Cartesian product, we transport its dual chain along $\pi$.
The pullback and orbit projection were defined in
\eqref{cg:quotient-maps}; the corresponding operation on dual chains
is their linear adjoint.

\begin{definition}[Pullback coefficients and the norm map]\label{pre:pullback}
For a coefficient system $\shf$ on $B$, its pullback has stalk
$(\pi^*\shf)_{\tilde s}=\shf_{\pi(\tilde s)}$ and copies the
restriction map of each quotient incidence. The norm map on dual
chains is $\pi^!:=\tau^{\mathsf T}$. In coefficient coordinates,
\begin{equation}\label{pre:norm-coordinates}
 (\pi^!z)_{\tilde s}=z_{\pi(\tilde s)},\qquad
 \langle\pi^!z,a\rangle_{\widetilde B}=\langle z,\tau a\rangle_B.
\end{equation}
Here each coefficient $z_s$ is a functional on the whole stalk
$\shf_s$; the formula applies to scalar and vector stalks alike.
\end{definition}

\begin{lemma}[Transfer of the evaluation pairing]\label{pre:transferid}
For the quotient $\pi:\widetilde B\to B$ by $K$, the maps
$\pi^*,\tau,\pi^!$ commute with the respective differentials.
Pullback identifies quotient cochains with invariant cochains,
and the norm identifies quotient dual chains with invariant dual
chains. Since $|K|=Q^2$ is odd,
\begin{equation}\label{pre:transfer-id}
 \tau\pi^*=\mathrm{id},\qquad
 \langle\pi^!z,\pi^*a\rangle_{\widetilde B}=\langle z,a\rangle_B.
\end{equation}
In particular, for $a_h\in\mathcal C_h^1$, $h=1,2,3$, and any
$z\in C_3(B;\shf_W^\vee)$,
\begin{equation}\label{pre:upstairs-cup}
 \langle z,a_1\smile a_2\smile a_3\rangle_B
 =\langle\pi^!z,\pi^*a_1\smile\pi^*a_2\smile\pi^*a_3\rangle_{\widetilde B}.
\end{equation}
\end{lemma}
A coordinate proof is given in \cref{app:transfer-proof}.

The norm map sends $\Lambda_\ell$ to
$\pi^!\Lambda_\ell\in
Z_3(\widetilde B;(\pi^*\shf_W)^\vee)$ by copying each dual
coefficient to all lifts of its cube. Applying
\eqref{pre:upstairs-cup}, for $f_h\in\mathcal C_h^1$ we obtain
\begin{equation}\label{cg:pairing-upstairs}
 \begin{aligned}
 T_\ell(f_1,f_2,f_3)
 &=\langle\Lambda_\ell,f_1\smile f_2\smile f_3\rangle_B\\
 &=\langle\pi^!\Lambda_\ell,
       \pi^*f_1\smile\pi^*f_2\smile\pi^*f_3
   \rangle_{\widetilde B}.
 \end{aligned}
\end{equation}
The fiber multiplicity contributes no scalar factor because
$Q^2=1$ in $F$. For $G_1,G_2,G_3\in\mathcal V$, expanding
these functions in the character basis expresses the upstairs
cup product of the cocycles $\Phi_h(G_h)$ as a sum of the
tensor products in \eqref{cg:factorize}.
Downstairs, the marked cube formula gives the following direct
evaluation.

\begin{lemma}[Pairing on the descended cocycles]\label{cg:keyidentity}
For $G_1,G_2,G_3\in\mathcal V$, evaluating the cup product of
$\Phi_1(G_1)$, $\Phi_2(G_2)$, and $\Phi_3(G_3)$ with $\Lambda$
gives the sheet array whose value at $x\in H$ is
$G_1(x)G_2(x)G_3(x)$:
\begin{equation}\label{cg:pairing}
 \begin{aligned}
 \big\langle\Lambda,
 \Phi_1(G_1)\smile\Phi_2(G_2)\smile\Phi_3(G_3)\big\rangle
 &=\gb{G_1\odot G_2\odot G_3}\\
 &=\sum_{x\in H}G_1(x)G_2(x)G_3(x)\gb x.
 \end{aligned}
\end{equation}
Consequently, for every $F$-linear $\ell:R\to F$,
\begin{equation}
 T_\ell(\Phi_1(G_1),\Phi_2(G_2),\Phi_3(G_3))
 =\ell(\gb{G_1\odot G_2\odot G_3}).
\end{equation}
\end{lemma}
\begin{proof}
Put $a=\gb{G_1\odot G_2\odot G_3}\in R$.
Applying $\Lambda$ to \eqref{cg:marked-cube-product} and using
\eqref{cg:top-normalization} gives
\begin{equation}
 \begin{aligned}
 \big\langle\Lambda,
 \Phi_1(G_1)\smile\Phi_2(G_2)\smile\Phi_3(G_3)\big\rangle
 &=\Lambda\big((\bar b_1\otimes_R\bar b_2\otimes_R\bar b_3)a\big)\\
 &=a.
 \end{aligned}
\end{equation}
Composing with $\ell$ proves the scalar identity by the definition
of $T_\ell$. \Cref{app:avgx} records the corresponding evaluation
upstairs using multiplication in the group algebra.
\end{proof}

For any $\F_2$-linear map $\varphi:F\to\F_2$, the diagonal
physical operation with phase $(-1)^{\varphi(T_\ell(x_1,x_2,x_3))}$
preserves the full CSS code spaces. By \cref{cg:invariance}, this
phase is constant on each product of three coboundary cosets
in the cocycle spaces, so every term in a computational logical
basis state receives the same phase.

\subsection{Information sets and addressable logical non-Clifford gates}
\label{cg:infoset-sec}\label{sec:addressable}

To obtain a diagonal logical pairing from \cref{cg:keyidentity},
we need basis functions whose values distinguish individual sheet
coordinates. The character basis with the functional
$\ell_H=\sum_{x\in H}\ev_x$ instead gives

\begin{equation}\label{cg:phasematch}
 \begin{aligned}
 &T_{\ell_H}\big(\Phi_1(\chi_{\alpha_1}),
       \Phi_2(\chi_{\alpha_2}),\Phi_3(\chi_{\alpha_3})\big)\\
 &\qquad=\sum_{x\in H}\chi_{\alpha_1+\alpha_2+\alpha_3}(x)
 =\one_{\alpha_1+\alpha_2+\alpha_3=0},
 \end{aligned}
\end{equation}
using character orthogonality and $Q=1$ in $F$. We therefore choose
an interpolation basis on a set of independent evaluation coordinates.

\begin{definition}[Information set and interpolation basis]
\label{cg:information-set-definition}
Write $\srk=|\mathcal A|=\dim_F\mathcal V$.
An information set for $\mathcal V$ is a subset $\calS\subseteq H$
with $|\calS|=\srk$ such that the restriction map
$\operatorname{res}_{\calS}:\mathcal V\to F^{\calS}$ is an isomorphism.
For $z\in \calS$, let $F_z$ be the unique function in $\mathcal V$
whose restriction is the coordinate indicator of $z$:
\begin{equation}\label{cg:deltas}
 F_z(z')=\one_{z=z'}\qquad(z,z'\in \calS).
\end{equation}
For a selected subset $I\subseteq \calS$, define
\begin{equation}
 \ell_I=\sum_{w\in I}\ev_w:R\to F,
 \qquad T_I:=T_{\ell_I},
 \qquad \Lambda_{\ell_I}=\sum_{w\in I}\Lambda_w.
\end{equation}
The set $I$ specifies the selected logical triples.
\end{definition}

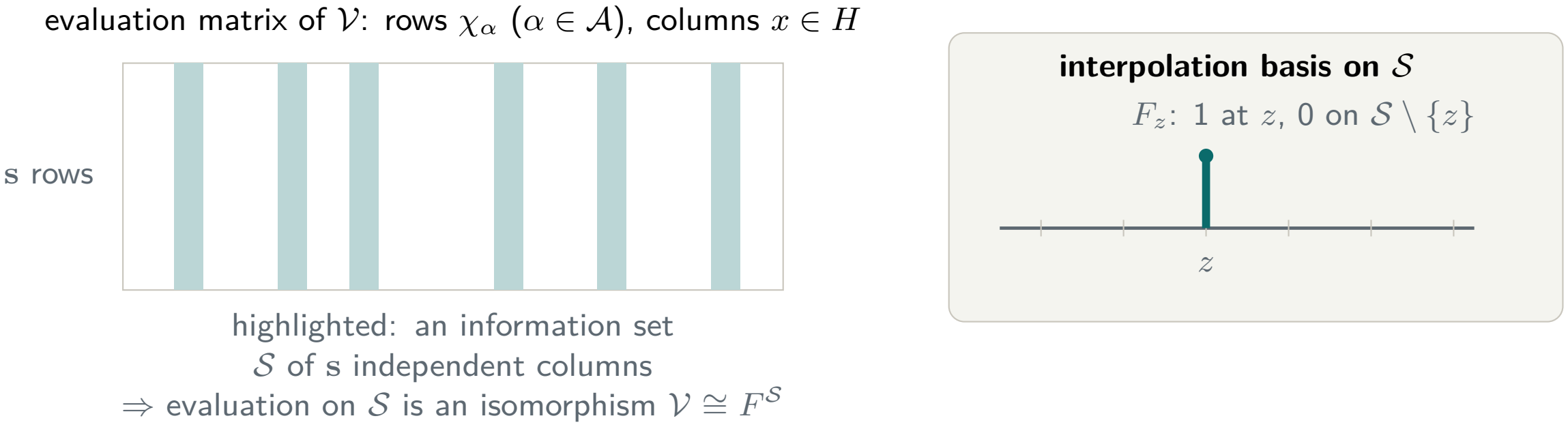
\begin{figure}[htbp]
\centering
\begin{tikzpicture}[x=1cm,y=1cm]
\draw[linec] (0,0) rectangle (6.4,2.2);
\foreach \xx in {0.5,1.5,2.2,3.6,4.6,5.7}
  \fill[accentc!28] (\xx,0) rectangle ++(0.28,2.2);
\node[smk,above=3pt] at (3.2,2.2) {evaluation matrix of $\mathcal V$: rows $\chi_\alpha$ ($\alpha\in\mathcal A$), columns $x\in H$};
\node[sml,anchor=east] at (-0.15,1.1) {$\srk$ rows};
\node[sml,text width=6.6cm,align=center] at (3.2,-0.75)
 {highlighted: an information set $\calS$ of $\srk$ independent columns\\
  $\Rightarrow$ evaluation on $\calS$ is an isomorphism $\mathcal V\cong F^{\calS}$};
\node[fbox,text width=5.6cm,minimum height=2.8cm,anchor=north west] (dp) at (8.0,2.5) {};
\node[smk] at (10.8,2.15) {\textbf{interpolation basis on $\calS$}};
\draw[mutedc,line width=0.9pt] (8.5,0.6) -- (13.1,0.6);
\foreach \xx in {8.9,9.7,10.5,11.3,12.1,12.9} \draw[linec] (\xx,0.52) -- (\xx,0.68);
\draw[accentc,line width=2.2pt] (10.5,0.6) -- (10.5,1.3);
\fill[accentc] (10.5,1.3) circle(2pt);
\node[sml,below=2pt] at (10.5,0.52) {$z$};
\node[sml,anchor=east] at (13.25,1.68) {$F_z$: 1 at $z$, 0 on $\calS\setminus\{z\}$};
\end{tikzpicture}
\caption{Choosing independent columns of the character evaluation
matrix gives an information set $\calS$. Changing the row basis makes
the restriction to $\calS$ the identity matrix, producing the functions
$F_z$ of \eqref{cg:deltas}. Their values outside $\calS$ need not vanish.}
\label{fig:infoset}
\end{figure}

An information set always exists. The character evaluation matrix
$G=(\chi_\alpha(x))_{\alpha\in\mathcal A,x\in H}$ has full row
rank $\srk$. Choose $\srk$ independent columns, indexed by $\calS$,
and write $M_{\calS}=G|_{\calS}$ using any fixed ordering of these columns.
The rows of $M_{\calS}^{-1}G$, now indexed by $\calS$, are exactly the
functions $F_z$. The indicator identity holds on $\calS$; the functions
may be nonzero elsewhere.
The labels $z\in \calS$ therefore specify evaluation coordinates,
without implying that $\Phi_h(F_z)$ is supported on a single sheet
(\cref{fig:infoset}).

\begin{theorem}[Fixed logical triples and subset selection]
\label{cg:diagonal}
Fix the codes, the normalized maps $\lambda_i$, and an information
set $\calS$ for $\mathcal V$. In each block $h$, the $\srk$ classes
$[\Phi_h(F_z)]\in H^1(\mathcal C_h)$, indexed by $z\in \calS$,
are linearly independent. For every $I\subseteq \calS$ and all
$x,y,z\in \calS$,
\begin{equation}
 \begin{aligned}
 T_I(\Phi_1(F_x),\Phi_2(F_y),\Phi_3(F_z))
 &=\big\langle[\Lambda_{\ell_I}],
       [\Phi_1(F_x)]\smile[\Phi_2(F_y)]\smile[\Phi_3(F_z)]
       \big\rangle\\
 &=\one_{x=y=z\in I}.
 \end{aligned}
\end{equation}
The cocycles $\Phi_h(F_z)$ and the full code check matrices are
independent of $I$. In particular, each block encodes at least
$\srk$ logical qudits.
\end{theorem}
\begin{proof}
By \cref{cg:keyidentity} and \eqref{cg:deltas},
\begin{equation}
 \begin{aligned}
 T_I(\Phi_1(F_x),\Phi_2(F_y),\Phi_3(F_z))
 &=\sum_{w\in I}F_x(w)F_y(w)F_z(w)\\
 &=\sum_{w\in I}\one_{x=w}\one_{y=w}\one_{z=w}
 =\one_{x=y=z\in I}.
 \end{aligned}
\end{equation}
For independence in block $1$, take $I=\calS$ and suppose that
$\sum_{z\in \calS}c_z\Phi_1(F_z)$ is a coboundary, with $c_z\in F$.
For any $z_0\in \calS$, evaluate $T_{\calS}$ on this linear combination,
$\Phi_2(F_{z_0})$, and $\Phi_3(F_{z_0})$. Coboundary invariance
makes the result zero, whereas the diagonal identity makes it
$c_{z_0}$. Hence every coefficient vanishes. The same argument
in the other two inputs proves independence in each block.
The dimension bound follows. Only the functional $\ell_I$ depends
on $I$.
\end{proof}

The block length $N_0=12nk^2t^3|H|$ has the form required by
\cref{gp:polynomial-length}. Hence, for any prescribed
$0<\zeta<1$, choosing the family there gives at least
$N_0^{1-\zeta}$ linearly independent cohomology classes in each block
for all sufficiently large $t$.

To express the logical action, extend the classes
$[\Phi_h(F_w)]$, $w\in \calS$, to a basis of $H^1(\mathcal C_h)$
defining the logical qudits. Write $a_{h,w}\in F$ for the computational
basis labels of the selected qudits in block $h$, and let
$\varphi:F\to\F_2$ be $\F_2$-linear. Initialize every other logical
qudit to $\ket0$. On this subspace, \cref{cg:diagonal} gives the logical phase
\begin{equation}
 (-1)^{\varphi(\sum_{w\in I}a_{1,w}a_{2,w}a_{3,w})}
 =\prod_{w\in I}(-1)^{\varphi(a_{1,w}a_{2,w}a_{3,w})}.
\end{equation}
Thus $I=\{w\}$ selects the triple labelled by $w$, while a general
subset $I$ selects the corresponding product of phases.
The choice of $\varphi$ and the conversion to qubits are
specified in \cref{db:binary}; \cref{addr:layer} gives the resulting
action within one fixed transversal CCZ layer.

Golowich, Tamo, and Zhu~\cite[Claim~3.11]{GTZ} also use coordinate indicators. They choose an
extendable set of Tanner code edges, meaning that arbitrary values
on those edges extend to a codeword. They choose codewords with
indicator values on that set, then take tensor products of the
associated graph cocycles to distinguish tuples of edges.
Here the functions $F_z$ distinguish the sheets in $\calS$.
The marked base edges fix the support of the cup product and the
normalization of the functionals $\lambda_i$.

\subsection{Existence of the normalized functionals}
\label{cg:dual-existence-sec}

We prove \cref{cg:dual-existence} by choosing the halfedge
multipliers so that each Fourier component of the graph coboundary
matrix is injective after deletion of the marked edge row.

\paragraph{Fourier transform over the coefficient field.}\label{pre:fouriersec}
We construct the functionals $\lambda_i$ one character component at
a time. Fourier decomposition again reduces maps on the cover to
matrices on the base graph; here their entries lie in $F$, so the
required linear equations are solved over the code's coefficient field.
An $F$-valued character is a homomorphism $H\to F^\times$.
Write $\widehat H=\operatorname{Hom}(H,F^\times)$. For our group
$H=(F^\times)^t$, these are the characters of
\cref{gp:field-characters}, indexed additively by
$\alpha\in(\Z/M\Z)^t$:
\begin{equation}\label{pre:character-indices}
 \chi_{\alpha+\beta}(x)=\chi_\alpha(x)\chi_\beta(x),
 \qquad \chi_0(x)=1,
 \qquad \chi_{-\alpha}(x)=\chi_\alpha(x)^{-1}.
\end{equation}
Thus $0$ indexes the trivial character. For $\mathfrak r=\sum_xa_x\gb x\in R$,
define its Fourier coordinates and the elements $e_\alpha\in R$ by
\begin{align}
 \Phi(\mathfrak r)_\alpha=\widehat{\mathfrak r}(\alpha)
   &:=\sum_{x\in H}a_x\chi_\alpha(x),\label{pre:fourier-transform}\\
 e_\alpha&:=Q^{-1}\sum_{x\in H}\chi_\alpha(x)^{-1}\gb x.
   \label{pre:fourier-idempotents}
\end{align}
Here $Q^{-1}=1$ in $F$, since $Q$ is odd. These definitions and the
following identities apply to any finite abelian group of odd order
whose exponent divides $|F|-1$, where the exponent is the least positive
integer $M$ such that $x^M=1_H$ for every $x\in H$.

\begin{lemma}[{Fourier transform of $F[H]$}]\label{pre:fourier}
For $R=F[H]$ and the characters above:
\begin{enumerate}[label=(\roman*)]
\item $\Phi:R\to\prod_{\alpha\in\widehat H}F$ is a ring isomorphism.
\item Character orthogonality gives
$\sum_{x\in H}\chi_\alpha(x)\chi_\beta(x)^{-1}
=Q\one_{\alpha=\beta}$.
\item The elements $e_\alpha$ are mutually orthogonal idempotents,
with $\widehat e_\alpha(\beta)=\one_{\alpha=\beta}$,
$\sum_\alpha e_\alpha=1_R$, and
$\gb g\,e_\alpha=\chi_\alpha(g)e_\alpha$.
\item If $\Psi:R^a\to R^b$ is $R$-linear with matrix
$\Psi=\sum_x\Psi_x\gb x$, $\Psi_x\in F^{b\times a}$, then
\begin{equation}\label{pre:fourier-block}
 \Psi(v e_\alpha)=\big(\widehat\Psi(\alpha)v\big)e_\alpha,
 \qquad \widehat\Psi(\alpha)=\sum_x\Psi_x\chi_\alpha(x)
 \quad(v\in F^a).
\end{equation}
Thus kernels and images can be computed separately in each component.
\end{enumerate}
\end{lemma}
These standard identities~\cite{Serre77,LN97}, including the inverse
transform and the explicit kernel and image formulas, are proved in
\cref{app:fourierx}.

The normalization $\lambda_i(\bar b_i)=1_R$ in \eqref{cg:lambda}
requires value one in every Fourier component. Indeed, the uniform
sheet array and its Fourier coordinates are
\begin{align}
 e_0&=Q^{-1}\sum_{x\in H}\gb x=\sum_{x\in H}\gb x,
       \label{pre:trivial-idempotent}\\
 \widehat{1_R}(\alpha)&=1,\qquad
 \widehat e_0(\alpha)=\one_{\alpha=0}.\label{pre:identity-uniform}
\end{align}
Normalizing to $1_R$ therefore retains every component and, by
Fourier inversion, every sheet value of a marked edge array.

In the monomial bases, the matrix of $\delta^0_{i,W}$ has rows
indexed by base edges $e$ and columns by $(v,U_i^d)$, $0\le d<u$.
Its entry is zero unless $v\in\partial e$; otherwise it is
\begin{equation}\label{cg:incidence-entry}
 \gamma_{i,v}(c)\sigma_{i,v}(c)^3c^d\,\gb{g_{i,v}(c)},
 \qquad c=c(e).
\end{equation}
Here the scalar factor comes from the restriction map for $W_i$,
and $\gb{g_{i,v}(c)}$ acts on sheet labels by
$x\mapsto g_{i,v}(c)x$.
By \cref{pre:fourier}(i), in component $\alpha$, multiplication by $\gb{g_{i,v}(c)}$
becomes multiplication by the nonzero scalar
$\chi_\alpha(g_{i,v}(c))$. Thus the $tnQ\times2tuQ$ matrix in
sheet coordinates splits into $Q$ matrices
$\widehat\delta^0_{i,W}(\alpha)\in F^{tn\times2tu}$, indexed by
base edges and base vertex polynomial coefficients. We seek a row
functional in each component that vanishes on the matrix image
and takes value one on the marked edge basis vector.

The following rank condition suffices. Allowing separate
halfedge multipliers is essential to our rank argument: a single
nonzero scalar at each vertex would only rescale that vertex's
matrix columns, preserving rank.

\begin{condition}[Deleted row rank]
\label{cg:deleted-row-condition}
For every direction $i\in\{1,2,3\}$ and Fourier index
$\alpha\in(\mathbb Z/M\mathbb Z)^t$, deleting the marked row
$b_i$ from $\widehat\delta^0_{i,W}(\alpha)$ leaves a matrix
with trivial kernel.
\end{condition}

\begin{proposition}[Dual functionals from deleted row rank]
\label{cg:dual-construction}
Under \cref{cg:deleted-row-condition}, for each direction $i$
there exists an $R$-linear functional $\lambda_i$ satisfying
\eqref{cg:lambda}.
\end{proposition}
\begin{proof}
Fix a direction $i$, and let $\one_{b_i}\in F^{tn}$ be the
coordinate vector of the marked edge. In each Fourier component,
this vector lies outside the image
of the full coboundary matrix. Indeed, if
$\widehat\delta^0_{i,W}(\alpha)v=\one_{b_i}$, the assumed trivial
kernel of the matrix with row $b_i$ deleted forces $v=0$,
contradicting $\widehat\delta^0_{i,W}(\alpha)v=\one_{b_i}$.
Since $\one_{b_i}$ lies outside the matrix image, every vector in
the subspace
$\operatorname{im}\widehat\delta^0_{i,W}(\alpha)\oplus F\one_{b_i}$
has a unique expression $w+c\one_{b_i}$, with $w$ in the matrix
image and $c\in F$. Define a linear functional on this subspace
by $w+c\one_{b_i}\mapsto c$. Extend a basis of this subspace
to a basis of $F^{tn}$, assigning the functional value zero on
the additional basis vectors. The resulting row vector
$\widehat\lambda_i(\alpha)\in F^{tn}$ therefore satisfies
$\widehat\lambda_i(\alpha)\widehat\delta^0_{i,W}(\alpha)=0$
and $\big(\widehat\lambda_i(\alpha)\big)_{b_i}=1$.
For the Fourier idempotents $e_\alpha$ of
\cref{pre:fourier}, define
$\lambda_i(v e_\alpha)=\big(\widehat\lambda_i(\alpha)v\big)e_\alpha$
for $v\in F^{tn}$,
and define its value on an arbitrary cochain by summing over
$\alpha$. The resulting map is $R$-linear, and
$\lambda_i\delta^0_{i,W}=0$ because this composition vanishes
in every Fourier component.
Moreover, $1_R=\sum_\alpha e_\alpha$ and
$\widehat{\bar b_i}(\alpha)=\one_{b_i}$, so normalization at the
marked edge gives
$\lambda_i(\bar b_i)=1_R$.
\end{proof}

By \cref{pre:dualtanner}, each row $\widehat\lambda_i(\alpha)$
is a Tanner codeword on the base graph for the duals of the local
codes in that Fourier component, with value $1$ at $b_i$.
The marked edge remains in the graph; we delete its row only
to test the rank.

\paragraph{Why random halfedge multipliers suffice.}
We now show that the halfedge multipliers can be chosen to satisfy
\cref{cg:deleted-row-condition}.
A vector in the kernel of the coboundary matrix with its marked row
deleted assigns a local polynomial to each vertex. Its two
rescaled endpoint evaluations agree on every unmarked edge. The MDS distance forces each nonzero local polynomial
to have many nonzero evaluations. Base graph expansion then forces
any such assignment to occupy many vertices. Finally, independent
halfedge multipliers make the required agreements unlikely.
The following count makes this argument uniform over all assignments.

\begin{lemma}[Injectivity after deleting a row]\label{cg:rank}
Let an $n$-regular bipartite multigraph on $2t$ vertices have
second-largest adjacency eigenvalue at most $\beta n$. At each
vertex fix a linear MDS code over $F$, of length $n$ and dimension
$u<n/2$. Write $q=|F|$ and assume $0\le\beta<1-u/n$.
Independently multiply each
halfedge coordinate by a uniform element of $F^\times$.
Choose a generator basis for each local code. The edge agreement
matrix maps the $2tu$ vertex coefficients to the $tn$ differences
between rescaled endpoint values, one per edge.
Delete any one row fixed before these multipliers are drawn, and set
\begin{equation}
 \kappa_{\mathrm r}=1-u/n-\beta,\qquad
 s_0=\lceil2t\kappa_{\mathrm r}\rceil,\qquad
 \vartheta=q^{-(n/2-u)}\left(\frac q{q-1}\right)^{n/2}.
\end{equation}
The probability that the remaining matrix has a nonzero kernel is
at most $2^{tn+1}q^2\vartheta^{s_0}$. The same bound holds in the
presence of arbitrary fixed nonzero halfedge factors.
\end{lemma}
\begin{proof}
Fixed halfedge factors can be absorbed into the prescribed MDS codes.
If $\vartheta\ge1$, the stated upper bound is at least one.
Assume henceforth that $\vartheta<1$.

Consider a nonzero assignment in the kernel of the matrix with its
marked row deleted. Let $V_{\mathrm{act}}$ be the vertices assigned
a nonzero local codeword, and put
$s_{\mathrm{act}}=|V_{\mathrm{act}}|$. A nonzero local MDS word has at least
$n-u+1$ nonzero coordinates, so at least $n-u$ remain after the
row deletion. On an undeleted edge, either both endpoint values
are nonzero or both vanish. Every nonzero undeleted incidence at
an active vertex therefore lies on an edge internal to $V_{\mathrm{act}}$. Hence
\begin{equation}
 (n-u)s_{\mathrm{act}}\le 2|E(V_{\mathrm{act}})|
 \le \frac{n s_{\mathrm{act}}^2}{2t}+\beta n s_{\mathrm{act}},
 \qquad s_{\mathrm{act}}\ge s_0.
\end{equation}
For the second inequality, write the indicator vector of
$V_{\mathrm{act}}$ as its average value times the all-ones
vector plus a vector whose entries sum to zero. On the zero-sum
subspace, the adjacency quadratic form is bounded above by
$\beta n$ times the squared norm; this vector has squared norm
at most $s_{\mathrm{act}}$.

Fix which $e$ undeleted edges have nonzero endpoint values, and
record separately whether each endpoint value on the deleted edge
is nonzero. Nonzero rescaling preserves support, so this also fixes
the support of the local codewords before rescaling. If $b\le2$
of the deleted incidences are nonzero, the total halfedge support
has size $2e+b$.
At an active vertex with allowed support of size $\ell_v$, the
MDS property makes the $n-\ell_v$ zero-coordinate constraints
independent whenever a nonzero word is feasible. The space of local
codewords supported on the allowed $\ell_v$ coordinates has dimension
$u-(n-\ell_v)=\ell_v-(n-u)$. Thus the number of assignments of
local codewords with this support pattern is at most
\begin{equation}
 q^{\sum_{v\in V_{\mathrm{act}}}(\ell_v-(n-u))}
 =q^{2e+b-(n-u)s_{\mathrm{act}}}.
\end{equation}
For each fixed assignment of local codewords before rescaling, an
agreement equation on an edge with nonzero endpoint values holds
with probability $(q-1)^{-1}$. These equations involve disjoint
pairs of random halfedge multipliers and are independent.
A union bound over these assignments therefore bounds the
probability of a kernel assignment with this support pattern by
\begin{equation}
 \begin{aligned}
 q^{2e+b-(n-u)s_{\mathrm{act}}}(q-1)^{-e}
 &=q^{e+b-(n-u)s_{\mathrm{act}}}
       \left(\frac q{q-1}\right)^e\\
 &\le q^2q^{-(n/2-u)s_{\mathrm{act}}}
       \left(\frac q{q-1}\right)^{n s_{\mathrm{act}}/2}
 =q^2\vartheta^{s_{\mathrm{act}}},
 \end{aligned}
\end{equation}
where $b\le2$ and $e\le n s_{\mathrm{act}}/2$.
There are at most $2^{tn-1}$ choices of
undeleted support and four choices of deleted edge support.
Using $s_{\mathrm{act}}\ge s_0$ and $\vartheta<1$ completes the union bound.
\end{proof}

\begin{proof}[Proof of \cref{cg:dual-existence}]
The character factors $\chi_\alpha(g_{i,v}(c))$ and the multipliers
$\sigma_{i,v}(c)^3$ are fixed and nonzero in each Fourier component.
The bound in \cref{cg:rank} allows such halfedge factors, so it
applies with $\beta=\lambda_0$, the normalized spectral bound
for the base multigraph.

For the fixed marked edges $b_i$, sample candidate halfedge
multipliers $\gamma_{i,v}(c)$ independently and uniformly from
$F^\times$, reusing each value on every lift of its base incidence.
A union bound over the three directions and the $Q=M^t$ Fourier
components bounds the probability that \cref{cg:deleted-row-condition}
fails by
\begin{equation}\label{cg:rankprob}
 3M^t2^{tn+1}q^2\vartheta^{\lceil2t\kappa_{\mathrm r}\rceil},
 \qquad\kappa_{\mathrm r}=1-u/n-\lambda_0.
\end{equation}
The events in different Fourier components need not be independent.
With the local parameters fixed and $t\to\infty$, the bound
tends to zero if
\begin{equation}\label{cg:rankcondition}
 \log M+n\log2-2\kappa_{\mathrm r}(n/2-u)\log q
       +\kappa_{\mathrm r}n\log\frac q{q-1}<0.
\end{equation}
Indeed, when $\vartheta<1$, the logarithm of
\eqref{cg:rankprob} is at most $\log(6q^2)$ plus $t$ times
the left side of \eqref{cg:rankcondition}.

By the assumed bound $\lambda_0\le1/14$, together with
$k=\lfloor n/7\rfloor$ and $u=3k-2$, we have
$\kappa_{\mathrm r}\ge1/2$ and $n/2-u\ge n/14$.
The left side of \eqref{cg:rankcondition} is consequently at most
\begin{equation}
 \log M+n\log2-\frac n{14}\log q
             +n\log\frac q{q-1}.
\end{equation}
For the fields in \eqref{gp:construction-fields}, $n$ grows
exponentially in the local parameter $r$, while
$\log M<\log q=\Theta(r^3)$. For sufficiently large $r$, the negative term
$-(n/14)\log q$ dominates the other terms. Choose such an $r$ and
hold it fixed as $t$ grows; this choice also ensures $\vartheta<1$.
Thus, after fixing such an $r$, for every sufficiently large
$t$ there are halfedge multipliers satisfying
\cref{cg:deleted-row-condition}. The bound is uniform in the
fixed slopes, shifts, and marked edges. For each such choice,
\cref{cg:dual-construction} supplies the required maps $\lambda_i$.
\end{proof}

%% file: sections/distance.tex
\section{Distance of the qudit codes}\label{db:sec}

We bound both Pauli distances using the cubical complex arguments of
Dinur, Lin, and Vidick~\cite[Sections~5--8]{DLV}.
We retain the notation of \cref{sec:prelim,gp:section,cg:section}
and follow their notation for new quantities where possible, to
make their arguments easier to compare with ours.

A cocycle of minimum
weight in its class cannot be reduced by adding a local coboundary.
Product expansion then forces its support to contain many nearby
faces, and spectral expansion forces the total support to be large.

For the other Pauli distance, they show that every cycle with
sufficiently small support is a boundary, so it cannot represent
a nontrivial logical operator. This means constructing a
higher-dimensional chain whose boundary is the given cycle;
such a chain is called a \emph{filling}.
They first construct fillings locally, then adjust them to agree
on overlaps. The obstruction to making them agree is a cocycle
for the dual local codes. When the original cycle is sufficiently
small, the cocycle distance bound implies that this obstruction
is a coboundary. This supplies the correction needed to combine
the local fillings into a global filling.

To apply these arguments with local check matrices that depend on the base vertex,
\cref{db:star-robustness} establishes uniform local robustness, and
\cref{db:dlv-compatibility} verifies the compatibility needed for
their cocycle and local filling arguments.

A walk on the vertices of $B$ that moves only in direction $i$
leaves the other two base vertices fixed. The graph of these
moves therefore splits into copies of $\Gamma_i$, one for each
pair of fixed base vertices. There are $4t^2$ such pairs, so
each component contains only a fraction of order $t^{-2}$ of
the vertices.
\Cref{db:walk-estimate} accounts for these component sizes when
applying the walk estimates of~\cite{DLV}. The resulting $t^{-2}$ factor
in the relative distance bound is of order $(\log N_0)^{-2}$,
since $t=\Theta(\log N_0)$.

\subsection{Local coefficients and the distance theorem}\label{db:setup}

Retain $B,F,H,Q,t,n,k$ and the color sets $S_i$ from
\cref{cg:section}. The matrices below express its coefficient systems
in the specified polynomial bases. In applying the results of~\cite{DLV},
their complex dimension is $3$ and their cochain degree is our $j$;
$t$ remains the number of base vertices on each side, and $k$
remains the local code dimension.

\begin{definition}[Local coefficients and dual local codes]
\label{db:coefficients}\label{db:dual-coefficients}
Fix a code block $h\in\{1,2,3\}$. For each direction $i$ and
base vertex $v\in V$, write $h_{i,v}^{\mathsf T}:F^k\to F^{S_i}$ for the
combined restriction map of $\shf_{i,\{h\}}$ in the polynomial
basis $1,U_i,\ldots,U_i^{k-1}$ of \cref{cg:graph-sec}.
Its image is the local code $C_{i,\{h\}}(v)$.
The matrix notation $h_{i,v}$ follows~\cite{DLV} and its dependence on the
fixed block $h$ is suppressed. In that convention, $h_{i,v}$ is
a parity-check matrix for our dual local code
$C_{i,\{h\}}(v)^\perp=\ker h_{i,v}$.
Choose a matrix of full row rank $h_{i,v}^{\perp}:F^{S_i}\to F^{n-k}$
with this row space. Following \cref{pre:sheaf}, replace each graph
local code with its dual, using $(h_{i,v}^{\perp})^{\mathsf T}$ as its generator
map. The external product of these systems descends to a coefficient
system $\shf_h^\perp$ on $B$, just as for $\shf_h$.

For a face $f$, let $I(f)\subseteq\{1,2,3\}$ be its active
directions, meaning the directions spanned by $f$.
If $i\notin I(f)$, let $v_i(f)$ be its base vertex in
direction $i$. In the specified bases, the stalks are
\[
 (\shf_h)_f\cong\bigotimes_{i\notin I(f)}F^k,
 \qquad (\shf_h^\perp)_f=\bigotimes_{i\notin I(f)}F^{n-k},
\]
with tensor factors ordered by direction and empty tensor product $F$.
Write $f\prec g$ for an immediate incidence: $f\subseteq g$ and
$\dim g=\dim f+1$. For such an incidence activating direction $i$
with color $c\in S_i$, the restriction $(\shf_h)_{f\to g}$ applies
the row indexed by $c$ of $h_{i,v_i(f)}^{\mathsf T}$ in factor $i$ and the identity
elsewhere. The restriction map for $\shf_h^\perp$
uses $(h_{i,v_i(f)}^\perp)^{\mathsf T}$ instead.
\end{definition}

Activating a different direction leaves $v_i(f)$ unchanged, and
the matrices are independent of the sheet. Thus restriction maps around
a square commute for both coefficient systems.

\begin{definition}[Chains, fillings, and distances]
\label{db:block-weight}\label{db:filling}\label{db:local-filling}
Let $Y=B$ or the auxiliary complex of \cref{db:auxiliary}, with
the same coefficient systems on each copy of $B$.
Use the cochains $C^j(Y;\shf_h)$ and linear dual chains
$C_j(Y;\shf_h^\vee)$ of \cref{pre:kronsec}, so
\[
 C^j(Y;\shf_h)=\bigoplus_{f\in Y(j)}(\shf_h)_f,
 \qquad
 C_j(Y;\shf_h^\vee)=\bigoplus_{f\in Y(j)}(\shf_h)_f^\vee.
\]
In the specified bases, the boundary is
$\partial_{h,j}=(\delta_h^{j-1})^{\mathsf T}$.
On $B$, these are the cochain spaces $\mathcal C_h^j$ and
differentials $\delta_h^j$ of \cref{cg:complexes}.
Write $\delta_h^\perp$ and $\partial_h^\perp$ for the maps of the
dual local code system $\shf_h^\perp$; the distinction between
$\perp$ and linear duality $\vee$ is as in \cref{pre:kronsec}.

The support and block weight are
$\operatorname{supp}(x)=\{f:x(f)\ne0\}$ and
$\bwt{x}=|\operatorname{supp}(x)|$. One block is the full vector on a
face. A \emph{filling} of a cycle $x$ is a chain
$z\in C_{j+1}(Y;\shf_h^\vee)$ with $\partial_{h,j+1}z=x$; a cycle has a filling
exactly when it is a boundary. The block systolic and cosystolic
distances are
\[
 \begin{aligned}
 \mu_{\mathrm{syst}}^Y(j)
 &=\min_{x\in\ker\partial_{h,j}\setminus\operatorname{im}\partial_{h,j+1}}\bwt{x},
 \\[2pt]
 \mu_{\mathrm{cosyst}}^Y(j)
 &=\min_{x\in\ker\delta_h^j\setminus\operatorname{im}\delta_h^{j-1}}\bwt{x}.
 \end{aligned}
\]
The fixed block index $h$ is suppressed in these distances.
An empty minimum is $+\infty$; the subscript ``dual'' denotes the
dual coefficient system. A \emph{local filling} satisfies the same
boundary equation inside a local chain complex, with its local
boundary map. Such fillings are not required to agree on overlaps.
\end{definition}

Hamming weight over $F$ is at least block weight. Under the check
convention of \cref{cg:css-parameters}, the block cosystolic and
systolic distances in degree $1$ therefore give lower bounds for
the $X$ and $Z$ distances, respectively.

Product expansion is used in the convention of \eqref{eq:pe}:
for local codes of length $n$, an array $A$ in the sum of the
directional code spaces admits a decomposition satisfying
$\rho n\sum_i|A_i|_i\le |A|$, where $|A|$ is its scalar Hamming
weight, as in \eqref{eq:pe}.

\begin{theorem}[Distance transfer]\label{db:distance}
Suppose the graph covers $\Gamma_1,\Gamma_2,\Gamma_3$ are simple
and connected, and their normalized biadjacency matrices have
all singular values other than the leading value $1$ at most $\xi$. For each code block
$h\in\{1,2,3\}$ and each base vertex tuple $(v_1,v_2,v_3)\in V^3$,
assume that the tuples
\[
 \begin{gathered}
 (C_{1,\{h\}}(v_1),C_{2,\{h\}}(v_2),C_{3,\{h\}}(v_3))
 \quad\text{and}\\
 (C_{1,\{h\}}(v_1)^\perp,C_{2,\{h\}}(v_2)^\perp,
  C_{3,\{h\}}(v_3)^\perp)
 \end{gathered}
\]
and all their nonempty subtuples are $\rho$-product expanding,
for a common $\rho>0$.
There is $\kappa=\kappa(\rho)\in(0,1]$, independent of $t,n,Q$
and $F$, such that, if
\begin{equation}\label{db:spectral-target}
 \xi\le\frac{\kappa^3}{18\cdot2^{18}},
\end{equation}
then the smaller Pauli distance $d_{\mathrm{fld}}$ of each
length-$N_0$ CSS code over $F$ satisfies
\begin{equation}\label{db:distance-bound}
 d_{\mathrm{fld}}\ge c_\kappa\frac{N_0}{t^2n^7},
 \qquad c_\kappa=\frac{\kappa^3}{7776\cdot2^{24}}.
\end{equation}
The distance bound holds for the full CSS code for all logical qubits.
\end{theorem}

\subsection{Local robustness and expansion of the face walks}
\label{db:local-robustness}\label{db:walks}

Fix a face $f$. In each direction where $f$ has a vertex factor,
every edge extending $f$ is incident to that vertex and uses its
restriction map. The cochain complex on faces containing $f$
is therefore a tensor product of fixed local maps, to which
the local results of~\cite{DLV} apply.

\begin{definition}[Local minimality and robustness]
\label{db:local-minimality}\label{db:robustness-def}
The \emph{upper star} of a face $f$ consists of the faces containing
it; write $Y_{\ge f}(j)=\{u\in Y(j):u\supseteq f\}$.
The cochain complex on the upper star of $f$ consists of
cochains supported on faces containing $f$, with the inherited
coboundary maps.
A cochain $x\in C^j(Y;\shf_h)$, $j\ge1$, is \emph{locally minimal} if,
for every vertex $v$ and every $y\in C^{j-1}(Y;\shf_h)$ supported on
$Y_{\ge v}(j-1)$, one has $\bwt{x+\delta_h^{j-1}y}\ge\bwt{x}$.
This is local co-minimality in~\cite[Definition~2.2]{DLV}.

For a local tensor product cochain complex with differential $d$, a cochain
$y$ is \emph{minimal} if it has minimum block weight among cochains
with coboundary $dy$. The complex is \emph{$\kappa$-robust} at a
degree if every minimal $y$ in that degree satisfies
$\bwt{dy}\ge\kappa n\bwt{y}$; this is~\cite[Definition~5.6]{DLV}.
\end{definition}

\begin{lemma}[Uniform local robustness]\label{db:star-robustness}
Under the product expansion hypothesis of \cref{db:distance}, all
cochain complexes on upper stars in both coefficient systems are
$\kappa$-robust below their top degree, for a common
$\kappa=\kappa(\rho)\in(0,1]$. The conclusion also holds with
vector-valued coefficients, with weight counting the union of face
supports.
\end{lemma}
\begin{proof}
For $f\in Y(p)$, the cochain complex on its upper star is
\begin{equation}
 \bigotimes_{i\notin I(f)}
 \left(F^k\xrightarrow{h_{i,v_i(f)}^{\mathsf T}}F^{S_i}\right),
\end{equation}
with $F^k$ in degree $0$, $F^{S_i}$ in degree $1$, and the total
degree shifted by $p$. This identification follows because each
inactive base vertex stays fixed as other directions are activated.
The dual system replaces $F^k$ by $F^{n-k}$ and uses
$(h_{i,v_i(f)}^\perp)^{\mathsf T}$ instead.
By~\cite[Lemma~5.3]{DLV} and duality, these complexes are exact below
their top degree. Thus minimizing among cochains with a fixed
coboundary is equivalent to minimizing modulo local coboundaries.
At the last differential, an input is a decomposition into local
codewords on lines, and its block weight counts the nonzero lines.
Thus \eqref{eq:pe} gives robustness at that degree.
Lemma~5.13 of~\cite{DLV} gives all lower degrees, and Proposition~5.14
of the same work allows vector-valued coefficients over the same field. Applying
these results to both local code systems and all subtuples gives
a common $\kappa\le1$ depending only on $\rho$ and the at most
three directions.
\end{proof}

A walk using only direction-$i$ edges leaves the other two
base vertices fixed. We therefore apply the spectral estimate
separately on each connected component, retaining its size in
the bound. To use the face walk estimates of~\cite{DLV}, we introduce an
auxiliary complex consisting of eight copies of $B$, with the
coefficient systems copied from $B$. Its systolic and cosystolic
distances are consequently the same as those of $B$.

\begin{definition}[Auxiliary complex and face walks]
\label{db:auxiliary}\label{db:face-walk}
Using the common base vertex set $V=V_-\sqcup V_+$ of
\cref{gp:construction}, set
\begin{equation}
 \Omega=H\times V^3,
 \qquad |\Omega|=8t^3Q.
\end{equation}
Thus $\Omega$ is the vertex set of $B$ in the coordinates of
\cref{cg:balanced}. For $c\in S_i$, let $s_{i,c}$ be the involution on $\Omega$ given
by the direction-$i$ color matching and its voltage. Involutions
in different directions commute: their base vertex actions are in
different coordinates, and they multiply the sheet label by commuting
elements of $H$.
Let $Y$ be the permutation cubical complex
of~\cite[Section~3.1]{DLV}: its vertices are
$\Omega\times\{0,1\}^3$, an edge in direction $i$ applies
$s_{i,c}$ and flips bit $i$, and higher faces are the resulting
commuting squares and cubes, with their direction and color labels.
Give $Y$ the coefficient systems of \cref{db:coefficients}.

For $0\le\ell\le j<3$, the face walk $W^{(j,\ell)}$ first chooses
a uniformly random $\ell$-face of the current $j$-face. It then
chooses uniformly a direction inactive in that $\ell$-face and a
color, and moves to the opposite $\ell$-face of the resulting
$(\ell+1)$-cube. Finally, it chooses uniformly a $j$-face containing
that opposite face. This is exactly
\cite[Definition~6.6]{DLV}.
\end{definition}

\begin{lemma}[Expansion of graph and face walks]
\label{db:components}\label{db:walk-estimate}
The complex $Y$ is a disjoint union of eight copies of $B$. Put
$r_*=1/(8t^2)$. For direction $i$, let $\mathsf P_i$ be the normalized
adjacency on $\Omega\times\{0,1\}$ with edges
$(\omega,b)\leftrightarrow(s_{i,c}\omega,1-b)$; this is the
bipartite double cover of the graph of direction-$i$ edges on $\Omega$.
Each connected component contains a fraction $r_*$ of the double
cover's vertices.
Under the connectivity and spectral hypotheses of
\cref{db:distance}, every real vector $x$ satisfies
\begin{equation}\label{db:double-cover}
 \langle x,\mathsf P_ix\rangle\le\xi\|x\|_2^2+
                    \frac{\|x\|_1^2}{2r_*|\Omega|}.
\end{equation}
Consequently, for $0\le\ell\le j<3$ and $A\subseteq Y(j)$,
\begin{equation}\label{db:walk}
 \langle\one_A,W^{(j,\ell)}\one_A\rangle
 \le\xi|A|+\binom3\ell2^{2-\ell}n^\ell
                       \frac{|A|^2}{r_*|Y(j)|}.
\end{equation}
Here $\one_A$ is the indicator of $A$, and inner products and
$\ell_p$ norms use counting measure.
\end{lemma}
\begin{proof}
In each direction, the sum modulo two of the added bit and the
original bipartite side is invariant. The three sums distinguish
eight copies of $B$. The graph in direction $i$ on $\Omega$ has
$4t^2$ components, each a copy of $\Gamma_i$ with $2tQ$ vertices. Its double
cover has two copies of each such component. On a component $C$,
the normalized adjacency has all eigenvalues other than $1$
bounded above by $\xi$; the eigenvalue $-1$ is allowed. Decomposing
$x_C$ into its mean and a zero-sum vector therefore gives
\begin{equation}
 \langle x_C,\mathsf P_ix_C\rangle\le
 \xi\|x_C\|_2^2+\frac{\|x_C\|_1^2}{2tQ}.
\end{equation}
Sum over components, use $\sum_C\|x_C\|_1^2\le\|x\|_1^2$
and $2tQ=2r_*|\Omega|$, and obtain \eqref{db:double-cover}.
For the face walk, apply the proof of~\cite[Lemma~6.12]{DLV}
with this quadratic form estimate. After fixing the active colors
and the unchanged side bits, its translation operator is a copy
of $\mathsf P_i$; the full operator is a direct sum of such copies and an
average over the inactive directions. Apply \eqref{db:double-cover}
to each copy and average the resulting inequalities. The incidence
averages in that proof and the count
$|Y(\ell)|=\binom3\ell2^{3-\ell}n^\ell|\Omega|$
then give \eqref{db:walk}.
\end{proof}

\subsection{From local estimates to Pauli distance}
\label{db:duality}

The cocycle bound follows from local robustness and the face walk
estimate. For the cycle bound, we also check that local fillings
can be compared on overlaps when the coefficients depend on the
base vertex.

\begin{proposition}[Systolic and cosystolic bounds]
\label{db:dlv-compatibility}\label{db:cosystolic-prop}
\label{db:systole-transfer}\label{db:local-view-exactness}
Under the hypotheses of \cref{db:distance}, both coefficient systems
on $Y$ satisfy, for $j=1,2$,
\begin{equation}\label{db:cosystoles}
 \mu_{\mathrm{cosyst}}^Y(j)\ge
 A_\kappa\frac{r_*|Y(j)|}{n^3},
 \qquad A_\kappa=\frac{\kappa^3}{18\cdot2^{24}}.
\end{equation}
For the original chain complex,
\begin{equation}\label{db:systoles}
 \mu_{\mathrm{syst}}^Y(1)
 \ge (6n)^{-3}\mu_{\mathrm{cosyst,dual}}^Y(2).
\end{equation}
These are the conclusions of~\cite[Propositions~7.1 and~8.1]{DLV}
for the present coefficient systems.
\end{proposition}
\begin{proof}
The local steps of~\cite[Claims~7.4--7.6]{DLV} apply in each
upper star by \cref{db:star-robustness}. Their cancellation step
uses the cocycle equation with the same incidence maps as the local
restrictions; the remaining support counts depend only on the
cubical geometry. Thus the proof of their Proposition~7.1 applies
when the matrices vary with the base vertex. Together with
\cref{db:star-robustness,db:walk-estimate}, it gives
the lower bound $(1-\xi C_1)r_*|Y(j)|/C_2$ for every nonzero
locally minimal cocycle, where the constants in that proposition
satisfy $C_1\le9\cdot2^{18}/\kappa^3$ and
$C_2\le9\cdot2^{24}n^3/\kappa^3$.
The spectral condition makes $1-\xi C_1\ge1/2$. A minimum-weight
representative modulo coboundaries is locally minimal, proving
\eqref{db:cosystoles} for each system.

For the cycle bound, use the local views of
\label{db:local-views}
\cite[Definition~8.2]{DLV}. These record, for each face, a chain on
the faces containing it. A local boundary checks the filling equation within
each view, and a second map compares the views on overlaps.
For $f\in Y(p)$ and $0\le p\le j\le3$, set
\begin{equation}
 L_j(f)=\bigoplus_{u\in Y_{\ge f}(j)}(\shf_h)_u^\vee,
 \qquad T^{p,j}=\bigoplus_{f\in Y(p)}L_j(f).
\end{equation}
A vector $y\in T^{p,j}$ assigns to each $p$-face $f$ a view $y(f)$
of the $j$-faces containing it; $y(f)[u]\in(\shf_h)_u^\vee$ is its value on
such a $j$-face $u$. Its weight counts nonzero views $y(f)$.
In the specified dual bases, the local boundary
$\partial_L:T^{p,j}\to T^{p,j-1}$ sums
$(\shf_h)_{u\to w}^{\mathsf T}y(f)[w]$ over immediate cofaces $w\succ u$ within the
view. The comparison
map $\Delta_j:T^{p,j}\to T^{p+1,j}$ is
\begin{equation}
 (\Delta_jy)(f')[u]
   =\sum_{\substack{f\prec f'\\\dim f=p}}y(f)[u]
 \qquad\bigl(u\in Y_{\ge f'}(j)\bigr),
\end{equation}
where the incidences in this sum are immediate and maps outside the
displayed degree range are zero. As in~\cite[Lemma~8.5]{DLV},
$\partial_L\Delta_j=\Delta_{j-1}\partial_L$: both orders sum the
same incidence maps over the same pairs of faces.

The chain complex $L_\bullet(f)$ is the tensor product of the
surjections $h_{i,v_i(f)}$, shifted by $p$. By
\cite[Lemmas~5.3--5.4]{DLV}, it is exact below degree three, and its
top kernel is the image of
\begin{equation}
 E_f=\bigotimes_{i\notin I(f)}(h_{i,v_i(f)}^\perp)^{\mathsf T}:
 (\shf_h^\perp)_f\longrightarrow L_3(f).
\end{equation}
For fixed $j$, the horizontal complex $(T^{\bullet,j},
\Delta_j)$ is a direct sum, over $j$-faces $u$, of the cellular
cochain complexes of the cubes $u$ with coefficients $(\shf_h)_u^\vee$.
It is exact in positive degrees; its kernel in degree $0$ consists
of the consistent views of global $j$-chains, as in
\cite[Claim~8.3 and Lemma~8.4]{DLV}.

Let $E:C^p(Y;\shf_h^\perp)\to T^{p,3}$ be the direct sum of the maps $E_f$.
On an incidence activating direction $i$, the unchanged
tensor factors use the same matrices, while factor $i$ is evaluated
at the incidence color. Hence
\begin{equation}\label{db:dual-compatibility}
 \Delta_3E=E\delta_h^\perp,
 \qquad \partial_LE=0.
\end{equation}
Each $E_f$ is injective, so $E$ preserves block weight.
These are the exactness and compatibility properties used in
\cite[Section~8.2]{DLV}: the obstruction to making local fillings
agree is a cocycle for $\shf_h^\perp$. If its block weight is below
the cosystolic distance, it is a coboundary, so the fillings can be adjusted to give a global
filling. Their incidence counts give \eqref{db:systoles} by
Proposition~8.1 with dimension three and degree $1$.
\end{proof}

\begin{proof}[Proof of \cref{db:distance}]
The cell counts in \eqref{cg:length} give
\begin{equation}\label{db:counts}
 N_0=k^2|B(1)|=12t^3nk^2Q\le12t^3n^3Q,
 \qquad |B(2)|=6t^3n^2Q.
\end{equation}
The complexes on the eight components of $Y$ are copies of the
complex on $B$, so their systolic and cosystolic distances equal
those on $B$. Substituting $|Y(j)|=8|B(j)|$ and
$r_*=1/(8t^2)$ into \cref{db:dlv-compatibility} gives
\begin{equation}
 \mu_{\mathrm{cosyst}}^B(1)\ge\frac{12A_\kappa tQ}{n^2},
 \qquad
 \mu_{\mathrm{syst}}^B(1)
 \ge (6n)^{-3}\frac{A_\kappa|B(2)|}{t^2n^3}
 =\frac{A_\kappa tQ}{36n^4}.
\end{equation}
Scalar weight is at least block weight. Using
$N_0\le12t^3n^3Q$, both Pauli distances are therefore at least
\begin{equation}
 \frac{A_\kappa}{432}\frac{N_0}{t^2n^7},
 \qquad \frac{A_\kappa}{432}
 =\frac{\kappa^3}{7776\cdot2^{24}}=c_\kappa.
\end{equation}
The hypotheses and constants are uniform over the three blocks.
\end{proof}

The factor $t^{-2}$ comes from the component fraction $r_*$.
Along the family of \cref{gp:theorem}, $Q=M^t$ with
fixed $M>1$, so \eqref{cg:length} gives $t=O(\log N_0)$.
With the field, local codes, and product expansion parameter fixed,
the distance bound becomes $d_{\mathrm{fld}}=\Omega(N_0/\log^2N_0)$.

%% file: sections/code_params.tex
\section{Qubit codes and proofs of the main results}\label{sec:parameters}

We combine the addressable logical CCZ gates of \cref{cg:diagonal} with the
distance bound of \cref{db:distance} to obtain qubit codes with
transversal implementations of these gates. The conversion has constant overhead when
the field and local parameters are fixed.

\subsection{Conversion to qubits and transversal CCZ gates}\label{db:binary}

We first represent each physical qudit by qubits.
A qubit may participate in several gates.
Repetition encoding assigns a different member of an encoded repetition block to each occurrence,
making the gates transversal. These are
the alphabet and locality reductions of Golowich and
Lin~\cite[Lemmas~3.44--3.45]{GL}. We choose the copies using a common
list of potential gates for all $I\subseteq \calS$, so the codes and
gate locations do not depend on the selected logical triples.

For block $h$, let $N_0$ and $\calK_{0,h}$ be its length and
encoded dimension over $F$, as in \cref{cg:css-parameters}, and let
$D_X^F,D_Z^F$ be its Pauli distances. Choose $F$-linear generating
sets for the $X$- and $Z$-check spaces, with each check vector of
weight at most $w$ and each physical qudit contained in at
most $v$ check vectors across both sets. We call this bound the
\emph{combined column degree}.

\begin{definition}[Binary expansion and the trace pairing]
\label{db:trace-coordinates}
Use $\mu=[F:\F_2]$ and the trace $\Tr$ from
\eqref{pre:field-trace}, so $F=\F_{2^\mu}$ and $\mu=\log_2q$.
Choose an $\F_2$-basis $\theta_1,\ldots,\theta_\mu$ of $F$ and its
trace dual basis $\theta_1^*,\ldots,\theta_\mu^*$, so
$\Tr(\theta_i\theta_j^*)=\one_{i=j}$. Encode each physical qudit by
\[
 \ket{x}\longmapsto\ket{x_1,\ldots,x_\mu},
 \qquad x=\sum_{j=1}^{\mu}\theta_jx_j.
\]
Express $X$ labels in $(\theta_j)$ and $Z$ labels in $(\theta_j^*)$.
\end{definition}

\begin{proposition}[Alphabet reduction]\label{db:scalar-restriction}
This encoding gives a binary CSS code with
\[
 N_{\mathrm{out}}=\mu N_0,\qquad
 \calK_{\mathrm{out},h}=\mu\calK_{0,h},
 \qquad D_X^{\mathrm{out}}\ge D_X^F,\quad
 D_Z^{\mathrm{out}}\ge D_Z^F.
\]
Its generators can be chosen with weight at most $\mu w$ and
combined column degree at most $\mu v$.
\end{proposition}
\begin{proof}
Restriction of scalars gives the length, dimension, and both
distance bounds by~\cite[Lemma~3.44]{GL}. The trace dual bases
identify the field Pauli commutation pairing with the binary dot
product. Replace each field check vector by its $\mu$ basis multiples.
Each resulting generator has weight at most $\mu w$, and each
qubit meets at most $\mu v$ generators.
\end{proof}

\begin{definition}[Trace phase and monomial occurrences]
\label{db:trace-phase}
Choose $\gamma_{\mathrm{tr}}\in F$ with
$\Tr(\gamma_{\mathrm{tr}})=1$ and put
$\varphi(s)=\Tr(\gamma_{\mathrm{tr}}s)$, so $\varphi(1)=1$.
Thus $\varphi$ restricts to the identity on $\F_2$, as needed for
the logical qubit CCZ phase.
For $I\subseteq \calS$, use the form $T_I$ from
\cref{cg:information-set-definition} to define the binary phase
$(-1)^{\varphi(T_I)}$.

A \emph{monomial occurrence} is a term in the geometric expansion of
$T_I$ before equal monomials are collected. Fix the list indexed by
the cube, cup product path, and three field coefficient indices,
retaining zero coefficients. Only the coefficients depend on $I$.
By \cref{cg:invariance}, each physical qudit belongs to at most
\[
 \tau_{\mathrm{occ}}=6n^2k^4\le6n^6
\]
occurrences.
\end{definition}

\begin{lemma}[Binary expansion of the phase]\label{db:binary-phase}
The phase $(-1)^{\varphi(T_I)}$ is a product of binary CCZ gates.
There is a common list of potential binary gate occurrences for all
$I$ in which each qubit occurs at most $\mu^2\tau_{\mathrm{occ}}$ times.
\end{lemma}
\begin{proof}
For a field monomial $cxyz$, with coefficient $c\in F$ and
computational basis labels $x,y,z\in F$, the physical basis gives
\begin{equation}\label{db:trace-expansion}
 \varphi(cxyz)=\sum_{i,j,l=1}^{\mu}
       \varphi(c\theta_i\theta_j\theta_l)x_i y_j z_l.
\end{equation}
Each nonzero binary coefficient specifies a CCZ gate across the three
blocks. Append the three binary indices to the common occurrence
list, retaining zero coefficients. Each qubit has
at most $\tau_{\mathrm{occ}}$ field occurrences and $\mu^2$ choices for the other
two indices. Repeated monomials contribute modulo two.
\end{proof}

\begin{lemma}[Repetition concatenation]\label{db:repetition}
Let a binary CSS code have parameters $N_{\mathrm{out}},\calK_{\mathrm{out}}$,
Pauli distances $D_X^{\mathrm{out}},D_Z^{\mathrm{out}}$, generator
weight at most $w_{\mathrm{out}}$, and combined column degree at most
$v_{\mathrm{out}}$. For any integer $R_{\mathrm{rep}}\ge1$, encoding
each qubit by $\ket{x}\mapsto\ket{x}^{\otimes R_{\mathrm{rep}}}$,
$x\in\F_2$, gives a CSS code with
\[
 N=R_{\mathrm{rep}}N_{\mathrm{out}},\quad \calK=\calK_{\mathrm{out}},\quad
 D_X=R_{\mathrm{rep}} D_X^{\mathrm{out}},\quad D_Z=D_Z^{\mathrm{out}},
\]
generator weight at most $\max\{R_{\mathrm{rep}}w_{\mathrm{out}},2\}$ and combined
column degree at most $v_{\mathrm{out}}+2$.
\end{lemma}
\begin{proof}
Use the construction of~\cite[Lemma~3.45]{GL}: impose
$Z_rZ_{r+1}$ for $1\le r<R_{\mathrm{rep}}$ within each repetition
block, lift outer $X$ operators to all copies, and lift outer $Z$
operators to the first copies. The cited lemma gives the length,
dimension, and $X$ distance. Modulo the repetition checks, a $Z$
operator is determined by its parity in each block; representing
these parities on the first copies gives the stated $Z$ distance.
These generators give the weight bound, and each copy
meets at most two additional repetition checks.
\end{proof}

\begin{theorem}[Fixed qubit codes and transversal CCZ gates]
\label{db:qubit-conversion}
Apply alphabet reduction and repetition concatenation to each of
the three codes over $F$, with
\begin{equation}\label{db:repetition-length}
 R_{\mathrm{rep}}=6\mu^2n^6.
\end{equation}
The resulting codes and the gate locations of a transversal CCZ
layer can be fixed independently of $I$. Activating a subset of
those locations implements $(-1)^{\varphi(T_I)}$ on encoded states.
For each code block $h$,
\begin{align}\label{db:binary-parameters}
 N&=\mu R_{\mathrm{rep}}N_0,& \calK_h&=\mu\calK_{0,h},&
 d&\ge d_{\mathrm{fld}},\\
 w_{\mathrm{bin}}&\le\max\{\mu R_{\mathrm{rep}}w,2\},&
 v_{\mathrm{bin}}&\le\mu v+2.\notag
\end{align}
Here $d$ is the smaller Pauli distance, and $w_{\mathrm{bin}}$ and
$v_{\mathrm{bin}}$ bound the binary generator weight and combined
column degree, respectively.
In particular, \cref{cg:css-parameters} gives $w\le4n$ and
$v\le2k+2n\le4n$, hence
\[
 w_{\mathrm{bin}}\le\max\{4\mu R_{\mathrm{rep}}n,2\},
 \qquad v_{\mathrm{bin}}\le4\mu n+2.
\]
\end{theorem}
\begin{proof}
Every qubit belongs to at most
$\mu^2\tau_{\mathrm{occ}}\le R_{\mathrm{rep}}$ potential occurrences by
\cref{db:binary-phase}. For each qubit, assign different
repetition copies to these occurrences. Each occurrence then uses
one copy in each block, and no copy belongs to two occurrences.
Relabeling the qubits within each block places the three qubits of
each occurrence at the same position. The common occurrence
list thus fixes a transversal CCZ layer independently of $I$.

For a given $I$, apply a gate exactly when its coefficient in
\eqref{db:trace-expansion} is one. All repetition copies have the
same computational basis value, so the resulting phase is precisely
$(-1)^{\varphi(T_I)}$, including cancellation between repeated terms.
Linearity gives the action on arbitrary encoded states, and
code space preservation follows from \cref{cg:invariance}.
The parameter bounds follow from
\cref{db:scalar-restriction,db:repetition,db:distance}.
\end{proof}

To obtain logical qubit CCZ gates, retain the qubit associated
with $1\in F$ in each selected logical qudit.

\input{sections/addressable}

\subsection{Final parameters and proof of the main theorem}
\label{sec:final-parameters}

The uniform product expansion hypothesis lets us choose the
spectral target before fixing the field. We then keep all local
parameters fixed as the base graphs grow.

\begin{proof}[Proof of \cref{thm:main}]
Fix $0<\eps<1$ and one power $P\ge2$ of two. Apply
\cref{ass:pe} with rate gap $1/8$, obtaining $\rho>0$.
The local robustness bound of \cref{db:star-robustness} gives
$\kappa=\kappa(\rho)>0$.
Choose positive $\lambda_0,\delta,\eta$ with
\begin{equation}
 \lambda_0\le1/14,\qquad
 \lambda_0+\delta+\eta<\kappa^3/(18\cdot2^{18}).
\end{equation}
Next choose $d_0,D$ satisfying
\cref{gp:base-degree-condition,gp:occupancy-condition}, with strict
slack in both inequalities. Neither constant depends
on a subfield extension degree.

Increase the odd integer $r$ until all of the following hold:
the roots and evaluation sets of \cref{gp:palettes} exist;
$k=\lfloor n/7\rfloor\ge n/8$;
the degree conclusion of \cref{gp:frobenius} holds;
the rank inequality \eqref{cg:rankcondition} holds; and
\begin{equation}\label{eq:entropy-target}
 6D\epsilon_E<\eps/4.
\end{equation}
Each condition holds for all sufficiently large $r$.
For the last one use $\epsilon_E\to0$ from \eqref{gp:digit-density}; for the rank condition
use its negative term of order $n\log q$.
Fix this $r$, the coefficient field $F$, and successful roots and
evaluation sets. Thus $q,M,n,k,\mu=\log_2q$, and $R_{\mathrm{rep}}$
are constants depending at most on $\eps$.

Let $t$ run through the unbounded sequence of base graph sizes
in \cref{gp:analysis-parameters}. Each base graph has $t$ vertices
on each side, and the covering group is $H=(F^\times)^t$.
For sufficiently large $t$, choose slopes giving the simple
expanding covers of \cref{gp:spectrum}, then shifts satisfying
\cref{gp:selection}. The bound \eqref{cg:rankprob} is uniform in
these choices, so successful halfedge multipliers can be chosen next.

The local codes $C_{i,\{h\}}(v)$ are diagonal rescalings of
$\RS(S_i,k)$, and their duals are generalized RS codes of
dimension $n-k$. Both dimensions lie between $n/8$ and
$7n/8$. Every tuple of local codes needed for the distance argument is
therefore covered by \cref{ass:pe}, regardless of its vertex-dependent
shifts or multipliers.

\Cref{cg:diagonal} gives $\calK_{0,h}\ge\srk=|\mathcal A|$
and $\srk$ matched logical triples with independently selectable
phases. From \eqref{gp:selection-size} and
\eqref{eq:entropy-target}, for all sufficiently large $t$,
\begin{equation}
 \log_M \srk\ge t-6tD\epsilon_E
             \ge(1-\eps/2)t.
\end{equation}
After conversion to qubits,
\begin{equation}\label{eq:final-length}
 N=12\mu R_{\mathrm{rep}}nk^2t^3M^t
   =\Theta_\eps(t^3M^t),\qquad
 \log_M N=t+O_\eps(\log t).
\end{equation}
Consequently
\begin{equation}
 \log_M \srk-(1-\eps)\log_M N
 \ge \frac\eps2t-O_\eps(\log t)\longrightarrow\infty.
\end{equation}
Thus the retained binary logical yield is at least $N^{1-\eps}$,
and the full binary dimension, which is at least $\mu\srk$, satisfies
the same lower bound.

For both Pauli types, \cref{db:distance} and
\eqref{db:binary-parameters} give
\begin{equation}
 d\ge\frac{c_\kappa}{\mu R_{\mathrm{rep}}n^7}\frac{N}{t^2}
   =\Omega_\eps\!\left(\frac{N}{(\log N)^2}\right).
\end{equation}
The check bounds in \eqref{db:binary-parameters} are constant.
\Cref{addr:layer} gives the required fixed transversal layer for
every subset, with complementary logical qubits initialized
to $\ket0$.
\end{proof}

\subsection{Magic state applications}\label{sec:magic-state-applications}

\begin{proof}[Proof of \cref{thm:distill}]
Fix a sufficiently large member of \cref{thm:main}, with block length
$N$ and $D_{\mathrm{dist}}:=\min_h D_{Z,h}=\Omega_\eps(N/\log^2N)$. We use the
postselected distillation routine of~\cite{BH}, in the form used
in~\cite{GL,LLL}, with ideal Clifford operations. Initialize the
selected logical qubits to $\ket+$ and the complementary ones
to $\ket0$. For independent input CCZ states of infidelity at most $p$,
Clifford twirling makes the noise a stochastic $Z$-type Pauli error.
Implement each location of $\mathcal L_{\calS}$ (\cref{addr:layer}) by
gate teleportation through one such state. A faulty input produces
a $Z$ error on at most one qubit in each block, since the CCZ
implementation is transversal.

Measure the $X$-type stabilizers and accept only the trivial syndrome.
With fewer than $D_{\mathrm{dist}}$ faulty locations, the error in each block has
weight below its $Z$ distance. It therefore either gives a nontrivial
syndrome or is a stabilizer and acts trivially on encoded states.
An accepted logical error requires at least $D_{\mathrm{dist}}$ faulty locations.
For the fixed code, a union bound gives probability
$O_N(p^{D_{\mathrm{dist}}})$ of an accepted logical error, while the
acceptance probability is $1-O_N(p)$. Conditional on acceptance,
ideal Clifford decoding therefore yields $\srk$ states with joint
infidelity $O_N(p^{D_{\mathrm{dist}}})$ relative to
$\ket{\CCZ}^{\otimes\srk}$.

Each round uses $|\mathcal L_{\calS}|\le N$ inputs. Iterate the same routine,
supplying the inputs to each invocation from independent earlier
batches, as in~\cite[Section~VI]{BH}. For sufficiently small initial
$p$, depending on the fixed code, the error decreases under iteration.
The error bounds are summable, so the product of the
acceptance probabilities over all rounds is positive. Rejected
runs therefore contribute only a constant factor to the expected overhead.
Apart from this factor, each round multiplies the input cost per output by
$|\mathcal L_{\calS}|/\srk$, while the error bound raises the previous
error to the power $D_{\mathrm{dist}}$. This gives the overhead exponent
\begin{equation}
 \gamma=\frac{\log(|\mathcal L_{\calS}|/\srk)}{\log D_{\mathrm{dist}}}
 \le\frac{\eps\log N}{\log N-2\log\log N-O_\eps(1)}
 =\eps\,(1+o(1)).
\end{equation}
Keeping $N$ fixed and iterating rounds gives expected overhead
$O(\log^\gamma(1/\delta))$ per output at target marginal infidelity $\delta$.
The $o(1)$ term tends to zero as $N$ grows along the family, so
choosing $\eps$ small gives arbitrarily small exponents.
\end{proof}

\begin{proof}[Proof of \cref{thm:fountain}]
Initialize the $\srk$ selected logical qubits of each block to
$\ket+$ and the complementary ones to $\ket0$. This logical
stabilizer state can be prepared by Clifford operations.
Apply the at most $N$ gates at $\mathcal L_{\calS}$ using ideal physical
CCZ gates. By \cref{addr:layer}, the output is the encoded state
$\bigotimes_{w\in \calS}\CCZ\ket{+\!+\!+}_w$, giving
$\srk\ge N^{1-\eps}$ logical CCZ states. They remain in the full
CSS codes, with the Pauli distances of \cref{thm:main}.
\end{proof}

\subsection{Construction from fixed local data}\label{sec:construction}

\begin{proposition}[Construction from fixed local data]\label{prop:algorithm}
Fix $0<\eps<1$. Suppose the field $F$ and subfields $K_i$,
the parameters $P,d_0,D,r,\lambda_0,\delta,\eta$, the root family
$A_{\mathrm{root}}$, and the color sets $B_{i,j}$ partitioning
the evaluation sets $S_i$ are given, satisfying the conditions
imposed in the proof of \cref{thm:main}.
These inputs remain fixed as $N$ grows.
A randomized algorithm constructs the check matrices, selected
logical cochains $\Phi_h(F_z)$ for $h\in\{1,2,3\}$ and $z\in \calS$,
and the transversal gate description in expected time
$N^{O_\eps(1)}$.
\end{proposition}
\begin{proof}
For the base graphs, use the bipartite two-lift construction of
\cite[Theorems~5.3 and~5.5]{MSS}. Our choice of $d_0$ gives
$2\sqrt{d_0-1}/d_0<\lambda_0$. At each lift step, enumerate
signings until the normalized signed adjacency matrix has norm
below a fixed rational number between these two bounds. Exact
rational positive definiteness tests certify the norm bound.
At size $t=O_\eps(\log N)$ there are
$2^{O(d_0t)}=N^{O_\eps(1)}$ choices. Find perfect matchings by
standard bipartite matching algorithms.

Sample slopes and check $|Z_\alpha|\le2t\delta$ in each direction
for every nontrivial character, together with simplicity.
By \cref{gp:uniform-exceptional-set,gp:simple-covers}, the successful
proportion tends to one; \cref{gp:block-norm-bound} then gives the
spectral bound. Choose the shifts by conditional
expectation of $|\mathcal A|$. For each next shift, enumerate its
$M$ possibilities and all $Q$ characters; the conditional probability
of the remaining constraints is an explicit power of $|E|/M$.
Retain a choice that does not lower the expectation. This constructs
the bound in \eqref{gp:selection-size} deterministically.

Sample multipliers and check the three deleted edge matrices in all
$Q$ finite field Fourier blocks. By \eqref{cg:rankprob}, the
successful proportion tends to one. Each matrix has $O(tn)$ rows and
columns. Field arithmetic has constant bit cost for the fixed local
data. Gaussian elimination constructs the maps $\lambda_i$, the
information set $\calS$, and the functions $F_z$ of
\cref{cg:information-set-definition}. Evaluating the local
polynomials gives $\Phi_h(F_z)$. Finally, expand the trilinear form
using the binary basis and assign repetition copies as in
\cref{db:qubit-conversion}. Since $Q=M^t\le N$, enumerating all
characters and forming these arrays takes polynomial time in $N$.
Both sampling steps have bounded expected repetitions
for sufficiently large family members.
\end{proof}

\appendix

%% file: sections/addressable.tex
\begin{corollary}[A fixed transversal layer for every subset]\label{addr:layer}
\label{db:logical-qubit-gates}
Fix the three qubit codes of \cref{db:qubit-conversion} and extend
the cohomology classes of \cref{cg:diagonal}, indexed
by $\calS$, to bases defining the logical qudits. In each selected logical qudit,
choose a binary basis beginning with $1\in F$ and retain its
first qubit.
Initialize all complementary logical qubits to $\ket0$, including
the other $\mu-1$ qubits in each selected logical qudit.

Let $\mathcal L$ be the fixed set of physical CCZ locations in the
transversal layer of \cref{db:qubit-conversion}. For $I\subseteq \calS$,
let $\mathcal L_I$ consist of the locations whose binary coefficient
in the expansion of $\varphi(T_I)$ is one. Applying these gates
implements
\begin{equation}\label{addr:unitary}
 \prod_{w\in I}\overline{\CCZ}_w
\end{equation}
transversally, in depth one, where $\overline{\CCZ}_w$ couples the chosen logical
qubits at address $w$ in the three blocks. The codes, logical bases,
and set $\mathcal L$ are independent of $I$, and $|\mathcal L|\le N$.
\end{corollary}
\begin{proof}
The logical binary bases can be chosen independently of the
physical basis in \cref{db:trace-coordinates}. On the initialized
subspace, the selected logical values $a_{h,w}$ lie in $\F_2$ and
all other logical qudits have computational basis label zero. By \cref{cg:diagonal}
and $\varphi(1)=1$, the phase is
\begin{equation}\label{db:logical-binary-phase}
 (-1)^{\varphi(\sum_{w\in I}a_{1,w}a_{2,w}a_{3,w})}
 =(-1)^{\sum_{w\in I}a_{1,w}a_{2,w}a_{3,w}},
\end{equation}
which is the logical CCZ action in \eqref{addr:unitary}.
\Cref{db:qubit-conversion} implements this phase at the locations
$\mathcal L_I$.
Each location uses a distinct qubit from each length-$N$ block,
so $|\mathcal L|\le N$.
\end{proof}

The initialization restricts logical inputs without adding
stabilizer checks. The dimensions and distances in
\eqref{db:binary-parameters} therefore describe the full codes,
while \eqref{addr:unitary} specifies their action on the initialized
subspace.

%% file: sections/appendix_reed_solomon.tex
\section{Reed--Solomon codes, Schur products, and information sets}
\label{app:classical}

This appendix records the classical coding facts used in the construction;
the facts hold over any finite field $F$.
A linear code of length $n$ over $F$ is a subspace $C\subseteq F^S$,
where $S$ is a coordinate set of size $n$. For Reed--Solomon codes,
$S\subseteq F$ is the evaluation set, whose elements are the edge colors
in the graph construction. The dimension is $k=\dim C$; the minimum distance
$d(C)=\min\{|c|:0\ne c\in C\}$ counts nonzero coordinates; a generator
matrix $G\in F^{k\times S}$ has row space $C$; the dual
$C^\perp=\{x\in F^S:x\cdot c=0\ \forall c\in C\}$ has dimension $n-k$,
where $x\cdot c=\sum_{s\in S}x_sc_s$.

\subsection{Reed--Solomon codes and the MDS property}

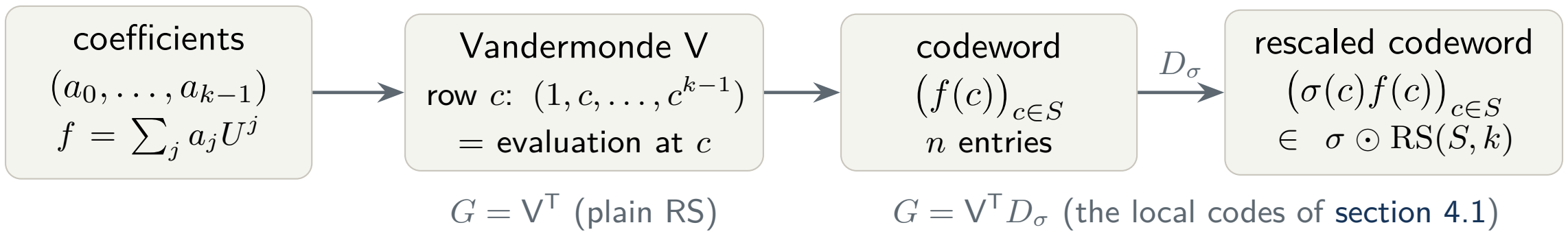
\begin{figure}[htbp]
\centering
\begin{tikzpicture}[x=1cm,y=1cm]
\node[fbox,text width=2.6cm] (co) at (0,0)
 {coefficients\\[1pt] $(a_0,\dots,a_{k-1})$\\[1pt] {\footnotesize $f=\sum_j a_jU^j$}};
\node[fbox,text width=3.1cm] (vd) at (4.1,0)
 {Vandermonde $\mathsf V$\\[1pt] {\footnotesize row $c$: $(1,c,\dots,c^{k-1})$}\\[1pt]
  {\footnotesize $=$ evaluation at $c$}};
\node[fbox,text width=2.5cm] (cw) at (8.0,0)
 {codeword\\[1pt] $\big(f(c)\big)_{c\in S}$\\[1pt] {\footnotesize $n$ entries}};
\node[fbox,text width=2.9cm] (sc) at (11.9,0)
 {rescaled codeword\\[1pt] $\big(\sigma(c)f(c)\big)_{c\in S}$\\[1pt]
  {\footnotesize $\in\ \sigma \odot \RS(S,k)$}};
\draw[flow] (co) -- (vd);
\draw[flow] (vd) -- (cw);
\draw[flow] (cw) -- node[sml,above]{$D_\sigma$} (sc);
\node[sml] at (4.1,-1.15) {$G=\mathsf V^{\mathsf T}$ (plain RS)};
\node[sml] at (10.0,-1.15) {$G=\mathsf V^{\mathsf T}D_\sigma$ (the local codes of \cref{cg:graph-sec})};
\end{tikzpicture}
\caption{Evaluation and rescaling of a Reed--Solomon code. Multiplying
the column of polynomial coefficients by $\mathsf V$ gives its evaluations;
multiplying by $D_\sigma\mathsf V$ gives the rescaled evaluations. The
corresponding row generator matrices are their transposes. Here
$D_\sigma=\diag(\sigma(c))_{c\in S}$ (\cref{pre:vrow}).}
\label{fig:rspipe}
\end{figure}

\begin{definition}\label{appA:rs}
For an evaluation set $S\subseteq F$ of size $n$ and $1\le k\le n$,
\begin{equation}
\RS(S,k)=\big\{(f(c))_{c\in S}:\ f\in F[U],\ \deg f<k\big\}\subseteq F^S .
\end{equation}
In the monomial basis $1,U,\dots,U^{k-1}$ the evaluation map is the
Vandermonde matrix $\mathsf V\in F^{n\times k}$, $\mathsf V_{c,j}=c^{\,j}$: row $c$ is
the evaluation functional $(1,c,c^2,\dots,c^{k-1})$ at $c$
(\cref{fig:rspipe}). Thus $\mathsf V$ acts on a column of polynomial coefficients;
with the row generator convention above, $G=\mathsf V^{\mathsf T}$.
\end{definition}

\begin{lemma}[Singleton bound; RS is MDS]\label{appA:mds}
Every $[n,k,d]$ code satisfies $d\le n-k+1$. The code $\RS(S,k)$ has
parameters $[n,k,n-k+1]$, and the following equivalent properties hold:
every $k$ coordinates determine the codeword; every $k\times k$ submatrix
of $\mathsf V$ is invertible; every nonzero codeword has at most $k-1$ zero
coordinates; any $n-k$ erasures are correctable; every $k$-subset of $S$
is an information set in the sense of \cref{appA:infoset} below.
\end{lemma}
\begin{proof}
Projection onto any $n-d+1$ coordinates is injective, since
two codewords agreeing there differ in at most $d-1<d$ positions, so
$k\le n-d+1$. For Reed--Solomon codes, a nonzero polynomial of degree $<k$ has at most
$k-1$ roots, so a nonzero codeword has weight at least $n-k+1$, matching
Singleton; the $k$ monomials are independent as functions on $S$ since
$k\le n$. Invertibility of the $k\times k$ submatrices is the Vandermonde
determinant $\prod_{c<c'}(c'-c)\ne0$, where the selected evaluation
points are given any fixed ordering. Lagrange interpolation then
recovers $f$ from any $k$ evaluations, proving the erasure statement.
\end{proof}

\subsection{Diagonal rescalings and duality}

\begin{definition}\label{appA:rescale}
For $\sigma\in(F^\times)^S$ and $C\subseteq F^S$, the rescaled code is
$\sigma \odot C=\{\sigma \odot c:c\in C\}$ with $(\sigma \odot c)(x)=\sigma(x)c(x)$,
equivalently $D_\sigma C$ for the invertible diagonal matrix
$D_\sigma=\diag(\sigma(x))$. Codes of the form $\sigma \odot \RS(S,k)$ are
called generalized Reed--Solomon (GRS). The scalars $\sigma(x)$
are their column multipliers.
\end{definition}

Rescaling preserves supports and weights, and hence the parameters
$[n,k,d]$, the MDS property, and information sets. The dual transforms
by the inverse rescaling:
\begin{equation}\label{appA:dualrescale}
(\sigma \odot C)^\perp=\sigma^{-1} \odot C^\perp,
\qquad\text{since }(\sigma\odot c)\cdot(\sigma^{-1}\odot x)
=c\cdot x.
\end{equation}

\begin{lemma}[GRS duality]\label{appA:grsdual}
$\RS(S,k)^\perp=u \odot \RS(S,\,n-k)$, where
$u(c)=\prod_{c'\in S\setminus\{c\}}(c-c')^{-1}$. Consequently the GRS
family is closed under duality, and
$(\sigma \odot \RS(S,k))^\perp=(\sigma^{-1}\odot u) \odot \RS(S,n-k)$.
Here $\RS(S,0)=\{0\}$ when $k=n$.
\end{lemma}
\begin{proof}
Both sides have dimension $n-k$, so it suffices to prove orthogonality:
for $\deg f<k$ and $\deg g<n-k$,
$\sum_{c\in S}u(c)f(c)g(c)=0$ because $\deg fg<n-1$ and, by Lagrange
interpolation on $S$ of a polynomial $h$ with $\deg h\le n-2$, the
coefficient of $U^{\,n-1}$ in
$\sum_ch(c)\prod_{c'\ne c}\frac{U-c'}{c-c'}$ is
$\sum_cu(c)h(c)$ and must vanish. The last claim combines this with
\eqref{appA:dualrescale}.
\end{proof}

\begin{example}\label{appA:f5}
Over $\F_5$ with $S=\{1,2,3,4\}$ and $k=2$: the dual multipliers are
$u=(4,3,2,1)$, and $\RS(S,2)^\perp=u \odot \RS(S,2)$. For instance
$u \odot (1,1,1,1)=(4,3,2,1)$ annihilates both generators of $\RS(S,2)$:
$4+3+2+1=10=0$ and $4\cdot1+3\cdot2+2\cdot3+1\cdot4=20=0$ in $\F_5$.
\end{example}

\subsection{Schur products}

\begin{definition}\label{appA:schurdef}
For $u,v\in F^S$, the Schur (coordinatewise) product is
$(u \odot v)(x)=u(x)v(x)$; for codes, $C \odot C'=\Span\{c \odot c':c\in C,\,c'\in C'\}$.
\end{definition}

\begin{lemma}[Schur products of RS codes]\label{appA:schur}
For $k+k'-1\le n$,
$\RS(S,k) \odot \RS(S,k')\subseteq\RS(S,\,k+k'-1)$, and with rescalings,
\begin{equation}
\big(\sigma \odot \RS(S,k)\big) \odot \big(\sigma' \odot \RS(S,k')\big)
\subseteq (\sigma\odot\sigma') \odot \RS(S,\,k+k'-1).
\end{equation}
\end{lemma}
\begin{proof}
For $\deg f<k$ and $\deg g<k'$, the pointwise product of their
evaluation vectors is the evaluation vector of $fg$. Polynomial degrees add,
$\deg fg\le (k-1)+(k'-1)<k+k'-1$. Rescalings multiply coordinatewise. For
$S=\{1,2,3,4\}\subset\F_5$, $k=k'=2$, the containment is an equality:
the products of $1,U$ span $1,U,U^2$, whose evaluations generate
$\RS(S,3)$.
\end{proof}

This containment supplies the local coefficient pairings used for
cup products in \cref{cg:section}. The local codes in the coefficient
systems of \eqref{cg:productcodes} are iterated Schur products, with
one factor for each block in $J$. Multiplying the polynomials whose
evaluations define these codewords gives degree $<r_J=|J|(k-1)+1$;
for all three blocks, this bound is $u=3k-2$.
For graph cocycles, \cref{pre:cupschur}
identifies the corresponding Schur products with evaluated cup
products. The requirement
$u<n/2$ in \cref{cg:rank} motivates the choice
$k=\lfloor n/7\rfloor$.

\subsection{Information sets and systematic form}

\begin{definition}\label{appA:infoset}
Let $C\subseteq F^S$ have dimension $k$ and generator matrix $G$. A set
$I\subseteq S$ with $|I|=k$ is an \emph{information set} if $G|_I$ is
invertible --- equivalently, the restriction $C\to F^I$ is an
isomorphism: every pattern on $I$ is realized by exactly one codeword. A
set $M\subseteq S$ is \emph{extendable} if $M\subseteq I$ for some
information set $I$ --- equivalently, the columns $\{G_x:x\in M\}$ are
linearly independent; equivalently, every pattern on $M$ is realized by
at least one codeword.
\end{definition}

\begin{figure}[htbp]
\centering
\begin{tikzpicture}[x=1cm,y=1cm]
\draw[linec] (0,0) rectangle (5.4,1.8);
\foreach \xx in {0.45,1.6,3.1,4.5}
  \fill[accentc!28] (\xx,0) rectangle ++(0.33,1.8);
\node[smk,above=2pt] at (2.7,1.8) {$G$ \ ($k\times n$)};
\node[sml,below=2pt] at (2.7,0) {highlighted: $I$, with $G|_I=M_I$ invertible};
\draw[flow] (6.0,0.9) -- (7.2,0.9);
\node[sml,above] at (6.6,1.0) {$M_I^{-1}\cdot$};
\draw[linec] (7.6,0) rectangle (13.0,1.8);
\foreach \xx/\yyy in {8.05/1.46,9.2/1.01,10.7/0.56,12.1/0.11}{
  \fill[accentc!28] (\xx,0) rectangle ++(0.33,1.8);
  \node[smk] at (\xx+0.165,\yyy+0.22) {$1$};}
\node[smk,above=2pt] at (10.3,1.8) {$G_{\mathrm{sys}}=M_I^{-1}G$: identity on $I$};
\node[sml,below=2pt] at (10.3,0) {row $z$: $F_z(z')=\one_{z=z'}$ on $I$};
\end{tikzpicture}
\caption{Information set and systematic form
(\cref{appA:infoset,appA:sysform}). Row-reducing onto the columns of $I$
puts a $1$ in each selected column of its own row and $0$ in the others;
the rows give the systematic basis $F_z$. Their values outside $I$ are fixed
by the code, while the diagonal logical pairing evaluates them only on
the selected coordinates in $I$.}
\label{fig:sysform}
\end{figure}
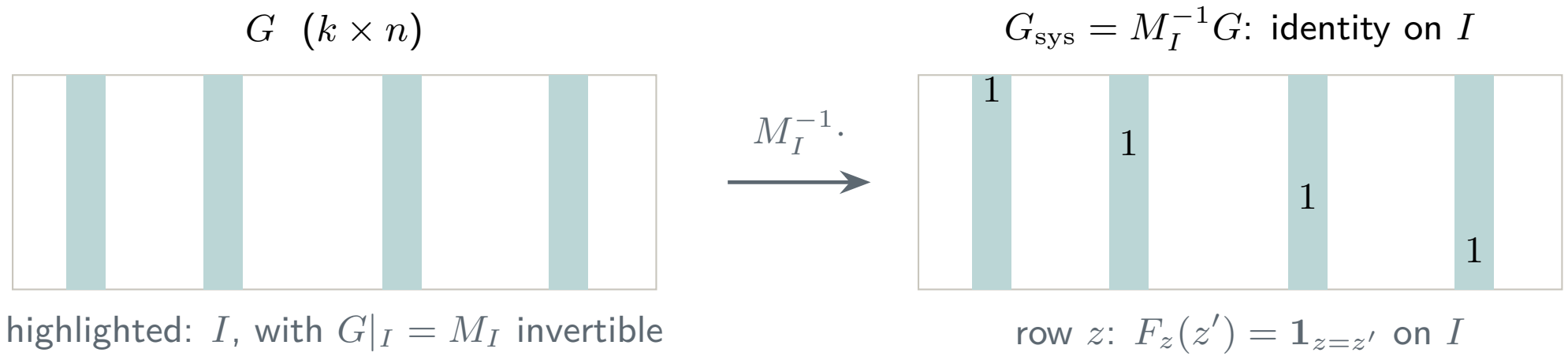

\begin{definition}\label{appA:sysform}
Given an information set $I$, the \emph{systematic generator matrix} is
$G_{\mathrm{sys}}=(G|_I)^{-1}G$, which restricts to the identity on the
columns of $I$. Its rows form the \emph{systematic basis}: row
$z$ (for $z\in I$) is the unique codeword $F_z$ with
$F_z(z')=\one_{z=z'}$ for $z'\in I$. Its values outside $I$ are
determined by membership in $C$
(\cref{fig:sysform}).
\end{definition}

For MDS codes every $k$-subset is an information set
(\cref{appA:mds}). For Reed--Solomon codes, the systematic basis
consists of evaluations of the Lagrange basis polynomials. For general
codes, including the character code $\calV$ of
\cref{cg:infoset-sec}, information sets exist whenever the generator
matrix has full row rank, and are found by Gaussian elimination (pivot
columns); a $k$-set fails exactly when some nonzero codeword vanishes on
all of it. These facts enter the construction in three places: any $k$
colors determine a local RS codeword (interpolation); the singleton
$\{b_i\}$ is an extendable set of the dual Tanner code in every Fourier
block, with $\widehat\lambda_i(\alpha)$ taking value one at that coordinate
(\cref{cg:dual-construction}); and \cref{cg:diagonal} chooses an
information set $\calS\subseteq H$ of $\calV$, whose systematic basis codewords $F_z$ give
independent cohomology classes $[\Phi_h(F_z)]$ in each block $h$.

\subsection{Line decompositions and product expansion}
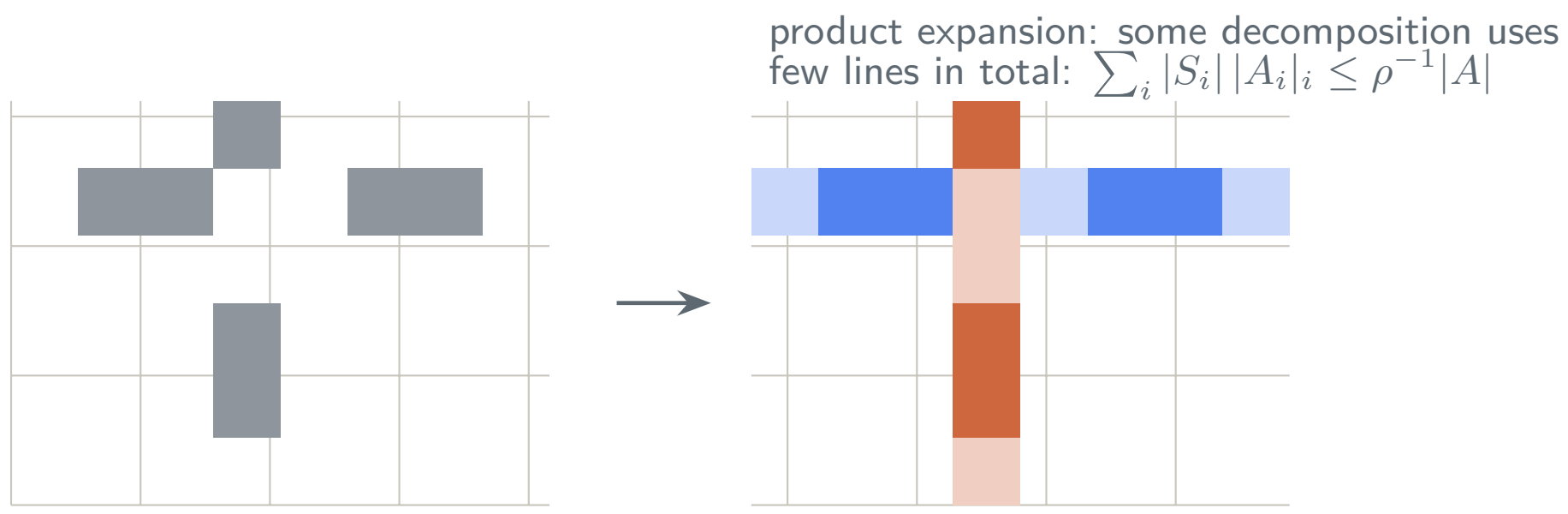
\begin{figure}[htbp]
\centering
\begin{tikzpicture}[x=0.52cm,y=0.52cm]
\draw[linec] (0,0) grid (8,6);
\foreach \x/\y in {1/4,2/4,5/4,6/4, 3/1,3/2,3/5}
  \fill[mutedc!70] (\x,\y) rectangle ++(1,1);
\node[sml] at (4,-0.9) {an array $A\in \mathcal Q_1+\mathcal Q_2$};
\node[sml] at (4,-1.6) {weight $|A|=7$};
\draw[flow] (9.0,3) -- (10.4,3);
\draw[linec] (11,0) grid (19,6);
\fill[dira!25] (11,4) rectangle (19,5);
\foreach \x in {12,13,16,17} \fill[dira!80] (\x,4) rectangle ++(1,1);
\fill[dirb!25] (14,0) rectangle (15,6);
\foreach \y in {1,2,5} \fill[dirb!80] (14,\y) rectangle ++(1,1);
\node[sml] at (15,-0.9) {$A=A_1+A_2$: one row line of $C_1$ (blue),};
\node[sml] at (15,-1.6) {one column line of $C_2$ (orange)};
\node[sml,anchor=west] at (11,7.0) {product expansion: some decomposition uses};
\node[sml,anchor=west] at (11,6.4) {few lines in total: $\sum_i|S_i|\,|A_i|_i\le\rho^{-1}|A|$};
\end{tikzpicture}
\caption{A decomposition into codewords supported on coordinate lines,
shown through its support. The blue row represents a codeword of $C_1$ and the orange
column a codeword of $C_2$. Their seven displayed nonzero entries have
disjoint supports, so their sum has weight seven and uses two active
lines. In general, product expansion guarantees a decomposition whose
total weighted number of active lines satisfies \eqref{eq:pe}, even
when other decompositions contain cancellations.}
\label{fig:pe}
\end{figure}

Product expansion, defined in \eqref{eq:pe}, bounds the number of
nonzero coordinate lines needed to represent an array. In two
directions these are rows and columns. Figure~\ref{fig:pe} illustrates
a line decomposition; in general the decomposition supplied by
product expansion may differ from the one in which the array was
originally presented, because line codewords can cancel.

%% file: sections/appendix_fourier_transform.tex
\section{Fourier decomposition and character twists}
\label{app:fourierx}

The proof below establishes the Fourier identities used to construct
the normalized functionals in \cref{cg:dual-existence-sec}. We then
interpret the component matrices as graph coboundaries with rescaled
restriction maps. This interpretation accounts for the signs of the
character indices in the sheet functions and in the Fourier blocks.

\subsection{Proof of the Fourier identities}

The coordinate formulas corresponding to \cref{pre:fourier}(i)--(iii)
are Fourier inversion, the two orthogonality identities, and the
idempotent relations. Write $\mathfrak r=\sum_xa_x\gb x\in R$:
\begin{equation}\label{pre:fourier-inverse}
 a_x=Q^{-1}\sum_{\alpha\in\widehat H}
               \widehat{\mathfrak r}(\alpha)\chi_\alpha(x)^{-1}.
\end{equation}
\begin{equation}\label{pre:character-orthogonality}
 \begin{aligned}
 \sum_{x\in H}\chi_\alpha(x)\chi_\beta(x)^{-1}
       &=Q\,\one_{\alpha=\beta},\\
 \sum_{\alpha\in\widehat H}\chi_\alpha(x)^{-1}\chi_\alpha(y)
       &=Q\,\one_{x=y}.
 \end{aligned}
\end{equation}
\begin{equation}\label{pre:idempotent-identities}
 \begin{gathered}
 e_\alpha e_\beta=\one_{\alpha=\beta}e_\alpha,
 \qquad \sum_\alpha e_\alpha=1_R,\\
 \gb g\,e_\alpha=\chi_\alpha(g)e_\alpha,
 \qquad \mathfrak r e_\alpha=\widehat{\mathfrak r}(\alpha)e_\alpha.
 \end{gathered}
\end{equation}
For an $R$-linear map $\Psi:R^a\to R^b$, write
$\widehat y(\alpha)\in F^b$ for the componentwise Fourier
transform of $y\in R^b$. Part~(iv) gives
\begin{equation}\label{pre:fourier-kernel-image}
 \begin{aligned}
 \ker\Psi
   &=\bigoplus_\alpha\bigl(\ker\widehat\Psi(\alpha)\bigr)e_\alpha,\\
 y\in\operatorname{im}\Psi
   &\quad\Longleftrightarrow\quad
   \widehat y(\alpha)\in\operatorname{im}\widehat\Psi(\alpha)
       \text{ for every }\alpha.
 \end{aligned}
\end{equation}

\begin{proof}[Proof of \cref{pre:fourier}]
The splitting field hypothesis implies that $\widehat H$ has
$Q$ elements and separates points: decompose $H$ into cyclic
factors and use the roots of unity in $F$ for each factor.
First prove (ii). If a character $\eta$ is nontrivial, choose
$h\in H$ with $\eta(h)\ne1$. Reindexing $x\mapsto hx$ gives
$\sum_{x\in H}\eta(x)=\eta(h)\sum_{x\in H}\eta(x)$,
so $\sum_{x\in H}\eta(x)=0$.
Apply this to $\eta=\chi_\alpha\chi_\beta^{-1}$ when
$\alpha\ne\beta$; when $\alpha=\beta$, every summand is one.
For the second orthogonality relation, if $x\ne y$, characters
separate points, so choose $\beta$ with
$\chi_\beta(x^{-1}y)\ne1$. Reindexing the sum by
$\alpha\mapsto\alpha+\beta$ forces it to vanish. If $x=y$,
all $Q$ summands are one.

The map in (i) preserves multiplication because
$\chi_\alpha(xy)=\chi_\alpha(x)\chi_\alpha(y)$.
Substitution of \eqref{pre:fourier-transform} into
\eqref{pre:fourier-inverse}, followed by the second relation in
\eqref{pre:character-orthogonality}, recovers $a_x$. This proves
that $\Phi$ is an isomorphism with the stated inverse.

For (iii), the first orthogonality relation gives
$\widehat e_\alpha(\beta)=\one_{\alpha=\beta}$. The identities in
\eqref{pre:idempotent-identities} now follow by componentwise
multiplication under $\Phi$.
For (iv), use $R$-linearity and the identity
$\gb x\,e_\alpha=\chi_\alpha(x)e_\alpha$ to obtain
\begin{equation}
 \Psi(v e_\alpha)
 =\sum_x\Psi_xv\,\gb x\,e_\alpha
 =\left(\sum_x\Psi_x\chi_\alpha(x)\right)v e_\alpha.
\end{equation}
The direct sums $R^a=\bigoplus_\alpha F^a e_\alpha$ and
$R^b=\bigoplus_\alpha F^b e_\alpha$ then give
\eqref{pre:fourier-kernel-image}.
\end{proof}

\begin{remark}[Orthogonality and resolution of the identity]
\label{appB:orth}\label{appB:resid}
The two identities in \eqref{pre:character-orthogonality} express
orthogonality with inverse character values over $F$. The corresponding
resolution of the identity is
\begin{equation}
 \mathfrak r=\sum_{\alpha\in\widehat H}\widehat{\mathfrak r}(\alpha)e_\alpha
 \qquad(\mathfrak r\in R),
\end{equation}
obtained by multiplying $\sum_\alpha e_\alpha=1_R$ by $\mathfrak r$ and
using \eqref{pre:idempotent-identities}. These are the identities
used in the bra--ket formulation of \cref{app:bloch}.
\end{remark}

\begin{example}[{$F=\F_4$, $H=\Z/3$}]\label{appB:f4}
Write $\F_4=\{0,1,\omega,\omega^2\}$ with $\omega^2=\omega+1$, so
$\mu_3=\{1,\omega,\omega^2\}$. The three characters of $\Z/3$ are
$\chi_k(j)=\omega^{kj}$, and the idempotents of $\F_4[\Z/3]$ are (the
$1/3$ prefactor equals $1$)
\begin{equation}
e_0=\gb{0}+\gb{1}+\gb{2},\quad
e_1=\gb{0}+\omega^2\gb{1}+\omega\gb{2},\quad
e_2=\gb{0}+\omega\gb{1}+\omega^2\gb{2}.
\end{equation}
One checks directly that $e_ke_l=\one_{k=l}e_k$, that
$e_0+e_1+e_2=\gb{0}=1_R$ (coefficient of $\gb{0}$: $1+1+1=3=1$; of $\gb{1}$ and
$\gb{2}$: $1+\omega+\omega^2=0$), and that $\gb{1}e_k=\omega^ke_k$. A circulant
$\mathfrak r=c_0\gb{0}+c_1\gb{1}+c_2\gb{2}$ acts on $Fe_k$ by
$\widehat{\mathfrak r}(k)=c_0+c_1\omega^k+c_2\omega^{2k}$. Thus the Fourier basis
diagonalizes convolution in this example just as it diagonalizes a
complex circulant matrix.
\end{example}

The sheet functions used as codewords have a useful description
in the idempotent basis:
\begin{equation}\label{pre:character-array}
 \gb{\chi_\alpha}
 =\sum_{x\in H}\chi_\alpha(x)\gb x
 =Q e_{-\alpha}=e_{-\alpha}.
\end{equation}
Indeed, \eqref{pre:group-translation} gives
\begin{equation}\label{pre:character-sheet-action}
 (\gb g\cdot\chi_\alpha)(x)
 =\chi_\alpha(g^{-1}x)
 =\chi_\alpha(g)^{-1}\chi_\alpha(x).
\end{equation}
Thus the sheet function $\chi_\alpha$ has eigencharacter
$\chi_{-\alpha}$. A right translation $x\mapsto\chi_\alpha(xg)$
has the scalar factor $\chi_\alpha(g)$; specifying the action
fixes this indexing convention.

\subsection{Character twists of graph coefficient systems}
Let $\pi:\tilde\Gamma\to\Gamma$ be a regular graph cover with
finite abelian deck group $H$. Choose $H$-equivariant sheet labels
on the vertex and edge fibers. For each incidence $v\in\partial e$, let
$g_{v,e}\in H$ be determined by the incidence rule
\begin{equation}\label{pre:incidence-sheets}
 (e,x)\text{ meets }(v,x/g_{v,e}).
\end{equation}
For a directed edge $e=(v,v')$, the usual edge voltage sending
tail sheets to head sheets is $g_{v,e}g_{v',e}^{-1}$.
The pullback $\pi^*\shf$ uses the base stalk and restriction map
at every lifted cell and incidence. Its direct image
$\pi_*\pi^*\shf$ has stalk
$\bigoplus_{x\in H}\shf_s$ at a base cell $s$; its elements are
arrays of coefficients indexed by the sheets.

For coefficient systems on the same graph, the stalkwise tensor
product is defined by
\begin{equation}\label{pre:stalkwise-tensor}
 (\shf\otimes\shf')_s=\shf_s\otimes\shf'_s,
 \qquad
 (\shf\otimes\shf')_{v\to e}
   =\shf_{v\to e}\otimes\shf'_{v\to e}.
\end{equation}
When both edge stalks are $F$, identify $F\otimes F$ with $F$
by $a\otimes b\mapsto ab$. This tensor product is defined on the
same base graph; the external product $\shf\boxtimes\shf'$ of
\cref{pre:crosssec} is defined on a Cartesian product of graphs.
A stalkwise tensor product of local code systems can have a
noninjective combined vertex restriction map. For example, the
nonzero tensor $U\otimes1-1\otimes U$ evaluates to zero on
every edge when both polynomial stalks contain $1$ and $U$.
Tensoring with the following rank-one system preserves injectivity.

\begin{definition}[Character twists]\label{pre:character-twist}
For $\alpha\in\widehat H$, define $L_\alpha$ to have stalk $F$
at every vertex and edge, with restriction maps
\begin{equation}\label{pre:local-system-restriction}
 (L_\alpha)_{v\to e}:F\longrightarrow F,
 \qquad a\longmapsto\chi_\alpha(g_{v,e})a.
\end{equation}
These maps are isomorphisms, so $L_\alpha$ is a rank-one
local system. Under the identifications $\shf_s\otimes F\cong\shf_s$,
the twisted coefficient system has restriction maps
\begin{equation}\label{pre:twisted-restriction}
 (\shf\otimes L_\alpha)_{v\to e}(a)
 =\chi_\alpha(g_{v,e})\shf_{v\to e}(a).
\end{equation}
The stalk dimensions are unchanged; the maps are rescaled
by nonzero incidence-dependent factors. In particular, the combined
vertex restriction map remains injective whenever it was injective for
$\shf$.
\end{definition}

\begin{proposition}[Rank-one local systems and Fourier blocks]
\label{pre:localsys}
Assume the finite abelian deck group $H$ has odd order and its
exponent divides $|F|-1$.
For a system of local codes $\shf$ on $\Gamma$, there are
isomorphisms
\begin{equation}\label{pre:cover-fourier}
 \begin{aligned}
 \pi_*\pi^*\shf
   &\cong\bigoplus_{\alpha\in\widehat H}\shf\otimes L_\alpha,\\
 H^\bullet(\tilde\Gamma;\pi^*\shf)
   &\cong\bigoplus_{\alpha\in\widehat H}
             H^\bullet(\Gamma;\shf\otimes L_\alpha).
 \end{aligned}
\end{equation}
Also $L_\alpha\otimes L_\beta\cong L_{\alpha+\beta}$.
The summand indexed by $\alpha$ is the eigenspace on which
$\gb g$ acts as $\chi_\alpha(g)$.
\end{proposition}
\begin{proof}
On each direct image stalk, let $\pi_\alpha$ be the action of the
idempotent $e_\alpha$:
\begin{equation}\label{pre:character-projector}
 \pi_\alpha=Q^{-1}\sum_{g\in H}\chi_\alpha(g)^{-1}\gb g.
\end{equation}
By \cref{pre:fourier}(iii), these operators satisfy
\begin{equation}\label{pre:character-projections}
 \pi_\alpha\pi_\beta=\one_{\alpha=\beta}\pi_\alpha,
 \qquad \sum_\alpha\pi_\alpha=I.
\end{equation}
Their images are the character eigenspaces. At a base cell $s$,
the $\alpha$-eigenspace consists of the arrays
$x\mapsto\chi_\alpha(x)^{-1}a_s$, with $a_s\in\shf_s$.
For the incidence in \eqref{pre:incidence-sheets}, restricting
such an array to the lifted edge gives
\begin{equation}\label{pre:twist-calculation}
 \begin{aligned}
 \shf_{v\to e}\bigl(\chi_\alpha(x/g_{v,e})^{-1}a_v\bigr)
 &=\chi_\alpha(x)^{-1}\chi_\alpha(g_{v,e})
                           \shf_{v\to e}(a_v).
 \end{aligned}
\end{equation}
Thus the coefficient of the common sheet pattern
$\chi_\alpha(x)^{-1}$ transforms by
\eqref{pre:twisted-restriction}, proving the first isomorphism.
The projectors commute with the coboundary, because the sheet
action does. The cochain complex therefore splits into these
summands, giving the cohomology decomposition.
Finally, the restriction map of $L_\alpha\otimes L_\beta$
has scalar factor
$\chi_\alpha(g_{v,e})\chi_\beta(g_{v,e})
=\chi_{\alpha+\beta}(g_{v,e})$, proving the last assertion.
\end{proof}

For the covers in \cref{gp:section}, $g_{v,e}=g_{i,v}(c(e))$
in graph factor $i$. Equation~\eqref{pre:twisted-restriction}
therefore gives exactly the finite field coboundary blocks used in
\cref{cg:dual-existence-sec}. By \eqref{pre:character-sheet-action}, the
graph cocycles used in \cref{cg:phidef} have sheet dependence
$\chi_\alpha$ and belong to the $L_{-\alpha}$ summands.

\begin{remark}[Block entries of equivariant maps]\label{appB:blocks}
An equivariant map decomposes into character blocks over either $F$ or
$\C$, with the coefficient field chosen for the calculation at hand.
The normalized adjacency blocks in \eqref{gp:block-action} use
complex-valued characters; their operator norms are bounded in
\cref{gp:spectrum}. The normalization $1/n$ in that formula is taken
in $\C$. The coboundary blocks $\widehat\delta^{\,0}_{i,W}(\alpha)$ in
\cref{cg:rank} instead use $F$-valued characters and are given by
\cref{pre:fourier}(iv). The normalized dual functional $\lambda_i$ is then
assembled from one $F$-valued row vector per block as in the proof of
\cref{cg:dual-construction}.
\end{remark}

\begin{remark}[Scalar field for the Fourier components]\label{appB:galois}
For $F=\F_q$ and $H=(F^\times)^t$, every character value already
lies in $F$, so the Fourier components require no extension of the
coefficient field. The alphabet $q$ is fixed before the covers grow;
the conversion to qubits in \cref{db:binary} therefore has constant
overhead.
\end{remark}

%% file: sections/appendix_bloch.tex
\section{Bloch theory and the logical gate form in Dirac notation}
\label{app:bloch}

The Fourier decomposition of our graph covers has the same algebraic
form as the Bloch decomposition of a periodic lattice. The base graph
plays the role of a unit cell, and the deck group $H$ acts by lattice
translations. This appendix expresses the dual functionals and logical
gate form in Dirac notation to explain this correspondence. The
pairings take values in a finite field; the state and measurement
language below describes an algebraic analogy.

\subsection{The crystal and its Brillouin zone}

Fix one graph factor $i$ and drop the index. The cover $\Gamma$ has
vertex copies $(v,z)$ and edge copies $(e,x)$ with sheet labels
$z,x\in H$: a crystal whose unit cell is the base multigraph and whose
lattice vector is the sheet label. Retain $H=(F^\times)^t$ and
$M=|F|-1$ from \cref{gp:construction}. The chosen generator
$\omega$ of $F^\times$ identifies the additive group
$(\Z/M\Z)^t$ with $H$ by
$\mathbf r\mapsto(\omega^{r_1},\ldots,\omega^{r_t})$.
Finiteness of $H$ corresponds to periodic (Born--von K\'arm\'an)
boundary conditions. The Brillouin zone is the finite dual group
$\hat H=\{\chi_\alpha:\alpha\in(\Z/M\Z)^t\}$: the index $\alpha$
plays the role of crystal momentum, and the Fourier blocks of
\cref{pre:localsys} are momentum sectors. The incidence labels
$g_v(c)$ specify how the edge and vertex sheets meet through
$(e,x)\leftrightarrow(v,x/g_v(c))$. For an edge oriented from $u$
to $v$, the voltage is $g_u(c)g_v(c)^{-1}$; its character value
is the corresponding translation phase. The stalks and restriction
maps are pulled back from the base graph, so the local constraints
are invariant under translations of the sheet labels.

The scalars form the finite field $F$ of characteristic two.
We use the symmetric $F$-bilinear pairing of the main text in place
of a Hermitian inner product. Transposition under this pairing
reverses the momentum index, as the Bloch basis below shows.
The normalization also simplifies:
$Q=|H|$ is odd, and an odd integer equals $1$ in characteristic two, so
$Q^{-1}=1$ in $F$ and the usual $1/\sqrt{Q}$ factors of Bloch theory
never appear. Since the characteristic does not divide $Q$,
the Fourier decomposition is nondegenerate (\cref{pre:fourier}).

\subsection{Wannier and Bloch bases in bra--ket notation}

The group algebra $R=F[H]$ is the finite field analogue of the
one-particle state space on the lattice. Define the \emph{Wannier basis} as the group basis in ket
notation,
\begin{equation}\label{appBl:wannier}
 \ket x:=\gb x\quad(x\in H),\qquad
 \braket{x}{y}=\one_{x=y},
\end{equation}
where the bracket is the symmetric $F$-bilinear form on $R$ declaring
the group basis orthonormal; the bra $\bra x$ is thus the coordinate
functional reading the coefficient of $\gb x$ --- that is,
\begin{equation}\label{appBl:evbra}
 \bra w=\ev_w
\end{equation}
as in \eqref{cg:coefficient-extraction}. A coefficient array
$\gb G=\sum_xG(x)\,\gb x$ becomes the position-space wavefunction
$\ket G=\sum_xG(x)\ket x$ with amplitudes $G(x)=\braket{x}{G}$.

Define the \emph{Bloch basis} by the idempotents of
\cref{pre:fourier},
\begin{equation}\label{appBl:bloch}
 \ket\alpha:=e_\alpha=\sum_{x\in H}\chi_\alpha(x)^{-1}\ket x,
 \qquad
 \bra\alpha:=\sum_{x\in H}\chi_\alpha(x)\bra x,
\end{equation}
(the prefactor $Q^{-1}$ of \cref{pre:fouriersec} equals $1$).
Here $\bra\alpha$ is the dual basis functional of $\ket\alpha$;
it is the transpose of $\ket{-\alpha}$ under the symmetric pairing.
The
character orthogonality of \cref{appB:orth} reads
$\braket{\alpha}{\beta}=\one_{\alpha=\beta}$, and the two resolutions of
the identity operator on $R$ are
\begin{equation}\label{appBl:resolutions}
 \sum_{x\in H}\ket x\bra x=\mathrm{id}_R
 =\sum_{\alpha}\ket\alpha\bra\alpha ,
\end{equation}
the second being the operator form of the ring identity
$1_R=\sum_\alpha e_\alpha$ (\cref{appB:resid}). As in physics, bra and
ket at the same momentum carry mutually inverse phases; here the
inverse phase replaces the complex conjugate.

Multiplication by $\gb g$ translates the group basis:
\begin{equation}\label{appBl:blochthm}
 \gb g\ket x=\ket{gx},\qquad
 \gb g\ket\alpha=\chi_\alpha(g)\,\ket\alpha ,
\end{equation}
the second identity being $\gb g\,e_\alpha=\chi_\alpha(g)e_\alpha$ from
\cref{pre:fouriersec} --- Bloch's theorem, with $\chi_\alpha(g)$ in the
role of $e^{\mathrm ik\cdot g}$. On coefficient functions, this
multiplication sends $G(x)$ to $G(g^{-1}x)$; it therefore corresponds
to $T_{g^{-1}}$ in the translation convention of \cref{gp:spectrum}.
An $R$-linear map commutes with these translations. An $R$-linear
endomorphism of $R$ is consequently diagonal in the Bloch basis, while a map
between free $R$-modules has one matrix block per momentum
(\cref{appB:blocks}). For the coboundary $\delta^0$ on one graph,
the block at $\alpha$ is the coboundary of the twisted coefficient
system $W\otimes L_\alpha$ of \cref{pre:localsys}, with incidence
factors $\chi_\alpha(g_v(c))$. These matrices are defined over $F$.
The adjacency operators in \cref{gp:spectrum} have an analogous
decomposition using complex-valued characters; their operator norms
are bounded uniformly over the nontrivial characters. The two
coefficient fields serve different purposes, as explained in
\cref{appB:blocks}.

The group basis vector of a character, viewed as a function on $H$, is
itself a Bloch ket at the opposite momentum:
\begin{equation}\label{appBl:chiket}
 \ket{\chi_\alpha}
 =\sum_{x\in H}\chi_\alpha(x)\ket x
 =e_{-\alpha}
 =\ket{-\alpha},
\end{equation}
because $Q^{-1}=1$. Thus the accepted characters of
\cref{gp:accepted-characters} span the selected momentum sectors
that form $\calV$.

\subsection{Pairings and scalar functionals in Dirac notation}

In the convention of \cref{pre:kron}, a chain is a functional on
cochains, and the pairing is written with the chain on the left. In Dirac's 
bra--ket language: \emph{cochains are kets, chains are bras}, and the
Kronecker pairing is the bracket, in the same order,
\begin{equation}\label{appBl:kronbracket}
 \braket{\lambda}{c}:=\inner{\lambda}{c}
 \qquad(\text{$\lambda$ a chain, $c$ a cochain}).
\end{equation}
A cycle is a bra annihilating the coboundary kets, and descent to
(co)homology is the statement that the bracket of a cycle-bra with a
cocycle-ket depends only on the two classes. The normalization of the
dual functionals (\cref{cg:lambda-sec}) becomes
$\braket{\lambda_i}{\bar b_i r}=r$: an $R$-valued bracket, $R$-valuedness
being translation equivariance of the bra. Its scalar components are
obtained by composing with position bras, and the per-address cycles of
\cref{cg:dualfamily} form a Wannier family,
$\bra{\Lambda_w}=\bra w\circ\Lambda$, the $H$-translates of
$\bra{\Lambda_{1_H}}$.

\paragraph{From the group algebra matrix to scalar dual chains.}
For the coefficient system $W_i$ of \eqref{cg:target-system}, the vertex
stalk is $F[U_i]_{<u}$, the edge stalk is $F$, and the restriction
at a color-$c$ incidence evaluates the polynomial at $c$ and multiplies
by $\gamma_{i,v}(c)\sigma_{i,v}(c)^3$.
Thus, for a color-$c$ edge $e=(v,v')$, a sheet $x\in H$, and a
0-cochain $z\in D_{i,W}^0$,
\begin{equation}\label{cg:cobound}
\begin{aligned}
 (\delta^0_{i,W}z)(e,x)
 ={}&\gamma_{i,v}(c)\sigma_{i,v}(c)^3
          z\big(v,x/g_{i,v}(c)\big)(c)\\
 &+\gamma_{i,v'}(c)\sigma_{i,v'}(c)^3
          z\big(v',x/g_{i,v'}(c)\big)(c).
\end{aligned}
\end{equation}
The matrix entry at the vertex monomial $U_i^\ell$ is
$\gamma_{i,v}(c)\sigma_{i,v}(c)^3c^\ell\gb{g_{i,v}(c)}$:
the scalar evaluates the monomial and the group element shifts
the sheet. If the entry of the row $\lambda_i$ at $e$ is
$\sum_y a_y\gb y$, then its scalar component
$\ev_{1_H}\circ\lambda_i$ has coefficient $a_{x^{-1}}$ at the
lifted edge $(e,x)$, since multiplication in $R$ gives
\begin{equation}\label{appBl:dual-sheet-coefficient}
 \ev_{1_H}\Big(\Big(\sum_{y\in H}a_y\gb y\Big)\gb x\Big)
 =a_{x^{-1}}.
\end{equation}

\paragraph{Scalar readouts.}
By \eqref{appBl:evbra}, an $F$-linear readout
$\ell=\sum_w\ell(\gb w)\,\ev_w$ is the bra
\begin{equation}\label{appBl:readout}
 \bra\ell=\sum_{w\in H}\ell(\gb w)\,\bra w,
 \qquad
 \ell(r)=\braket{\ell}{r};
\end{equation}
the subset readout $\ell_I$ is the sum of position bras
$\bra{\ell_I}=\sum_{w\in I}\bra w$ over the sites $I$, and the full-sum
readout is the zero-momentum bra,
$\bra{\ell_H}=\sum_w\bra w=\bra{\alpha{=}0}$.

The distinction between convolution and pointwise multiplication
also has a direct interpretation in this language. The ring product of $R$ is convolution: composition of
translations, pointwise in the Bloch basis
($e_\alpha e_\beta=\one_{\alpha=\beta}e_\alpha$). The pointwise product
of functions on $H$, which is what the trilinear form computes, is the
Hadamard product in the Wannier basis --- an interaction vertex
\emph{diagonal in position}. Summing the product uniformly over $H$
imposes momentum conservation, as expressed by the character identity
\eqref{cg:phasematch}: using \eqref{appBl:chiket} and
\eqref{appBl:readout},
\begin{equation}\label{appBl:phasematch}
 T_H\big(\chi_\alpha,\chi_\beta,\chi_\gamma\big)
 =\braket{\alpha{=}0}{\chi_\alpha\odot\chi_\beta\odot\chi_\gamma}
 =\braket{0}{-(\alpha+\beta+\gamma)}
 =\one_{\alpha+\beta+\gamma=0},
\end{equation}
three-wave mixing with momentum conservation --- the ``phase matching''
of nonlinear optics, whence the label of \eqref{cg:phasematch}.

The key identity (\cref{cg:keyidentity}) and the diagonal theorem
(\cref{cg:diagonal}) now take the following form.
For $F_1,F_2,F_3\in\mathcal V$, the cup product
of the cocycles $\Phi_h(F_h)$ is a ket whose bracket with the translation-equivariant bra
$\bra\Lambda$ is the position-space wavefunction of the pointwise
product, $\ket{F_1\odot F_2\odot F_3}$; composing with a readout bra,
\begin{equation}\label{appBl:key}
 T\big(\Phi_1(F_1),\Phi_2(F_2),\Phi_3(F_3)\big)
 =\braket{\ell}{F_1\odot F_2\odot F_3},
\end{equation}
and for the interpolation basis of \cref{cg:infoset-sec},
\begin{equation}\label{appBl:diag}
 T_I\big(\Phi_1(F_x),\Phi_2(F_y),\Phi_3(F_z)\big)
 =\sum_{w\in I}\braket{w}{F_x\odot F_y\odot F_z}
 =\sum_{w\in I}F_x(w)F_y(w)F_z(w)
 =\one_{x=y=z\in I}.
\end{equation}
The restrictions $F_z|_{\calS}$ are coordinate indicators, so
\eqref{appBl:diag} selects the logical triples indexed by $I$.
The functions may be nonzero outside $\calS$, and the cocycles
$\Phi_h(F_z)$ need not have disjoint physical supports
(\cref{cg:infoset-sec}). In the character basis, the full-sum
functional instead couples triples satisfying the phase matching
condition \eqref{appBl:phasematch}.
Gate selection (\cref{sec:addressable}) varies the bra:
the codes, the cocycles $\Phi_h(F_z)$, and the cup product are fixed,
and the subset $I$ enters through
$\bra{\Lambda_{\ell_I}}=\sum_{w\in I}\bra{\Lambda_w}$.

\subsection{Dual cycles and readouts in Dirac notation}
\label{appBl:tab1}

\Cref{tab:dictionary} collects the dual functionals and scalar cycles
of \cref{cg:lambda-sec}. In the notation of \eqref{appBl:kronbracket},
functional evaluation and chain pairing are both $\braket{\lambda}{u}$.
We derive these identities and the resulting trilinear form below.

\begin{table}[htbp]
\centering
\caption{The dual functionals and their scalar homology classes.
The maps $\ev_w$ extract sheet coefficients, and
$\langle\cdot,\cdot\rangle$ denotes Kronecker evaluation.}
\label{tab:dictionary}
{\small
\begin{tabular}{@{}ll@{}}
\toprule
Functional statement & Homological interpretation\\
\midrule
$\lambda_i\delta^0_{i,W}=0$ &
 $\ev_w\circ\lambda_i\in Z_1(\Gamma_i;W_i^\vee)$\\
$\lambda_i(\bar b_i a)=a$ &
 $\langle\lambda_i,[\bar b_i a]\rangle=a$\\
$\ev_w\lambda_i(\bar b_i\gb x)=\one_{w=x}$ &
 $\langle\ev_w\circ\lambda_i,[\bar b_i\gb x]\rangle=\one_{w=x}$\\
$\Lambda\delta_W^2=0$ &
 $\partial_3 \Lambda_w =0$\\
$\ell\circ\Lambda$ &
 $\Lambda_\ell=\sum_w\ell(\gb w)\Lambda_w$\\
$\ell_I=\sum_{w\in I}\ev_w$ &
 $\Lambda_{\ell_I}=\sum_{w\in I}\Lambda_w$\quad$(I\subseteq H)$\\
\bottomrule
\end{tabular}}
\end{table}

Throughout, $\bra w$ is the sheet bra \eqref{appBl:evbra} (the $\ev_w$
of the main text), and
$\bra w\lambda_i\ket u:=\bra w\big(\braket{\lambda_i}{u}\big)$ denotes
the scalar matrix element of an $R$-valued chain.

\paragraph{The cycle condition.} The functional statement
$\lambda_i\delta^0_{i,W}=0$ of \eqref{cg:lambda} says that the bra
$\bra{\lambda_i}$ annihilates every coboundary ket:
\begin{equation}\label{appBl:row1}
 \braket{\lambda_i}{\delta^0_{i,W}z}=0\qquad(\text{all }z).
\end{equation}
Since the boundary $\partial$ is the transpose of the coboundary
$\delta$ (\cref{pre:kron}), the adjoint identity
\eqref{pre:kronecker-adjoint}, $\langle\lambda,\delta a\rangle
=\langle\partial\lambda,a\rangle$, reads in brackets as
$\braket{\lambda}{\delta a}=\braket{\partial\lambda}{a}$: the
coboundary acts on kets as the boundary acts on bras. Vanishing of
\eqref{appBl:row1} for every $z$ is therefore equivalent to
$\partial_1\lambda_i=0$. Composing with sheet bras,
$\bra w\lambda_i\,\delta^0_{i,W}=0$: each scalar bra
$\bra w\lambda_i$ is a $1$-cycle, i.e.\ a dual Tanner codeword
(\cref{pre:dualtanner}) --- the pairing form of \cref{cg:eq19hom}.

\paragraph{Evaluation as a bracket.} The functional value
$\lambda_i(z)\in R$ and the $R$-valued pairing
$\langle \lambda_i,z\rangle$ of \cref{cg:eq19hom} are, by definition
\eqref{appBl:kronbracket}, the single object
$\braket{\lambda_i}{z}$. Its scalar components are the matrix
elements:
\begin{equation}\label{appBl:row2}
 \ev_w\big(\lambda_i(z)\big)
 =\bra w\lambda_i\ket z
 =\langle \ev_w\circ\lambda_i,\ z\rangle ,
\end{equation}
one equality reading the composite from the right (first evaluate the
chain, then take the $w$-coefficient), the other from the left (the
scalar chain $\bra w\lambda_i$ pairs with $z$); associativity of
composition is the entire content.

\paragraph{The normalization.} Writing the $R$-module action on
cochains inside the ket, $\ket{\bar b_ir}=r\ket{\bar b_i}$,
$R$-linearity of $\lambda_i$ moves the sheet array out of the
bracket:
\begin{equation}\label{appBl:row3}
 \lambda_i(\bar b_ir)
 =\braket{\lambda_i}{\bar b_ir}
 =r\,\braket{\lambda_i}{\bar b_i}
 =r\cdot1_R=r .
\end{equation}
By \eqref{appBl:row1} the bra $\bra{\lambda_i}$ is a cycle, so the bracket depends
only on the class, $\braket{\lambda_i}{\bar b_ir}
=\langle\lambda_i,[\bar b_ir]\rangle=r$. For the scalar version take
$r=\gb x$ and use Wannier orthonormality \eqref{appBl:wannier}:
\begin{equation}
 \bra w\lambda_i\ket{\bar b_i\gb x}
 =\braket{w}{x}
 =\one_{w=x},
\end{equation}
which is $\langle\ev_w\circ\lambda_i,[\bar b_i\gb x]\rangle=\one_{w=x}$
--- Eq.~\eqref{cg:scalar-normalization}.

\paragraph{Scalar top cycles.} A pure tensor in degree two has
one factor $z_i\in D_{i,W}^0$ and two factors
$z_j\in D_{j,W}^1$ for $j\ne i$. Since the graph complexes
have no degree two term, only the $i$th factor contributes to its
coboundary. The functional
$\Lambda=\lambda_1\otimes_R\lambda_2\otimes_R\lambda_3$
of \eqref{cg:Lambda} therefore satisfies
\begin{equation}\label{appBl:row4}
 \braket{\Lambda}{\delta_W^2(z_1\otimes_R z_2\otimes_R z_3)}
 =\lambda_i(\delta^0_{i,W}z_i)\prod_{j\ne i}\lambda_j(z_j)=0
\end{equation}
by \eqref{appBl:row1}. Such tensors span degree two, so
$\Lambda\delta_W^2=0$ and
$(\ell\circ\Lambda)\delta_W^2=0$ for every readout $\ell$.
Taking $\bra\ell=\bra w$ and
applying the adjoint identity as above,
$\bra w\Lambda\ket{\delta^2g}
=\braket{\partial_3(\bra w\Lambda)}{g}=0$ for all $g$, i.e.\
$\partial_3\Lambda_w=0$ for $\Lambda_w=\ev_w\circ\Lambda=\bra w\Lambda$:
the pairing form.

\paragraph{The readout cycle.} Insert the Wannier resolution
of the identity \eqref{appBl:resolutions} between the readout bra and
$\Lambda$:
\begin{equation}\label{appBl:row5}
 \bra\ell\Lambda
 =\bra\ell\Big(\sum_{w\in H}\ket w\bra w\Big)\Lambda
 =\sum_{w\in H}\braket{\ell}{w}\,\bra w\Lambda ,
\end{equation}
which is $\Lambda_\ell=\sum_w\ell(\gb w)\,\Lambda_w$ of
\eqref{cg:Lambdaell}, since $\braket{\ell}{w}=\ell(\gb w)$: the
expansion of a readout cycle in the Wannier family of
\cref{cg:dualfamily} is literally a resolution of the identity.

\paragraph{The trilinear form.} For cocycles $f_h$ in the three
input complexes, composing the preceding identities gives
\begin{equation}\label{appBl:row6}
 T(f_1,f_2,f_3)=(\ell\circ\Lambda)(f_1\smile f_2\smile f_3)
 =\bra\ell\Lambda\ket{f_1\smile f_2\smile f_3}
 =\braket{\Lambda_\ell}{f_1\smile f_2\smile f_3}.
\end{equation}
By \eqref{appBl:row4} the bra $\bra{\Lambda_\ell}$ is a cycle, and by the
Leibniz rule a triple cup product with one coboundary input is itself
a top coboundary, so the bracket depends only on the three classes:
$T(f_1,f_2,f_3)=\langle\Lambda_\ell,[f_1]\smile[f_2]\smile[f_3]\rangle$ --- the
Kronecker form \eqref{cg:kronform}.

\paragraph{Subset readouts.} The sum of position bras
$\bra{\ell_I}=\sum_{w\in I}\bra w$ gives, by \eqref{appBl:row5} restricted to the
sites $I$, $\Lambda_{\ell_I}=\sum_{w\in I}\Lambda_w$; \eqref{appBl:row6} with
$\ell=\ell_I$ then gives
$T_I=\langle\Lambda_{\ell_I},\,\cdot\,\rangle$ --- the
subset selection form used in \cref{cg:diagonal}.

\subsection{The dictionary}

\begin{table}[htbp]
\centering
\small
\begin{tabular}{@{}p{0.47\textwidth}p{0.47\textwidth}@{}}
\toprule
\textbf{This manuscript} & \textbf{Bloch theory}\\
\midrule
base multigraph $\overline\Gamma_0$ &
unit cell\\
deck group $H=(F^\times)^t$, $Q=|H|=M^t$ &
lattice translation group (periodic boundary conditions)\\
sheet label $x\in H$ of a lifted cell &
lattice vector of the cell's crystal copy\\
character $\chi_\alpha$; Fourier block at $\alpha$ &
crystal momentum $k$; momentum sector of the Brillouin zone $\hat H$\\
group algebra $R=F[H]$ &
one-particle state space on the lattice\\
group basis $\gb x=\ket x$ &
Wannier state at cell $x$\\
idempotent $e_\alpha=\ket\alpha$ &
Bloch state at momentum $\alpha$\\
$R\cong\prod_\alpha F$ \ (\cref{pre:fourier}) &
Bloch decomposition; one band label per momentum\\
multiplication by $\gb g$ &
translation by $g$ of the position basis\\
$\gb g\,e_\alpha=\chi_\alpha(g)e_\alpha$ &
Bloch's theorem: translations act by momentum phases\\
$R$-linear map; $R$-valued pairing &
translation-invariant operator; equivariant linear functional\\
incidence label $g_v(c)$ &
translation from a vertex sheet to an incident edge sheet\\
block of $\delta^0$ at $\alpha$: twist $W\otimes L_\alpha$
(\cref{pre:localsys}) &
Bloch block of a translation-invariant map, with incidence phases
$\chi_\alpha(g_v(c))$\\
spectral bound for all $\alpha\neq0$ \ (\cref{gp:spectrum}) &
uniform spectral bound over the Brillouin zone away from $k=0$\\
$\ev_w=\bra w$; readout $\ell_I=\sum_{w\in I}\ev_w$ &
position bra at cell $w$; sum of position bras on $I$\\
full readout $\ell_H$ &
zero-momentum bra $\bra{k{=}0}$\\
ring product of $R$ \ / \ pointwise product on $F^H$ &
convolution (composition of translations) \ / \ position-diagonal
interaction vertex\\
$T_H(\chi_\alpha,\chi_\beta,\chi_\gamma)=\one_{\alpha+\beta+\gamma=0}$
\ \eqref{cg:phasematch} &
momentum conservation in three-wave mixing (phase matching)\\
interpolation basis $F_z$, $z\in \calS$ \ (\cref{cg:infoset-sec}) &
logical register indexed by $\calS$, with independently selectable triples\\
accepted characters $\chi_\alpha$, $\alpha\in\calA$ &
Bloch-localized logical register ($\ket{\chi_\alpha}=\ket{-\alpha}$)\\
per-address cycles $\Lambda_w=\ev_w\circ\Lambda$ \ (\cref{cg:dualfamily}) &
family of linear functionals obtained by translating $\bra{\Lambda_{1_H}}$\\
gate selection via $\Lambda_{\ell_I}$ \ (\cref{sec:addressable}) &
fixed cochains, classically chosen linear functional\\
balanced product: shared sheet $x=x_1x_2x_3$ &
relative translations of the three crystals gauged away; one
center-of-mass lattice coordinate\\
$Q$ odd; $Q^{-1}=1$ in $F$ &
no normalization factors; complete band decomposition (semisimplicity)\\
\bottomrule
\end{tabular}
\caption{The Bloch theory dictionary. Left: objects of the main text.
Right: their analogues for a periodic lattice. The character index
is $\alpha\in(\Z/M\Z)^t$. Brackets are $F$-bilinear; in the Bloch
basis, $\bra\alpha$ is the transpose of $\ket{-\alpha}$ under the
symmetric pairing. These are algebraic correspondences, with no
probabilistic interpretation of the finite field functionals.}
\label{tab:bloch}
\end{table}

The Fourier block arguments in \cref{cg:rank,gp:section,appB:blocks}
correspond to working at fixed crystal momentum, with estimates
uniform in the character index as $t$ grows. For the logical
construction, characters make the local code constraints explicit,
while the interpolation basis and the functionals $\ell_I$ allow
independent selection of logical triples. Gaussian elimination on
the character evaluation matrix gives this interpolation basis
(\cref{cg:infoset-sec}).

%% file: sections/appendix_balanced_product_cup_product.tex
\section{The cup product in the coinvariants model}
\label{app:avgx}

The construction in \cref{cg:section} uses tensor products over $F$
to define cocycles on the Cartesian product of the graphs and
compute their cup products. Tensoring the graph complexes over $R$
gives the balanced complex used to evaluate the resulting pairing.
Here we verify the comparison in coefficient coordinates.

\subsection{Transfer of cochains and dual chains}\label{app:transfer-proof}
\begin{proof}[Proof of \cref{pre:transferid}]
The coefficient systems on $\widetilde B$ are pulled back from $B$:
restriction maps at all lifts of an incidence agree. Each lifted
cell has the same incidences as its image, with multiplicities
retained, so pullback commutes with coboundaries. Summing the same
incidence equations over each fiber shows that $\tau$ commutes
with coboundaries. Transposing gives the chain map identity for
$\pi^!=\tau^{\mathsf T}$.

An invariant cochain is constant on the lifts of each cell.
Its common value is the unique quotient coefficient pulling back
to it. The same argument on dual stalks identifies the image of
$\pi^!$ with the invariant dual chains, using
\eqref{pre:norm-coordinates}.

Every fiber has $|K|=Q^2$ elements. Consequently,
\begin{equation}
 (\tau\pi^*a)_s=\sum_{\tilde s\in\pi^{-1}(s)}a_s=|K|a_s=a_s,
\end{equation}
since $|K|\cdot1_F=1$. Taking the adjoint gives
\begin{equation}
 \langle\pi^!z,\pi^*a\rangle_{\widetilde B}
 =\langle z,\tau\pi^*a\rangle_B=\langle z,a\rangle_B.
\end{equation}
Finally, pullback preserves the cup product by \cref{pre:descent}.
Applying this identity to the last display proves
\eqref{pre:upstairs-cup}.
\end{proof}

\subsection{Invariants, coinvariants, and averaging}

Retain $T_J^\bullet$, $K$, and $\pi:\widetilde B\to B$ from
\cref{cg:product-complexes,cg:quotient-models}. For every nonempty
$J\subseteq\{1,2,3\}$, identify $(T_J^\bullet)_K$ with
$\mathcal C_J^\bullet$. Under this identification, the orbit
projection $\tau$ sends a tensor product of graph cochains over
$F$ to the corresponding tensor product over $R$.

\begin{lemma}[Averaging map]\label{appC:avg}
For $v\in T_J^\bullet$, the map
$\mathrm{avg}:(T_J^\bullet)_K\to(T_J^\bullet)^K$ defined by
$\mathrm{avg}([v])=\sum_{g\in K}g\cdot v$ is well defined.
Its inverse is the restriction of $\tau$ to invariant cochains.
\end{lemma}
\begin{proof}
The averaging argument in \cref{pre:invcoinv} applies to the
$K$-action, since $|K|=Q^2$ is odd. Its quotient projection is
$\tau$ under \eqref{cg:quotient-models}.
\end{proof}

Under the coinvariants identification, $\mathrm{avg}$ is precisely
$\pi^*$. In particular, for $u_i\in D_{i,J}^{p_i}$ with
$p_i\in\{0,1\}$,
\begin{equation}\label{appC:orbit-sum}
 \pi^*(u_1\otimes_R u_2\otimes_R u_3)
 =\sum_{(g_1,g_2,g_3)\in K}
       (\gb{g_1}u_1)\otimes_F(\gb{g_2}u_2)
                          \otimes_F(\gb{g_3}u_3).
\end{equation}
Transferring a sheet translation between tensor factors permutes
the summands, so the formula respects the balancing relations.
Applying $\tau$ returns the balanced tensor because all $Q^2$
summands have the same image and $Q^2=1$ in $F$.
For disjoint nonempty block sets $J,J'$ and degrees $p,q\ge0$
with $p+q\le3$, the cup product in the coinvariants model is
\begin{equation}
 a\smile b=\tau\bigl(\mathrm{avg}(a)\smile\mathrm{avg}(b)\bigr),
 \qquad a\in\mathcal C_J^p,\quad b\in\mathcal C_{J'}^q.
\end{equation}

\subsection{Evaluating the pairing on the Cartesian product}

The scalar transfer identity is \eqref{cg:pairing-upstairs}.
The next lemma expresses its $R$-valued counterpart through the
three graph functionals.

\begin{lemma}[Evaluation upstairs]\label{cg:upstairs}
Let $m_R:R\otimes_F R\otimes_F R\to R$ be multiplication in
the group algebra. For every $u\in\mathcal C_W^3$,
\begin{equation}
 \langle\Lambda,u\rangle
 =m_R\big((\lambda_1\otimes_F\lambda_2\otimes_F\lambda_3)
              (\pi^*u)\big).
\end{equation}
Interpreting the norm componentwise in $R$ gives, as maps
$T_{\{1,2,3\}}^3\to R$,
\begin{equation}\label{cg:normlambda}
 \pi^!\Lambda=\Lambda\circ\tau
 =m_R\circ(\lambda_1\otimes_F\lambda_2\otimes_F\lambda_3).
\end{equation}
\end{lemma}
\begin{proof}
For $u_i\in D_{i,W}^1$, the orbit projection sends
$u_1\otimes_F u_2\otimes_F u_3$ to
$u_1\otimes_R u_2\otimes_R u_3$. By the definition of $\Lambda$,
\begin{equation}
 (\Lambda\circ\tau)(u_1\otimes_F u_2\otimes_F u_3)
 =\prod_{i=1}^3\lambda_i(u_i).
\end{equation}
Pure tensors span the top cochain space, proving
\eqref{cg:normlambda}; the norm equality uses $\pi^!=\tau^{\mathsf T}$
from \cref{pre:pullback}. Composing with $\pi^*$ and using
$\tau\pi^*=\mathrm{id}$ gives the first identity.
\end{proof}

For character inputs $\alpha_1,\alpha_2,\alpha_3\in\mathcal A$,
put $\alpha=\alpha_1+\alpha_2+\alpha_3$. The graph factors
$A_i\in D_{i,W}^1$ in \eqref{cg:factorize} satisfy
$A_i=\bar b_i\gb{\chi_\alpha}$. Thus \cref{cg:upstairs} evaluates
their tensor product over $F$ as
$\prod_{i=1}^3\lambda_i(A_i)=\gb{\chi_\alpha}^{\,3}$.
Multiplication of these outputs in $R$ is convolution. The character
array is idempotent for this product: since $Q=1$ as a scalar in
$F$, for every $x\in H$,
\begin{equation}
 (\gb{\chi_\alpha}^{\,2})_x
 =\sum_{y\in H}\chi_\alpha(y)\chi_\alpha(y^{-1}x)
 =Q\chi_\alpha(x)=\chi_\alpha(x).
\end{equation}
Hence $\gb{\chi_\alpha}^{\,3}=\gb{\chi_\alpha}$, agreeing with
\eqref{cg:pairing} because the pointwise product of the three
input characters is $\chi_\alpha$.

\subsection{Translations of the scalar top cycles}

The scalar cycles defined in \eqref{cg:Lambda} have two further
properties used in the interpretation of \cref{app:bloch}.

\begin{proposition}[Independent translates of a top cycle]
\label{cg:dualfamily}\label{addr:family}
The cycles $\Lambda_w=\ev_w\circ\Lambda$, $w\in H$, are linearly
independent. Under the dual $H$-action, they satisfy
$g\cdot\Lambda_w=\Lambda_{gw}$ for all $g,w\in H$.
\end{proposition}
\begin{proof}
By \eqref{cg:top-normalization},
\begin{equation}
 \Lambda_w\big((\bar b_1\otimes_R\bar b_2\otimes_R\bar b_3)
                 \gb x\big)=\one_{w=x}.
\end{equation}
Evaluating a linear relation among the $\Lambda_w$ on these
cochains shows that every coefficient is zero.
For the translation statement, use the dual action
$(g\cdot z)(u)=z(g^{-1}\cdot u)$ and $R$-linearity of $\Lambda$:
\begin{equation}
 (g\cdot\Lambda_w)(u)
 =\ev_w(\gb{g^{-1}}\Lambda(u))
 =\ev_{gw}(\Lambda(u))
 =\Lambda_{gw}(u).
\end{equation}
\end{proof}